\documentclass[12pt]{article}
\usepackage{amsmath,amssymb,amsfonts,color,graphicx,cite,color}
\usepackage{latexsym}
\usepackage{epsfig,psfrag,rotating,soul}
\usepackage{rotfloat}

\input paperdef

\newcommand{\hbs}{\ensuremath{h \to \bar b s + b \bar s}}
\newcommand{\brhbs}{\ensuremath{\br(\hbs)}}

\graphicspath{{figs/}}

\evensidemargin \oddsidemargin
\allowdisplaybreaks

\begin{document}
\thispagestyle{empty}

\def\thefootnote{\fnsymbol{footnote}}

\begin{flushright}
\mbox{}
\end{flushright}

\vspace{0.5cm}

\begin{center}

\begin{large}
\textbf{The Quark Flavor Violating Higgs Decay \boldmath{\hbs}\\[.5em] 
in the MSSM}
\end{large}

\vspace{1cm}

{\sc
M.E.~G{\'o}mez$^{1}$%
\footnote{email: mario.gomez@dfa.uhu.es}%
, S.~Heinemeyer$^{2,3}$%
\footnote{email: Sven.Heinemeyer@cern.ch}%
~and M.~Rehman$^{2}$%
\footnote{email: rehman@ifca.unican.es}%
\footnote{MulitDark Scholar}
 
}

\vspace*{.7cm}

{\sl
$^1$ Department of Applied Physics, University of Huelva, 21071 Huelva, Spain

\vspace*{0.1cm}

\vspace*{0.1cm}
$^2$Instituto de F\'isica de Cantabria (CSIC-UC), E-39005 Santander, Spain

\vspace*{0.1cm}

$^3$Instituto de F\'isica Te\'orica, (UAM/CSIC), Universidad
  Aut\'onoma de Madrid,\\ Cantoblanco, E-28049 Madrid, Spain
}

\end{center}

\vspace*{0.1cm}

\begin{abstract}
\noindent

We study the quark flavor violating Higgs-boson decay \hbs\
in the Minimal Supersymmetric Standard Model (MSSM). The decay is
analyzed first in a model independent, and in a second step in the
minimal flavor violationg (MFV) Constrained MSSM. 
The experimental constraints from $B$-Physics observables (BPO) and
electroweak precision observables (EWPO) are also calculated and imposed
on the parameter space. It is shown that in some cases the EWPO restrict
the flavor violating parameter space stronger than the BPO.
In the model independent analysis values of \order{10^{-4}} can be found
for \brhbs. In the MFV CMSSM such results can only be obtained in very
restricted parts of the parameter space.
The results show that it is not excluded to observe the decay \hbs\ in
the MSSM at future $e^+e^-$ colliders.
\end{abstract}

\def\thefootnote{\arabic{footnote}}
\setcounter{page}{0}
\setcounter{footnote}{0}

\newpage



\section{Introduction}

Supersymmetry (SUSY) is one of the most intriguing ideas over the last
30 years of high energy physics. One of the major goals of the large
hadron collider (LHC) and future colliders is to discover SUSY (or any
other sign of physics beyond the Standard Model (SM)). So far this
search was unsuccessful for SUSY particles as for any other BSM model.
Another way to learn about SUSY is to study the indirect effects of the
SUSY particles on SM observables. Flavor Changing Neutral Current
(FCNC) processes offer a unique prospective in this regard. In the SM
FCNC processes are absent at tree level and can only occur
at one-loop level. The only source of FCNC's in the SM is the CKM matrix
and these processes are highly supressed due to GIM
cancellations~\cite{GIM}. On the other hand, in the Minimal
Supersymmetric Standard Model (MSSM)~\cite{mssm}, possible misalignment
between the quark and scalar quark mass matrices is another source which
can dominate the SM contribution by several orders of magnitude. Any
possible experimental deviation from the SM prediction for FCNS's would
be a clear evidence of new physics and potentially a hint for MSSM. 

Within the MSSM, flavor mixing can occur in the scalar fermion sector due
to the possible presence of soft SUSY-breaking parameters in the
respective mass matrices, which are off-diagonal in flavor space (mass
parameters as well as trilinear couplings). This yields many new sources
of flavor (and $\cp$-) violation, which potentially lead to 
large non-standard effects in flavor processes in conflict with
experimental bounds from low-energy flavor observables involving
strange, charm or bottom mesons~\cite{HFAgroup}.
An elegant way to solve the above problems (in general BSM models) is
provided by the Minimal Flavor Violation (MFV)
hypothesis~\cite{MFV1,MFV2}, where flavor (and $\cp$-) violation is
assumed to originate entirely from the CKM matrix. 
For example, in the MSSM the off-diagonality in the sfermion mass matrix
reflects the misalignment (in flavor space) between fermion and
sfermion mass matrices, that cannot be diagonalized simultaneously. 
One way to introduce this misalignment within the MSSM under the MFV
hypothesis is the following. Assuming no flavor violation at the Grand
Unification (GUT) scale, off-diagonal sfermion mass matrix entries can
be generated by Renormalization Group Equations (RGE) running to the
electroweak (EW) scale due to the presence of non-diagonal Yukawa
matrices in RGE’s.  
In this paper we will take into account both possibilities: the general
parametrization of flavor violation at the EW scale, as well as flavor
violation induced only by CKM effects in the RGE running from the GUT to
the EW scale.

MFV sceneraios are well motivated by the fact that low energy meson
physics puts tight constraints on the possible value of the FCNC
couplings, especially for the first and second generation squarks which
are sensitive to the data on $K^0-\bar{K}^0$ and $D^0-\bar{D}^0$
mixing. However, the third generation is less constrained, since present
data on $B^0-\bar{B}^0$ mixing still leaves some room for FCNCs. This
allows some parameter space for the more general scenerios focusing on 
the mixing between second and third generation (s)quarks.
One such example is the neutral higgs decay \hbs.
The SM contribution is highly suppressed
for this process but the SUSY-QCD quark-squark-gluino loop contribution
can enhance the MSSM contribuion by several orders of magnitude. Also the
SUSY-EW one loop contribution from quark-squark-chargino and
quark-squark-neutralino loop even though subdominent, can have sizable
effects on the \brhbs, where in particular the interfrence effects of
SUSY-QCD and SUSY-EW loop corrections can be relevant.
This decay in the framework of the MSSM has been analyzed in the
literature: the SUSY-QCD contributions for this decay were calculated 
in~\cite{HdecNMFV,SUSY-QCD}, and the SUSY-EW contributions using the mass
insertion approximation were calculated in~\cite{Demir}. Later
in~\cite{SUSY-EW} the SUSY-EW contributions and their interference
effects with the SUSY-QCD contribution were calculated using exact
diagonalization of the squark mass matrices. In all these
analysis, only LL mixing (see below for an exact definition) in the
squarks mass matrix was considered, and experimental constraints were
imposed only from \bsg. Most recently in~\cite{SUSY-EW-RR} also RR
mixing has been included. 
However mixing of the LR or RL elements 
of the mass matrix and constraints from other $B$-Physics observables
(BPO) or  potential other constraints were not taken into account 
(except in the most recent analysis in~\cite{SUSY-EW-RR}).

\smallskip
In this paper we will analyze the decay \hbs, evaluated at the full
one-loop level, by taking into account the experimental constraints
not only from $B$-Physics observable but also from the electroweak
precision observables (EWPO). In the scalar quark sector we will not only
consider the LL mixing, but also include the LR-RL and RR mixing for
our analysis of \brhbs. We will analyze this
decay first in a model independent approach (MI) where flavor mixing
parameters are put in by hand without any emphasis on the origin of this
mixing (but respecting the experimental bounds from BPO and EWPO).
In a second step we perform the analysis in the MFV Constrained MSSM
(CMSSM), where flavor mixing is generated by the
RGE running from GUT down to electroweak scale.

The paper is organized as follows: First we review the main features of
the MFV CMSSM and flavor mixing in the MSSM in \refse{sec:model_Setup}.  
The details about calculation and computational setup of the low energy
observables are given in \refse{sec:CompSetup}. The numerical results
are presented in \refse{sec:NResults}, where first we show the MI
analysis, followed by the results in MFV CMSSM. Our conclusions can be
found in \refse{sec:conclusions}.

\section{Model set-up}
\label{sec:model_Setup}

In this section we will first briefly review the MSSM and
parameterization of sfermion mixing at low energy. Subsequently, we will
give a brief recap of the CMSSM and the concept of MFV. 


\subsection{Flavor mixing in the MSSM}
\label{sec:FVMSSM}

In this section we give a brief description about how we parameterize
flavor mixing at the EW scale. We are using the same
notation as in \citeres{drhoLFV,arana-LFV,arana,arana-NMFV2,MFV-CMSSM}. 

\medskip
The most general hypothesis for flavor mixing assumes mass matrices for
the scalar quarks (we ignore flavor mixing in the slepton sector) that
are not diagonal in flavor space.
The superfields are rotated such that quark matrices are diagonal. The rotation is performed via 
the CKM matrix and the relavent terms in the soft SUSY-breaking Lagrangian (to be defined below)  get
rotated from the interaction eigenstate basis to what is known as the Super-CKM basis.

In the squarks sector we have two $6 \times 6$
mass matrices, based on the corresponding six Super-CKM
eigenstates,  
${\tilde U}_{L,R}$ with $U = u, c, t$ for up-type squarks and 
${\tilde D}_{L,R}$ with $D = d, s, b$ for down-type squarks.

The non-diagonal entries in these $6 \times 6$ general matrices for squarks
can be described in terms of a set of 
dimensionless parameters $\deFABij$ ($F=Q,U,D; A,B=L,R$; $i,j=1,2,3$, 
$i \neq j$) where  $F$ identifies the squark type, $L,R$ refer to the 
``left-'' and ``right-handed'' SUSY partners of the corresponding
fermionic degrees of freedom, and $i,j$ indexes run over the three
generations.  

One usually writes the $6\times 6$ non-diagonal mass matrices,  
${\mathcal M}_{\tilde u}^2$ and ${\mathcal M}_{\tilde d}^2$, referred to
the Super-CKM basis, being ordered respectively as $(\SupL, \SchaL,
\StopL, \SupR, \SchaR, \StopR)$,  $(\SdownL, \SstrL, \SbotL, \SdownR,
\SstrR, \SbotR)$ and write them in terms of left- and right-handed blocks
$M^2_{\tilde q \, AB}$ ($q=u,d$, $A,B=L,R$), which are non-diagonal
$3\times 3$ matrices,  
\begin{equation}
\cM_{\tilde q}^2 =\left( \begin{array}{cc}
M^2_{\tilde q \, LL} & M^2_{\tilde q \, LR} \\[.3em] 
M_{\tilde q \, LR}^{2 \, \dagger} & M^2_{\tilde q \,RR}
\end{array} \right), \qquad \tilde q= \tilde u, \tilde d~,
\label{eq:blocks-matrix}
\end{equation} 
 where:
 \begin{alignat}{5}
 M_{\tilde u \, LL \, ij}^2 
  = &  m_{\tilde U_L \, ij}^2 + \left( m_{u_i}^2
     + (T_3^u-Q_u\sw^2 ) M_Z^2 \cos 2\beta \right) \delta_{ij},  \notag\\
 M^2_{\tilde u \, RR \, ij}
  = &  m_{\tilde U_R \, ij}^2 + \left( m_{u_i}^2
     + Q_u\sw^2 M_Z^2 \cos 2\beta \right) \delta_{ij} \notag, \\
  M^2_{\tilde u \, LR \, ij}
  = &  \left< \cH_2^0 \right> {\cal A}_{ij}^u- m_{u_{i}} \mu \cot \beta \, \delta_{ij},
 \notag, \\
 M_{\tilde d \, LL \, ij}^2 
  = &  m_{\tilde D_L \, ij}^2 + \left( m_{d_i}^2
     + (T_3^d-Q_d \sw^2 ) M_Z^2 \cos 2\beta \right) \delta_{ij},  \notag\\
 M^2_{\tilde d \, RR \, ij}
  = &  m_{\tilde D_R \, ij}^2 + \left( m_{d_i}^2
     + Q_d\sw^2 M_Z^2 \cos 2\beta \right) \delta_{ij} \notag, \\
  M^2_{\tilde d \, LR \, ij}
  = &  \left< \cH_1^0 \right> {\cal A}_{ij}^d- m_{d_{i}} \mu \tb \, \delta_{ij}~,
\label{eq:SCKM-entries}
\end{alignat}
with, $i,j=1,2,3$, $Q_u=2/3$, $Q_d=-1/3$, $T_3^u=1/2$ and
$T_3^d=-1/2$. $M_{Z,W}$ denote the $Z$~and $W$~boson masses, with
$\sw^2 = 1 - \MW^2/\MZ^2$, 
and $(m_{u_1},m_{u_2}, m_{u_3})=(m_u,m_c,m_t)$, $(m_{d_1},m_{d_2},
m_{d_3})=(m_d,m_s,m_b)$ are the quark masses. $\mu$ is the
Higgsino mass term and $\tb = v_2/v_1$
with  $v_1=\left< \cH_1^0 \right>$ and $v_2=\left< \cH_2^0
\right>$ being the two vacuum expectation values of the corresponding
neutral Higgs boson in the Higgs $SU(2)_L$ doublets, 
$\cH_1= (\cH^0_1\,\,\, \cH^-_1)$ and $\cH_2= (\cH^+_2 \,\,\,\cH^0_2)$.

It should be noted that the non-diagonality in flavor comes
exclusively from the soft SUSY-breaking parameters, that could be
non-vanishing for $i \neq j$, namely: the masses  
$m_{\tilde U_L \, ij}^2$,
$m_{\tilde U_R \, ij}^2$, 
$m_{\tilde D_L \, ij}^2$, 
$m_{\tilde D_R \, ij}^2$ 
and the trilinear couplings, ${\cal A}_{ij}^q$.   
 
It is important to note that due to $SU(2)_L$ gauge invariance
the same soft masses $m_{\tilde Q \, ij}$ enter in both up-type and
down-type squarks mass matrices. 
The soft SUSY-breaking parameters for the up-type squarks differ from
corresponding ones for down-type squarks by a rotation with CKM
matrix. 
The sfermion mass matrices in terms of the $\deFABij$ are given as
\begin{equation}  
m^2_{\tilde U_L}= \left(\begin{array}{ccc}
 m^2_{\tilde Q_{1}} & \de_{12}^{QLL} m_{\tilde Q_{1}}m_{\tilde Q_{2}} & 
 \de_{13}^{QLL} m_{\tilde Q_{1}}m_{\tilde Q_{3}} \\
 \de_{21}^{QLL} m_{\tilde Q_{2}}m_{\tilde Q_{1}} & m^2_{\tilde Q_{2}}  & 
 \de_{23}^{QLL} m_{\tilde Q_{2}}m_{\tilde Q_{3}}\\
 \de_{31}^{QLL} m_{\tilde Q_{3}}m_{\tilde Q_{1}} & 
 \de_{32}^{QLL} m_{\tilde Q_{3}}m_{\tilde Q_{2}}& m^2_{\tilde Q_{3}} 
\end{array}\right)~,
\label{mUL}
\end{equation}
 
\noindent
\begin{equation}
m^2_{\tilde D_L}= V_{\rm CKM}^\dagger \, m^2_{\tilde U_L} \, V_{\rm CKM}~,
\label{mDL}
\end{equation}
 
\noindent 
\begin{equation}  
m^2_{\tilde U_R}= \left(\begin{array}{ccc}
 m^2_{\tilde U_{1}} & \de_{12}^{URR} m_{\tilde U_{1}}m_{\tilde U_{2}} & 
 \de_{13}^{URR} m_{\tilde U_{1}}m_{\tilde U_{3}}\\
 \de_{{21}}^{URR} m_{\tilde U_{2}}m_{\tilde U_{1}} & m^2_{\tilde U_{2}}  & 
 \de_{23}^{URR} m_{\tilde U_{2}}m_{\tilde U_{3}}\\
 \de_{{31}}^{URR}  m_{\tilde U_{3}} m_{\tilde U_{1}}& 
 \de_{{32}}^{URR} m_{\tilde U_{3}}m_{\tilde U_{2}}& m^2_{\tilde U_{3}} 
\end{array}\right)~,
\end{equation}

\noindent 
\begin{equation}  
m^2_{\tilde D_R}= \left(\begin{array}{ccc}
 m^2_{\tilde D_{1}} & \de_{12}^{DRR} m_{\tilde D_{1}}m_{\tilde D_{2}} & 
 \de_{13}^{DRR} m_{\tilde D_{1}}m_{\tilde D_{3}}\\
 \de_{{21}}^{DRR} m_{\tilde D_{2}}m_{\tilde D_{1}} & m^2_{\tilde D_{2}}  & 
 \de_{23}^{DRR} m_{\tilde D_{2}}m_{\tilde D_{3}}\\
 \de_{{31}}^{DRR}  m_{\tilde D_{3}} m_{\tilde D_{1}}& 
 \de_{{32}}^{DRR} m_{\tilde D_{3}}m_{\tilde D_{2}}& m^2_{\tilde D_{3}} 
\end{array}\right)~,
\end{equation}

\noindent 
\begin{equation}
v_2 {\cal A}^u  =\left(\begin{array}{ccc}
 m_u A_u & \de_{12}^{ULR} m_{\tilde Q_{1}}m_{\tilde U_{2}} & 
 \de_{13}^{ULR} m_{\tilde Q_{1}}m_{\tilde U_{3}}\\
 \de_{{21}}^{ULR}  m_{\tilde Q_{2}}m_{\tilde U_{1}} & 
 m_c A_c & \de_{23}^{ULR} m_{\tilde Q_{2}}m_{\tilde U_{3}}\\
 \de_{{31}}^{ULR}  m_{\tilde Q_{3}}m_{\tilde U_{1}} & 
 \de_{{32}}^{ULR} m_{\tilde Q_{3}} m_{\tilde U_{2}}& m_t A_t 
\end{array}\right)~,
\label{v2Au}
\end{equation}

\noindent 
\begin{equation}
v_1 {\cal A}^d  =\left(\begin{array}{ccc}
 m_d A_d & \de_{12}^{DLR} m_{\tilde Q_{1}}m_{\tilde D_{2}} & 
 \de_{13}^{DLR} m_{\tilde Q_{1}}m_{\tilde D_{3}}\\
 \de_{{21}}^{DLR}  m_{\tilde Q_{2}}m_{\tilde D_{1}} & m_s A_s & 
 \de_{23}^{DLR} m_{\tilde Q_{2}}m_{\tilde D_{3}}\\
 \de_{{31}}^{DLR}  m_{\tilde Q_{3}}m_{\tilde D_{1}} & 
 \de_{{32}}^{DLR} m_{\tilde Q_{3}} m_{\tilde D_{2}}& m_b A_b 
\end{array}\right)~.
\label{v1Ad}
\end{equation}

In all this work, for simplicity, we are assuming that all $\deFABij$
parameters are real, therefore, the hermiticity of 
$\cM_{\tilde q}^2$ implies $\del{FAB}{ij} = \del{FBA}{ji}$ 
and only the entries on and above the
diagonal need to be filled. The $\deFABij$ are located at the following 
places in the mass matrix:

$$
\left(\begin{array}{ccc|ccc}
\noent & \delta^{FLL}_{12} & \delta^{FLL}_{13} &
\noent & \delta^{FLR}_{12} & \delta^{FLR}_{13} \\
\noent & \noent & \delta^{FLL}_{23} &
\delta^{FRL*}_{12} & \noent & \delta^{FLR}_{23} \\
\noent & \noent & \noent &
\delta^{FRL*}_{13} & \delta^{FRL*}_{23} & \noent \\ \hline
\noent & \noent & \noent &
\noent & \delta^{FRR}_{12} & \delta^{FRR}_{13} \\
\noent & \noent & \noent &
\noent & \noent & \delta^{FRR}_{23} \\
\noent & \noent & \noent &
\noent & \noent & \noent
\end{array}\right)
$$

The next step is to rotate the squark states from the Super-CKM basis, 
${\tilde q}_{L,R}$, to the physical basis. 
If we set the order in the Super-CKM basis as above, 
$(\SupL, \SchaL, \StopL, \SupR, \SchaR, \StopR)$ and  
$(\SdownL, \SstrL, \SbotL, \SdownR, \SstrR, \SbotR)$, 
and in the physical basis as
${\tilde u}_{1,..6}$ and ${\tilde d}_{1,..6}$, respectively, these last
rotations are given by two $6 \times 6$ matrices, $R^{\tilde u}$ and
$R^{\tilde d}$,  
\BE
\VL  \tiu_{1} \\ \tiu_{2}  \\ \tiu_{3} \\
                                    \tiu_{4}   \\ \tiu_{5}  \\\tiu_{6}   \VR
  \; = \; R^{\tiu}  \VL \SupL \\ \SchaL \\\StopL \\ 
  \SupR \\ \SchaR \\ \StopR \VR ~,~~~~
\VL  \tid_{1} \\ \tid_{2}  \\  \tid_{3} \\
                                   \tid_{4}     \\ \tid_{5} \\ \tid_{6}  \VR             \; = \; R^{\tid}  \VL \SdownL \\ \SstrL \\ \SbotL \\
                                      \SdownR \\ \SstrR \\ \SbotR \VR ~,
\label{newsquarks}
\end{equation} 
yielding the diagonal mass-squared matrices for squarks as follows,
\BEA
{\rm diag}\{m_{\tiu_1}^2, m_{\tiu_2}^2, 
          m_{\tiu_3}^2, m_{\tiu_4}^2, m_{\tiu_5}^2, m_{\tiu_6}^2 
           \}  & = &
R^{\tiu}  \;  {\cal M}_{\tiu}^2   \; 
 R^{\tiu \dagger}    ~,\\
{\rm diag}\{m_{\tid_1}^2, m_{\tid_2}^2, 
          m_{\tid_3}^2, m_{\tid_4}^2, m_{\tid_5}^2, m_{\tid_6}^2 
          \}  & = &
R^{\tid}  \;   {\cal M}_{\tid}^2   \; 
 R^{\tid \dagger}    ~.
\EEA 


\subsection{The CMSSM and MFV}
\label{sec:cmssm}
 
The  MSSM is the simplest Supersymmetric structure we can build from the SM 
particle content. The general set-up for the soft SUSY-breaking
parameters is given by~\cite{mssm}\\ 
\begin{eqnarray}
\label{softbreaking}
-\cL_{\rm soft}&=&(m_{\tilde Q}^2)_i^j {\tilde {\cal Q}}^{\dagger i}
{\tilde {\cal Q}}_{j}
+(m_{\tilde U}^2)^i_j {\tilde {\cal U}}_{i}^* {\tilde {\cal U}}^j
+(m_{\tilde D}^2)^i_j {\tilde {\cal D}}_{i}^* {\tilde {\cal D}}^j
\nonumber \\
& &+(m_{\tilde L}^2)_i^j {\tilde {\cal L}}^{\dagger i}{\tilde {\cal L}}_{j}
+(m_{\tilde E}^2)^i_j {\tilde {\cal E}}_{i}^* {\tilde {\cal E}}^j
\nonumber \\
& &+m^2_{H_1}\cH_1^{\dagger} \cH_1
+m^2_{H_2}\cH_2^{\dagger} \cH_2
+(B \mu \cH_1 \cH_2
+ {\rm h.c.})
\nonumber \\
& &+ ( ({\bar A}^d)_{ij}\cH_1 {\tilde {\cal D}}_{i}^*{\tilde {\cal Q}}_{j}
+({\bar A}^u)_{ij}\cH_2 {\tilde {\cal U}}_{i}^*{\tilde {\cal Q}}_{j}
+({\bar A}^e)_{ij}\cH_1 {\tilde {\cal E}}_{i}^*{\tilde {\cal E}}_{j}
\nonumber \\
& & +\frac{1}{2}M_1 {\tilde B}_L^0 {\tilde B}_L^0
+\frac{1}{2}M_2 {\tilde W}_L^a {\tilde W}_L^a
+\frac{1}{2}M_3 {\tilde G}^a {\tilde G}^a + {\rm h.c.}).
\end{eqnarray}
Here we have used calligraphic capital letters for the sfermion fields in the
interaction basis with generation indices, 
\begin{eqnarray}
\tilde {\cal U}_{1,2,3}&=&\tilde u_R,\tilde c_R,\tilde t_R ;\quad 
\tilde {\cal D}_{1,2,3}=\tilde d_R,\tilde s_R,\tilde b_R ; \quad 
\tilde {\cal Q}_{1,2,3}=(\tilde u_L \, \tilde d_L)^T, (\tilde c_L\, \tilde s_L)^T, (\tilde t_L \, \tilde b_L)^T  \\
\tilde {\cal E}_{1,2,3}&=&\tilde e_R,\tilde \mu_R,\tilde \tau_R ; \quad  
\tilde {\cal L}_{1,2,3}=(\tilde \nu_{eL} \, \tilde e_L)^T, (\tilde \nu_{\mu L}\, \tilde \mu_L)^T, (\tilde \nu_{\tau L} \, \tilde \tau_L)^T
\end{eqnarray}
and all the gauge indices have been omitted.
Here, in accordance with \refse{sec:FVMSSM}, $m_{\tilde Q}^2$ and
$m_{\tilde L}^2$ are $3 \times 3$ 
matrices in family space (with $i,j$ being the
generation indeces) for the soft masses of the
left handed squark ${\tilde {\cal Q}}$ and slepton ${\tilde {\cal L}}$
$SU(2)$ doublets, respectively. $m_{\tilde U}^2$, $m_{\tilde D}^2$ and
$m_{\tilde E}^2$ contain the soft masses for right handed up-type squark
${\tilde {\cal U}}$,  down-type squarks ${\tilde {\cal D}}$ and charged
slepton ${\tilde {\cal E}}$ $SU(2)$ singlets, respectively. $\bar A^u$, 
$\bar A^d$ and $\bar A^e$ are the $3 \times 3$ matrices for the trilinear
couplings for up-type squarks, down-type 
squarks and charged slepton, respectively.
$m_{H_1}$ and $m_{H_2}$ contain the soft
masses of the Higgs sector. In the last line $M_1$, $M_2$ and $M_3$
define the bino, wino  and gluino mass terms, respectively, 
see \refeqs{v2Au}, (\ref{v1Ad}).

\medskip
Within the CMSSM the soft SUSY-breaking
parameters are assumed to be universal at the Grand Unification scale
$M_{\rm GUT} \sim 2 \times 10^{16} \gev$,
\begin{eqnarray}
\label{soft}
& (m_{\tilde Q}^2)_{i j} = (m_{\tilde U}^2)_{i j} 
 = (m_{\tilde D}^2)_{i j} = (m_{\tilde L}^2)_{i j}
 = (m_{\tilde E}^2)_{i j} = m_0^2\  \delta_{i j}, & \nonumber \\
& m_{H_1}^2 = m_{H_2}^2 = m_0^2, &\\
& m_{\tilde{g}}\ =\ m_{\tilde{W}}\ =\ m_{\tilde{B}}\ =\ m_{1/2}, &  
\nonumber \\
& (\bar A^u)_{i j}= A_0 e^{i \phi_A} (Y_U)_{i j},\ \ \ 
  (\bar A^d)_{i j}= A_0 e^{i \phi_A}
(Y_D)_{i j},\ \ \ 
(\bar A^e)_{i j}= A_0e^{i \phi_A} (Y_E)_{i j}. & \nonumber 
\end{eqnarray}
There is a common mass for all the scalars, $m_0^2$, a single gaugino 
mass, $m_{1/2}$, and all the trilinear soft-breaking terms are directly 
proportional to the corresponding Yukawa couplings in the superpotential 
with a proportionality constant $A_0 e^{i \phi_A}$, containing a
potential non-trivial complex phase.
The other phases can be redefined and
included in the phase of $\mu$ (for a review see for example
\cite{Ibrahim:2007fb}). However, they are very constrained by the electric
dipole moments(EDM's) of leptons and nucleons~\cite{Pospelov:2005pr}.

With the use of the Renormalization Group Equations (RGE) of the MSSM,
one can obtain the SUSY spectrum at the EW scale.
All the SUSY masses and mixings are then given as 
a function of $m_0^2$, $m_{1/2}$, $A_0$, and 
$\tb = v_2/v_1$, the ratio of the two vacuum expectation values (see
below). We require radiative symmetry breaking to fix $|\mu|$ and 
$|B \mu|$ \cite{rge,bertolini} with the tree--level Higgs potential.

By definition, this model fulfills the MFV hypothesis, since the only
flavor violating terms stem from the CKM matrix. 
The important point is that, even in a model with universal soft
SUSY-breaking terms at some high energy scale as the CMSSM, some
off-diagonality in the squark mass matrices appears at the EW scale. 
Working in the basis where the squarks are rotated parallel to the
quarks, the so-called  Super CKM (SCKM) basis, the squark mass
matrices are not flavor diagonal at the EW scale.
This is due to the fact that at $M_{\rm GUT}$ there exist two non-trivial 
flavor structures, namely the two Yukawa matrices for the up and down quarks, 
which are not simultaneously diagonalizable. This implies that 
through RGE evolution some flavor mixing leaks into the sfermion mass matrices.
In a general SUSY model the presence of new flavor structures
in the soft SUSY-breaking terms would generate large flavor mixing in
the sfermion mass matrices. However, in the CMSSM, the two Yukawa
matrices are the only source of flavor change. As always in the SCKM
basis, any off-diagonal entry  
in the sfermion mass matrices at the EW scale will be necessarily
proportional to a product of Yukawa couplings.

In \citere{MFV-CMSSM} it was shown that even under the MFV hypothesis in
the CMSSM non-negligible flavor violation effects can be induced at the
EW scale. Confronted with precision data from flavor observables or
electroweak precision observables, this can lead to important
restrictions of the CMSSM parameter space. These constraints will be
imposed on the SUSY parameters in our numerical analysis below. Details
about these observables and their calculation are given in the next
section. 

\section{Low-energy Observables}
\label{sec:CompSetup}

Here we briefly describe the calculations of the observables evaluated in this
work. We start with the evaluation of the flavor violating Higgs decay,
\hbs, and then give a short description of the precision observables
used to restrict the allowed parameter space.


\subsection{The flavor violating Higgs decay \boldmath{\hbs}}
\label{sec:hbs} 

\begin{figure}[ht!]
\begin{center}
\psfig{file=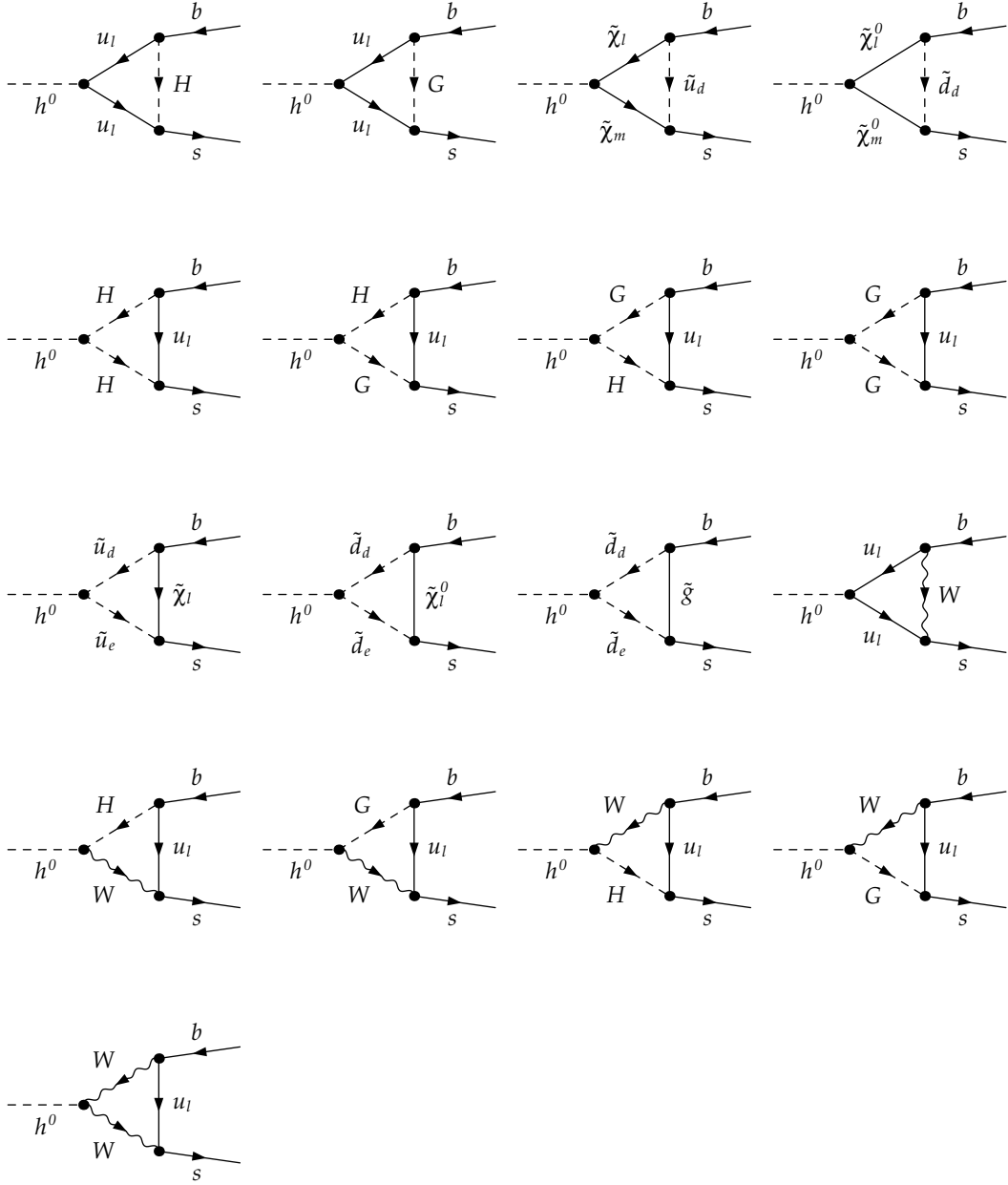}
\end{center}
\vspace{-2em}
\caption{Generic Feynman diagrams for the EW and QCD corrections to
  \hbs\ (vertex diagrams).}
\label{FD:FeynDiag1}
\end{figure}

\begin{figure}[ht!]
\begin{center}
\psfig{file=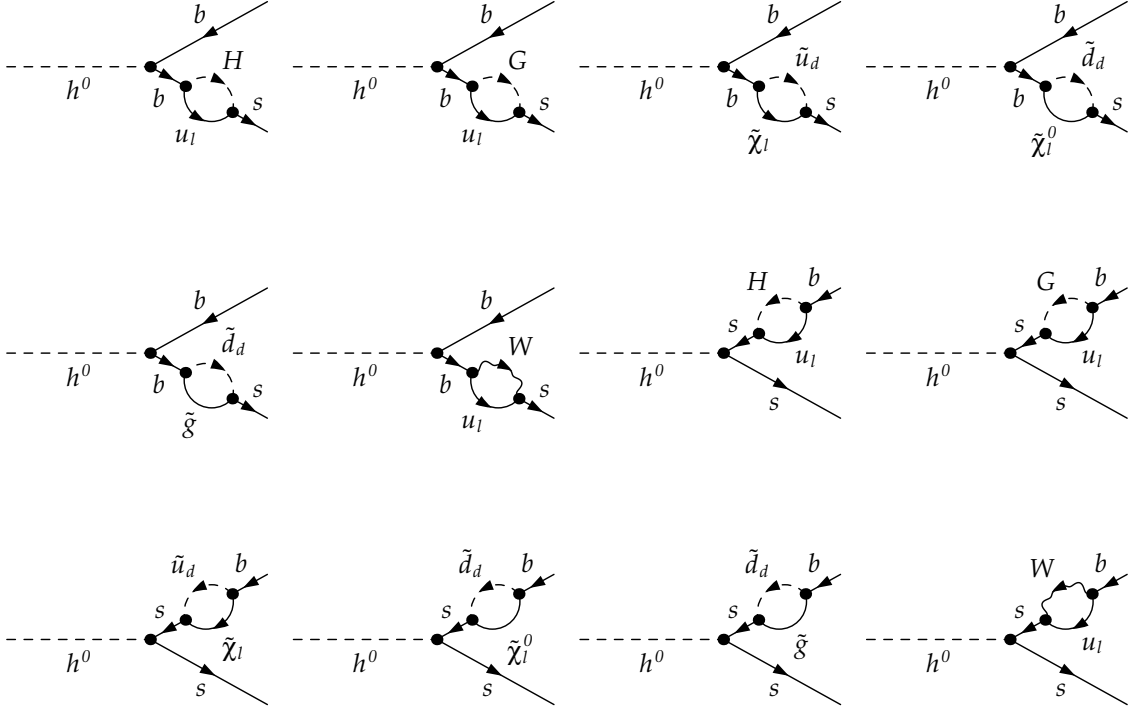}
\end{center}
\caption{Generic Feynman diagrams for the EW and QCD corrections to
  \hbs\ (self-energy contributions).
}
\label{FD:FeynDiag2}
\end{figure}

We start with the evaluation of the flavor violating Higgs decay. 
In SM the branching ratio \brhbs\ can be
at most of \order{10^{-7}}~\cite{HdecNMFV}, too small to have a
chance of detection at the LHC. But because of the strong FCNC gluino
couplings and the $\tb$-enhancement inherent to the MSSM Yukawa
couplings, we may expect several orders of magnitude increase of the
branching ratio as compared to the SM result, see 
\citere{HdecNMFV, SUSY-QCD}. We (re-)\,calculate full one-loop 
contributions from SUSY-QCD as well as SUSY-EW loops with the help of
the \fa~\cite{feynarts,famssm} and \fc~\cite{formcalc} packages.
The lengthy analytical results are not shown here. We take
into account mixing in the LL and RR part, as well as in the LR and
RL part of the mass matrix, contrary to 
\citeres{HdecNMFV,SUSY-QCD,Demir,SUSY-EW,SUSY-EW-RR}, where only 
the LL and RR mixing had been considred. For our numerical analysis we
define  
\begin{align}
\brhbs = \frac{\Ga(\hbs)}{\Ga_{h, {\rm tot}}^{\rm MSSM}}
\end{align}
where $\Gamma_{h, {\rm tot}}^{\rm MSSM}$ is the total decay width of the
light Higgs boson $h$ of the MSSM, as evaluated with
\fh~\cite{feynhiggs,mhiggslong,mhiggsAEC,mhcMSSMlong,Mh-logresum}. 
The contributing Feynman diagrams for the decay \hbs\ are shown in
\reffi{FD:FeynDiag1} (vertex corrections) and in \reffi{FD:FeynDiag2}
(self-energy corrections).

\medskip
Which BR might be detectable at the LHC or an $e^+e^-$ collider such as
the ILC can only be established by means of specific experimental
analyses, which, to our knowledge, do not exist yet. However, in the
literature it is expected to measure BR's at the level of $10^{-3}$ at
the LHC~\cite{HdecNMFV}.
In the clean ILC environment in general Higgs boson branching ratios
below the level of $10^{-4}$ can be observed, see
e.g.\ \citere{ILCreview} for a recent review. We will take this as a
rough guideline down to which level the decay \hbs\ could be observable.


\subsection{\boldmath{$B$}-physics observables}
\label{sec:bpo}

In order to determine which flavor mixing (i.e.\ which combination of
parameters) is still allowed by experimental data we evaluated flavor
precision observables and electroweak precision observables. Here we
start with the brief description of the evaluation of 
several $B$-physics observables (BPO):
\bsg, \bmm\ and \dmbs. 
Concerning \bsg\ included in the calculation are the most 
relevant loop contributions to the Wilson coefficients:
(i)~loops with Higgs bosons (including the resummation of large $\tb$
effects~\cite{Isidori:2002qe}),  
(ii)~loops with charginos and 
(iii)~loops with gluinos. 
For \bmm\ there are three types of relevant one-loop corrections
contributing to the relevant Wilson coefficients:
(i)~Box diagrams, 
(ii)~$Z$-penguin diagrams and 
(iii)~neutral Higgs boson $\phi$-penguin diagrams, where $\phi$ denotes the
three neutral MSSM Higgs bosons, $\phi = h, H, A$ (again large resummed
$\tb$ effects have been taken into account).
In our numerical evaluation there are included
what are known to be the dominant contributions to
these three types of diagrams \cite{Chankowski:2000ng}: chargino
contributions to box and $Z$-penguin diagrams and chargino and gluino
contributions to $\phi$-penguin diagrams.   
Concerning \dmbs, in the MSSM 
there are in general three types of one-loop diagrams that contribute:
(i)~Box diagrams, 
(ii)~$Z$-penguin diagrams and 
(iii)~double Higgs-penguin diagrams (again including the resummation of
large $\tb$ enhanced effects).
In our numerical evaluation there are included again what are known to be the
dominant contributions to these three types of diagrams in scenarios
with non-minimal flavor violation (for a review see, for instance,
\cite{Foster:2005wb}): gluino contributions to box 
diagrams, chargino contributions to box and $Z$-penguin diagrams, and 
chargino and gluino contributions to double $\phi$-penguin diagrams. 
More details about the calculations employed can be
found in \citeres{arana,arana-NMFV2}.
We perform our numerical calculation with the {\tt BPHYSICS} subroutine
taken from the {\tt SuFla} code~\cite{sufla} (with some additions and
improvements as detailed in \citeres{arana,arana-NMFV2}), which has
been implemented as a subroutine into (a private version of) \fh.
The experimental status and SM prediction of these observables
are given in the 
\refta{tab:ExpStatus-BPO}~\cite{hfag:rad,Misiak:2009nr,Chatrchyan:2013bka,Aaij:2013aka,Buras:2012ru,hfag:pdg,Buras:1990fn,Golowich:2011cx}%
\footnote{
Using the most up-to-date value of 
$\bmm = 2.9\pm 0.7 \times 10^{-9}$~\cite{bmm-CMS-LHCb}
would have had a minor impact on our analysis.
}

\begin{table}[htb!]
\BC
\renewcommand{\arraystretch}{1.5}
\begin{tabular}{|c|c|c|}
\hline
Observable & Experimental Value & SM Prediction  \\\hline
\bsg & $3.43\pm 0.22 \times 10^{-4}$ & $3.15\pm 0.23 \times 10^{-4}$ \\
\bmm & $(3.0)^{+1.0}_{-0.9} \times 10^{-9}$ & $3.23\pm 0.27 \times 10^{-9}$ \\
\dmbs & $116.4\pm 0.5 \times 10^{-10} \mev $ 
      & $(117.1)^{+17.2}_{-16.4} \times 10^{-10} \mev$   \\
\hline
\end{tabular}
\caption{Present experimental status of $B$-physics observables with
their SM prediction.}  
\label{tab:ExpStatus-BPO}
\renewcommand{\arraystretch}{1.0}
\EC
\end{table}


\subsection{Electroweak precision observables}
\label{sec:ewpo}

Electroweak precision observables (EWPO) that are known with an accuracy
at the per-mille level or better  
have the potential to allow a discrimination between quantum effects of 
the SM and SUSY models, see \citere{PomssmRep} for a review.  For example 
the $W$-boson mass $\MW$, whose present 
experimental value is \cite{LEPEWWG}
\begin{align}
\label{eq:EWPO-today}
\MW^{\text{exp,today}} &= 80.385 \pm  0.015 \gev~.
\end{align}
The experimental uncertanity will further be reduced~\cite{Baak:2013fwa} 
to
\begin{align}
\label{eq:EWPO-future}
\de\MW^{\text{exp,future}} \lsim 4\mev
\end{align}
at the ILC and at the GigaZ option of the ILC, respectively. 
Even higher precision could be expected from the FCC-ee, see, e.g.,
\citere{fcc-ee-paris}.

The $W$-boson mass can be evaluated from
\begin{align}
\MW^2 \KL 1 - \frac{\MW^2}{\MZ^2} \KR = 
\frac{\pi \al}{\wz \Gmu} (1 + \De r)
\end{align}
where $\al$ is the fine-structure constant and $\Gmu$ the Fermi 
constant.  This relation arises from comparing the prediction for muon 
decay with the experimentally precisely known Fermi constant.  
The one-loop contributions to $\De r$ can be written as
\begin{align}
\De r = \De\al - \frac{\cw^2}{\sw^2}\De\rho + (\De r)_{\text{rem}},
\end{align}
where $\De\al$ is the shift in the fine-structure constant due to the 
light fermions of the SM, $\De\al \propto \log(\MZ/m_f)$, and $\De\rho$ 
is the leading contribution to the $\rho$ parameter~\cite{rho} from 
(certain) fermion and sfermion loops (see below).  The remainder part 
$(\De r)_{\text{rem}}$ contains in particular the contributions from the 
Higgs sector.

\medskip
The SUSY contributions to $\MW$ can well be approximated with the
$\rho$-parameter approximation~\cite{PomssmRep,delrhoNMFV}. 
$\MW$ is affected by shifts in the 
quantity $\De\rho$ according to
\begin{align}
\label{eq:precobs}
\De\MW \approx \frac{\MW}{2}\frac{\cw^2}{\cw^2 - \sw^2} \De\rho\
\end{align}
The quantity $\De\rho$ is defined by the relation
\begin{align}
\De\rho = \frac{\Si_Z^{\text{T}}(0)}{\MZ^2} -
          \frac{\Si_W^{\text{T}}(0)}{\MW^2}
\label{eq:drho}
\end{align} 
with the unrenormalized transverse parts of the $Z$- and $W$-boson 
self-energies at zero momentum, $\Si_{Z,W}^{\text{T}}(0)$.  It 
represents the leading universal corrections to the electroweak 
precision observables induced by mass splitting between partners in 
isospin doublets~\cite{rho}. Consequently, it is sensitive to the 
mass-splitting effects induced by flavor mixing.
The effects from flavor violation in the squark and slepton sector,
entering via $\De\rho$ have been evaluated in
\citeres{delrhoNMFV,drhoLFV,MFV-CMSSM} and included in \fh. 
In particular, in \citere{delrhoNMFV} it has been shown that for the
squark contributions $\De\rho$ constitutes an excellent approximation to
$\De r$. We use \fh\ for our numerical evaluation, where $\MW$ is
evaluated as
\begin{align}
\MW^{\rm MSSM} &= \MW^{\SM} + \De\MW^{\rm MSSM}~,
\end{align}
where $\De\MW^{\rm MSSM}$ is calculated via
\refeq{eq:precobs}. \fh\ takes into account the full set of one-loop
squark contributions to $\De\rho$ (including NMFV 
effects~\cite{delrhoNMFV,drhoLFV}), as well as
the leading gluonic two-loop corrections~\cite{Drho2LQCD}. 
In \citere{MFV-CMSSM} it was shown that EWPO and in particular $\MW$ can
lead to relevant restrictions on the (C)MSSM parameter space in the
presence of intergenerational mixing in the squark sector.

The prediction of $\MW$ also suffers from various kinds of
theoretical uncertainties, parametric and intrinsic. Starting with the
parametric uncertainties, an 
experimental error of $1 \gev$ on $\mt$ yields a parametric uncertainty on
$\MW$ of about $6 \mev$, while the parametric uncertainties induced by the
current experimental error of the hadronic contribution to the shift in
the fine-structure constant, $\Delta\alpha_{\rm had}$, and by the
experimental error of $\MZ$ amount to about $2 \mev$ and $2.5 \mev$,
respectively. The uncertainty of the $\MW$ prediction caused by the
experimental uncertainty of the Higgs mass 
$\delta \Mh^{\rm exp} \lsim 0.3 \gev$ is
signifcantly smaller ($\approx 0.2 \mev$). The intrinsic 
uncertainties from unknown higher-order corrections in the case of no
flavor mixing have been estimated to be around (4.7-9.4)~MeV in the
MSSM \cite{Heinemeyer:2006px,Haestier:2005ja} depending on the SUSY mass
scale. We have added the parameteric uncertanities in quadrature and add
the result linearly to the uncertanity from 
the unknown higher order corrections in the case of no flavor mixing.
We assume additional 10\% uncertanity from the flavor mixing
contribution to $\De\rho^{\rm MSSM}$ and (via \refeq{eq:precobs}) 
add it linearly to the other uncertainties. 
\section{Numerical Results}
\label{sec:NResults}

In this section we present our numerical results. We start with the
model independent approach, where we do not specifiy the origin of the
flavor violating $\deFABij$, but take into account the existing limits
from BPO and (evaluate newly the ones from) EWPO. In a second step we
briefly investigate the results in the CMSSM.


\subsection{Model independent analysis}
\label{sec:NR_modelInd}

In the model independent analaysis we first define our set of input
paramters and discuss how they are restriced by BPO and EWPO introduced
above. In the allowed parameter space we evaluate \brhbs\ and show that
it might be detectable at future $e^+e^-$ colliders.


\subsubsection{Input Parameters} 

For the following numerical analysis we chose the MSSM parameter sets of 
\citere{drhoLFV, arana-LFV}.  This framework contains six specific 
points S1\dots S6 in the MSSM parameter space, all of which are 
compatible with present experimental data, including LHC searches and the  
measurements of the muon anomalous magnetic moment.  The values of the 
various MSSM parameters as well as the values of the predicted MSSM mass 
spectra are summarized in \refta{tab:spectra}.  They were evaluated with 
the program 
\fh~\cite{feynhiggs,mhiggslong,mhiggsAEC,mhcMSSMlong,Mh-logresum}.

For simplicity, and to reduce the number of independent MSSM input 
parameters, we assume equal soft masses for the sleptons of the first 
and second generations (similarly for the squarks), and for the left and 
right slepton sectors (similarly for the squarks).  We choose equal 
trilinear couplings for the stop and sbottom squarks and for the 
sleptons consider only the stau trilinear coupling; the others are set 
to zero.  We assume an approximate GUT relation for the gaugino 
soft-SUSY-breaking parameters.  The pseudoscalar Higgs mass $\MA$ and 
the $\mu$ parameter are taken as independent input parameters.  In 
summary, the six points S1\dots S6 are defined in terms of the 
following subset of ten input MSSM parameters:
\begin{align*}
&m_{\tilde L_1} = m_{\tilde L_2}\,, &
&m_{\tilde L_3}\,, &
&(\text{with~} m_{\tilde L_i} = m_{\tilde E_i},\ i = 1,2,3) \\
&m_{\tilde Q_1} = m_{\tilde Q_2} &
&m_{\tilde Q_3}\,, &
&(\text{with~} m_{\tilde Q_i} = m_{\tilde U_i} = m_{\tilde D_i},\ i = 1,2,3) \\
&A_t = A_b\,, &
&A_\tau\,, \\
&M_2 = 2 M_1 = M_3/4\,, &
&\mu\,, \\
&\MA\,, &
&\tb\,.
\end{align*}

The specific values of these ten MSSM parameters in \refta{tab:spectra} 
are chosen to provide different patterns in the various sparticle 
masses, but all leading to rather heavy spectra and thus naturally in 
agreement with the absence of SUSY signals at the LHC.  In particular, 
all points lead to rather heavy squarks of the first/second generation
and gluinos above $1500\gev$ and  
heavy sleptons above $500\gev$ (where the LHC limits would also permit 
substantially lighter sleptons).  The values of $\MA$ within the 
interval $(500,1500)\gev$, $\tb$ within the interval $(10,50)$ and a 
large $A_t$ within $(1000,2500)\gev$ are fixed such that a light Higgs 
boson $h$ within the LHC-favoured range $(123,127)\gev$ is obtained.

\begin{table}[htb!]
\centerline{\begin{tabular}{|c|c|c|c|c|c|c|}
\hline
 & S1 & S2 & S3 & S4 & S5 & S6 \\\hline
$m_{\tilde L_{1,2}}$& 500 & 750 & 1000 & 800 & 500 &  1500 \\
$m_{\tilde L_{3}}$ & 500 & 750 & 1000 & 500 & 500 &  1500 \\
$M_2$ & 500 & 500 & 500 & 500 & 750 &  300 \\
$A_\tau$ & 500 & 750 & 1000 & 500 & 0 & 1500  \\
$\mu$ & 400 & 400 & 400 & 400 & 800 &  300 \\
$\tb$ & 20 & 30 & 50 & 40 & 10 & 40  \\
$\MA$ & 500 & 1000 & 1000 & 1000 & 1000 & 1500  \\
$m_{\tilde Q_{1,2}}$ & 2000 & 2000 & 2000 & 2000 & 2500 & 1500  \\
$m_{\tilde Q_{3}}$ & 2000 & 2000 & 2000 & 500 & 2500 & 1500  \\
$A_t$ & 2300 & 2300 & 2300 & 1000 & 2500 &  1500 \\\hline
$m_{\til_{1\dots 6}}$ & 489--515 & 738--765 & 984--1018 & 474--802  & 488--516 & 1494--1507  \\
$m_{\tinu_{1\dots 3}}$ & 496 & 747 & 998 & 496--797 & 496 &  1499 \\
$m_{{\tilde\chi}_{1,2}^\pm}$ & 375--531 & 376--530 & 377--530 & 377--530  & 710--844 & 247--363  \\
$m_{{\tilde\chi}^0_{1\dots 4}}$ & 244--531 & 245--531 & 245--530 & 245--530  & 373--844 & 145--363  \\
$M_h$ & 126.6 & 127.0 & 127.3 & 123.1 & 123.8 & 125.1  \\
$M_H$ & 500 & 1000 & 999 & 1001 & 1000 & 1499  \\
$M_A$ & 500 & 1000 & 1000 & 1000 & 1000 & 1500  \\
$M_{H^\pm}$ & 507 & 1003 & 1003 & 1005 & 1003 & 1502  \\
$m_{\tilde u_{1\dots 6}}$& 1909--2100 & 1909--2100 & 1908--2100 & 336--2000 & 2423--2585 & 1423--1589  \\
$m_{\tilde d_{1\dots 6}}$ & 1997--2004 & 1994--2007 & 1990--2011 & 474--2001 & 2498--2503 &  1492--1509 \\
$m_{\tilde g}$ & 2000 & 2000 & 2000 & 2000 & 3000 &  1200 \\
\hline
\end{tabular}}
\caption{Selected points in the MSSM parameter space 
(upper part) and their corresponding spectra (lower part).  All 
dimensionful quantities are in GeV.}
\label{tab:spectra}
\end{table}

The large values of $\MA \geqslant 500 \gev$ place the Higgs sector of our 
scenarios in the so-called decoupling regime\cite{Haber:1989xc}, where 
the couplings of $h$ to gauge bosons and fermions are close to the SM 
Higgs couplings, and the heavy $H$ couples like the pseudoscalar $A$, 
and all heavy Higgs bosons are close in mass.  With increasing $\MA$, 
the heavy Higgs bosons tend to decouple from low-energy physics and the 
light $h$ behaves like the SM Higgs boson.  This type of MSSM Higgs sector 
seems to be in good agreement with recent LHC data~\cite{LHCHiggs}.  We 
checked with the code HiggsBounds~\cite{higgsbounds} that this is indeed 
the case (although S3 is right `at the border').

Particularly, the absence of gluinos at the LHC so far forbids too low 
$M_3$ and, through the assumed GUT relation, also a too low $M_2$.  This 
is reflected by our choice of $M_2$ and $\mu$ which give gaugino masses 
compatible with present LHC bounds.  Finally, we required that all our 
points lead to a prediction of the anomalous magnetic moment of the muon 
in the MSSM that can fill the present discrepancy between the SM
prediction and the experimental value.


\subsubsection{Experimental Constraints on \boldmath{$\deFABij$}}

In this section we will present the present experimental constraints on
the squark mixing parameters $\deFABij$ for the above mentioned MSSM
points S1\dots S6 defined in \refta{tab:spectra}. The experimental
constraints from BPO for the same set of parameters that we are using
were already calculated in \cite{arana-NMFV2} for one $\deFABij \neq0$ ,
which we reproduce here for completeness in the \refta{tab:boundsS1S6}.   

\begin{figure}[htb!]
\begin{center}
\psfig{file=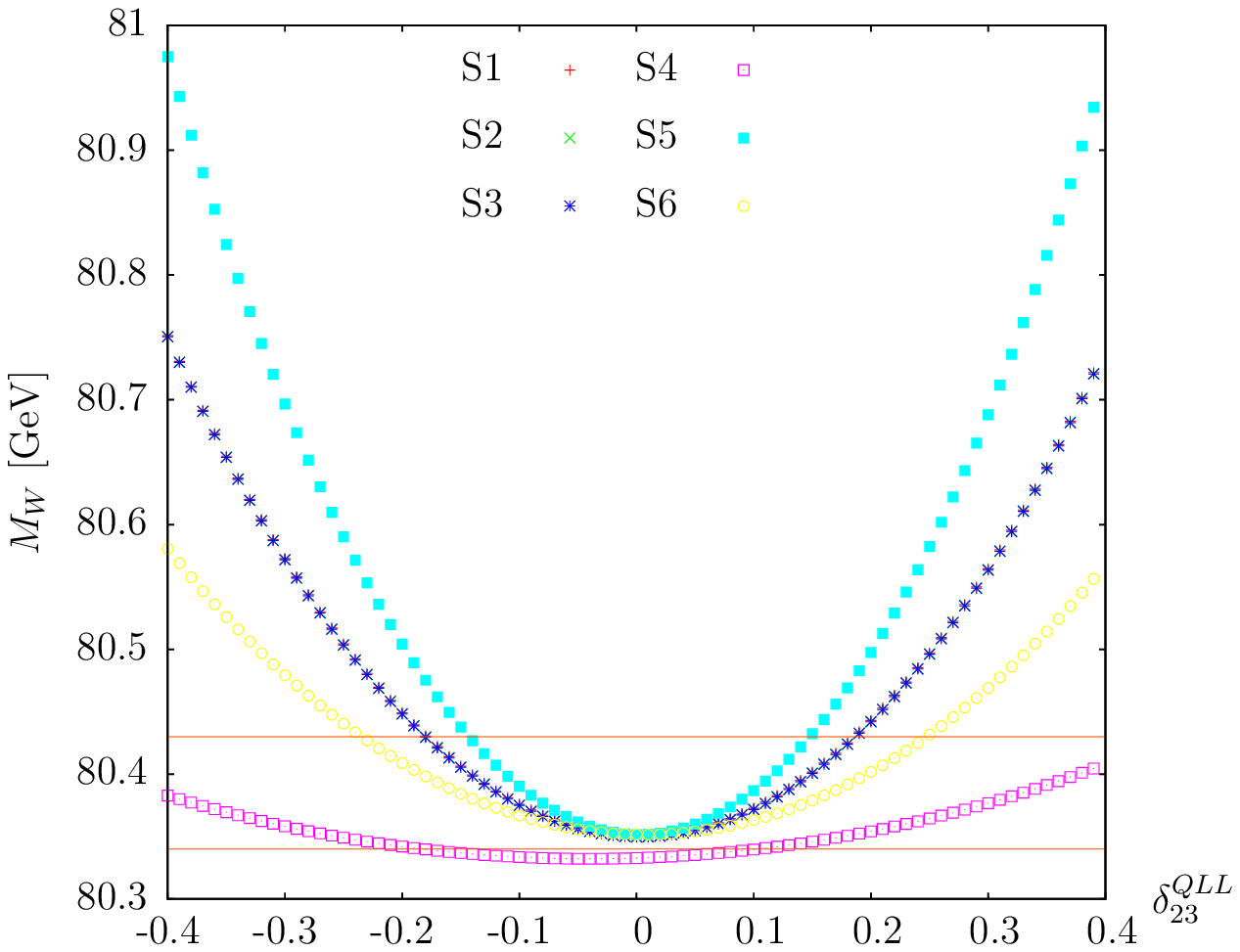,scale=0.57,angle=0,clip=}
\psfig{file=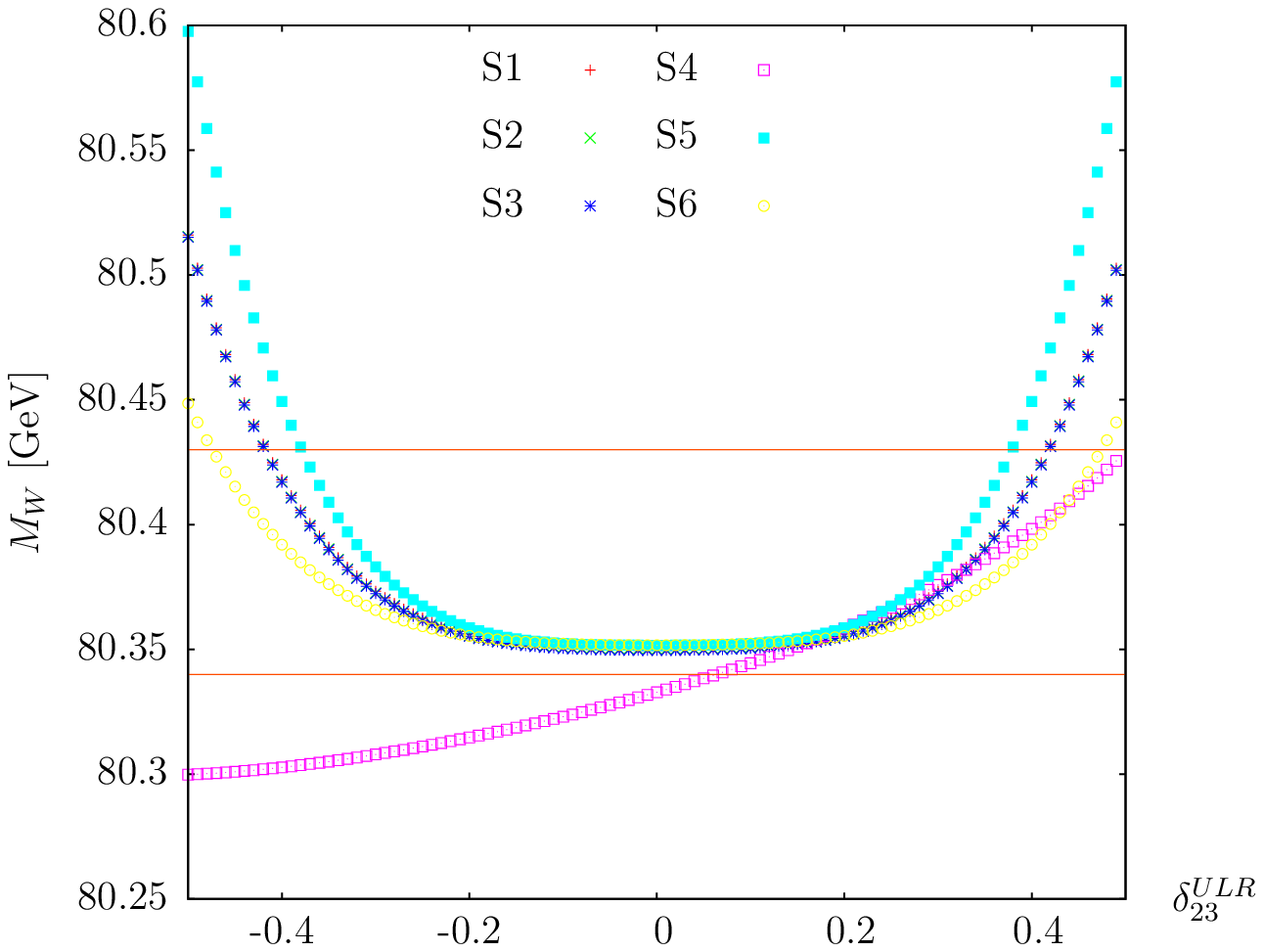,scale=0.57,angle=0,clip=}\\
\vspace{2.0cm}
\psfig{file=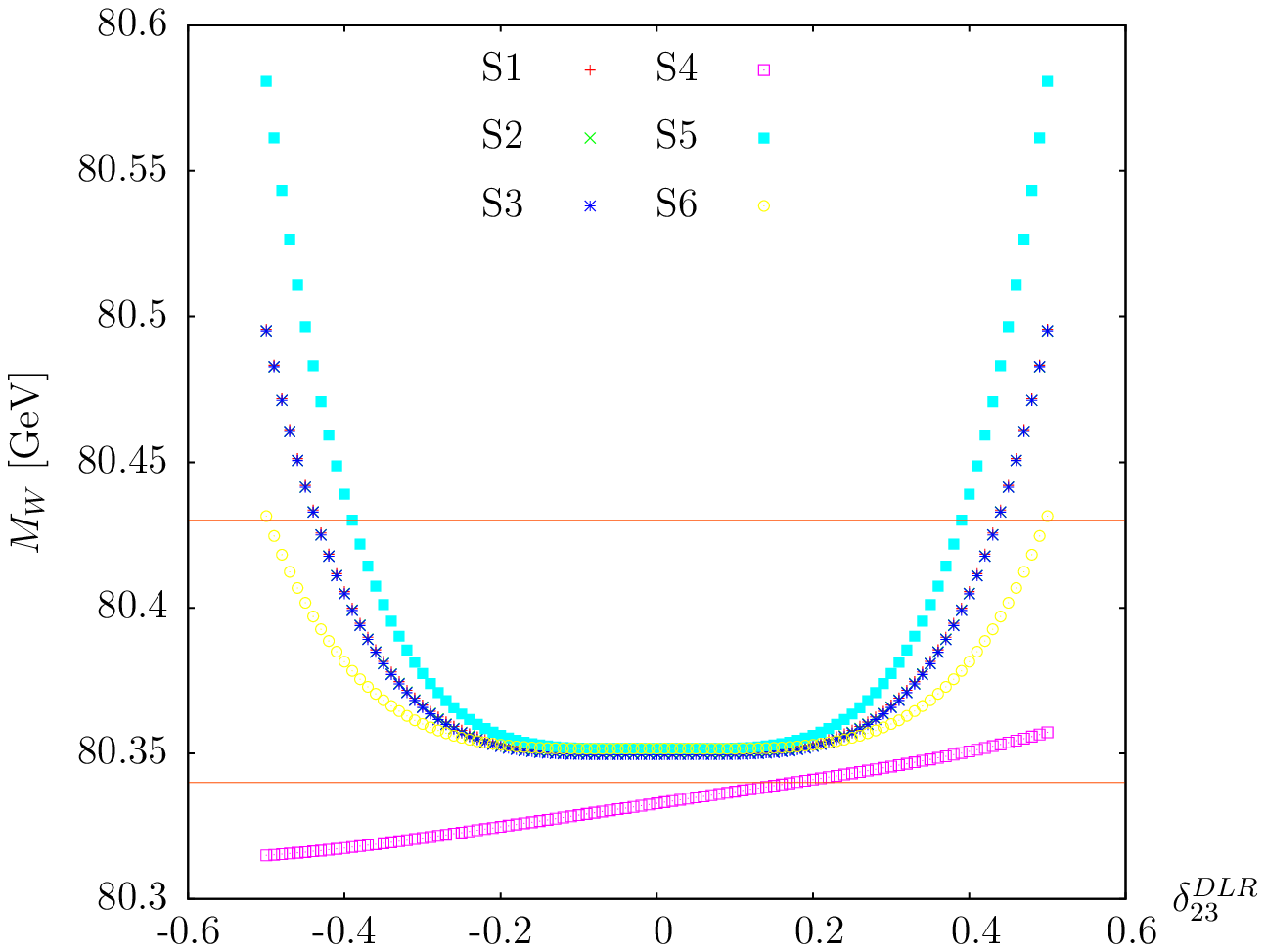,scale=0.56,angle=0,clip=}
\vspace{0.2cm}
\end{center}
\caption{$\MW$ as a function of $\del{QLL}{23}$ (upper left), 
$\del{ULR}{23}$ (upper right) and $\del{DLR}{23}$ (lower). }   
\label{Fig:DMW}
\end{figure}

We now turn our attention to the constraints from $\MW$. 
In \reffi{Fig:DMW} we show the
$\MW$ as a function of $\del{QLL}{23}$, $\del{ULR}{23}$ and
$\del{DLR}{23}$ in the scenarios S1 \ldots S6. 
The area between the orange lines shows the allowed 
value of $\MW$ with $3\sigma$ experimental uncertainty.
The corresponding constraints from $\MW$ on $\deFABij$, also taking
into account the theoretical uncertainties as described at the end of
\refse{sec:ewpo}, are shown in \refta{tab:EWPOboundsS1S6}. 
No constraints can be found on the $\del{RR}{ij}$, 
as their contribution to $\MW$ does not reach the MeV level, and consequently
we do not show them here. Furtheremore, the constraints on the
$\del{URL}{23}$ and $\del{DRL}{23}$ are similar to that of
$\del{ULR}{23}$ and $\del{DLR}{23}$ respectively and not shown
here. 

On the other hand, the constraints on $\del{QLL}{23}$ are modified by the
EWPO specially the region (-0.83:-0.78) for the point S5, which was
allowed by the BPO, is now excluded. The allowed intervals for the
points S1-S3 have also shrunk. However the point S4 was already excluded
by BPO, similarly the allowed interval for S6 do not get modified by
EWPO. The constraints on $\del{ULR}{23}$ and $\del{DLR}{23}$
are less restrictive then the ones from BPO except for the point S4
where the region (0.076:0.12) is excluded for $\del{DLR}{23}$ by
EWPO.

\renewcommand{\arraystretch}{1.10}
\begin{table}[htb!]
\begin{center}
\resizebox{9.0cm}{!} {
\begin{tabular}{|c|c|c|} \hline
 & & Total allowed intervals \\ \hline
$\delta^{QLL}_{23}$ & \begin{tabular}{c}  S1 \\ S2 \\ S3 \\ S4 \\ S5 \\ S6 \end{tabular} &  
\begin{tabular}{c} 
(-0.27:0.28) \\ (-0.23:0.23) \\ (-0.12:0.06) (0.17:0.19) \\ excluded \\ (-0.83:-0.78) (-0.14:0.14) \\ (-0.076:0.14) \end{tabular} \\ \hline
$\delta^{ULR}_{23}$  & \begin{tabular}{c}  S1 \\ S2 \\ S3 \\ S4 \\ S5 \\ S6 \end{tabular}    
& \begin{tabular}{c} 
(-0.27:0.27) \\ (-0.27:0.27) \\ (-0.27:0.27) \\ excluded \\ (-0.22:0.22) \\ (-0.37:0.37) \end{tabular}   \\ \hline
$\delta^{DLR}_{23}$  & \begin{tabular}{c}  S1 \\ S2 \\ S3 \\ S4 \\ S5 \\ S6 \end{tabular}    & 
\begin{tabular}{c} 
(-0.0069:0.014) (0.12:0.13) \\ (-0.0069:0.014) (0.11:0.13) \\ (-0.0069:0.014) (0.11:0.13) \\ (0.076:0.12) (0.26:0.30) \\ (-0.014:0.021) (0.17:0.19) \\ (0:0.0069) (0.069:0.076) \end{tabular}  \\ \hline
$\delta^{URL}_{23}$  & \begin{tabular}{c}  S1 \\ S2 \\ S3 \\ S4 \\ S5 \\ S6 \end{tabular}    & 
\begin{tabular}{c}
(-0.27:0.27) \\ (-0.27:0.27) \\ (-0.27:0.27) \\ excluded \\ (-0.22:0.22) \\ (-0.37:0.37) \end{tabular} 
  \\ \hline
$\delta^{DRL}_{23}$  & \begin{tabular}{c}  S1 \\ S2 \\ S3 \\ S4 \\ S5 \\ S6 \end{tabular}    &
\begin{tabular}{c} (-0.034:0.034) \\ (-0.034:0.034) \\ (-0.034:0.034) \\ excluded \\ (-0.062:0.062) \\ (-0.021:0.021) \end{tabular} 
  \\ \hline
$\delta^{URR}_{23}$ & \begin{tabular}{c}  S1 \\ S2 \\ S3 \\ S4 \\ S5 \\ S6 \end{tabular}   & \begin{tabular}{c} 
(-0.99:0.99) \\ (-0.99:0.99) \\ (-0.98:0.97) \\ excluded \\ (-0.99:0.99) \\ (-0.96:0.94)  \end{tabular}    \\ \hline
$\delta^{DRR}_{23}$  & \begin{tabular}{c}  S1 \\ S2 \\ S3 \\ S4 \\ S5 \\ S6 \end{tabular}    &
\begin{tabular}{c}  (-0.96:0.96) \\ (-0.96:0.96) \\ (-0.96:0.94) \\ excluded \\ (-0.97:0.97) \\ (-0.97:-0.94) (-0.63:0.64) (0.93:0.97)
\end{tabular}    \\ \hline
\end{tabular}}  
\end{center}
\caption{Present allowed (by BPO) intervals for the squark mixing parameters
$\deFABij$ for the selected S1-S6 MSSM points defined in
\refta{tab:spectra}\cite{arana-NMFV2}. 
}
\label{tab:boundsS1S6}
\vspace{-4em}
\end{table}
\renewcommand{\arraystretch}{1.55}

\renewcommand{\arraystretch}{1.1}
\begin{table}[htb!]
\begin{center}
\resizebox{9.0cm}{!} {
\begin{tabular}{|c|c|c|} \hline
 & & Total allowed intervals \\ \hline
$\delta^{QLL}_{23}$ & \begin{tabular}{c}  S1 \\ S2 \\ S3 \\ S4 \\ S5 \\ S6 \end{tabular} &  
\begin{tabular}{c} 
(-0.18:0.18) \\ (-0.18:0.18) \\ (-0.18:0.18) \\ (-0.53:-0.17)(0.10:0.45) \\ (-0.14:0.14) \\ (-0.23:0.23) \end{tabular} \\ \hline
$\delta^{ULR}_{23},\delta^{URL}_{23}$  & \begin{tabular}{c}  S1 \\ S2 \\ S3 \\ S4 \\ S5 \\ S6 \end{tabular}    
& \begin{tabular}{c} 
(-0.41:0.41) \\ (-0.41:0.41) \\ (-0.41:0.41) \\ (0.10:0.50) \\ (-0.39:0.39) \\ (-0.47:0.47) \end{tabular}   \\ \hline
$\delta^{DLR}_{23},\delta^{DRL}_{23}$  & \begin{tabular}{c}  S1 \\ S2 \\ S3 \\ S4 \\ S5 \\ S6 \end{tabular}    & 
\begin{tabular}{c} 
(-0.43:0.43) \\ (-0.43:0.43) \\ (-0.43:0.43) \\ (0.16:0.99) \\ (-0.39:0.39) \\ (-0.49:0.49) \end{tabular}  \\ \hline
\end{tabular}}  
\end{center}
\caption{Present allowed (by $\MW$) intervals for the squark mixing parameters
$\deFABij$ for the selected S1-S6 MSSM points defined in
\refta{tab:spectra}. 
}
\label{tab:EWPOboundsS1S6}
\end{table}
\renewcommand{\arraystretch}{1.55}


\subsubsection{\boldmath{\brhbs}}

\begin{figure}[ht!]
\begin{center}
\psfig{file=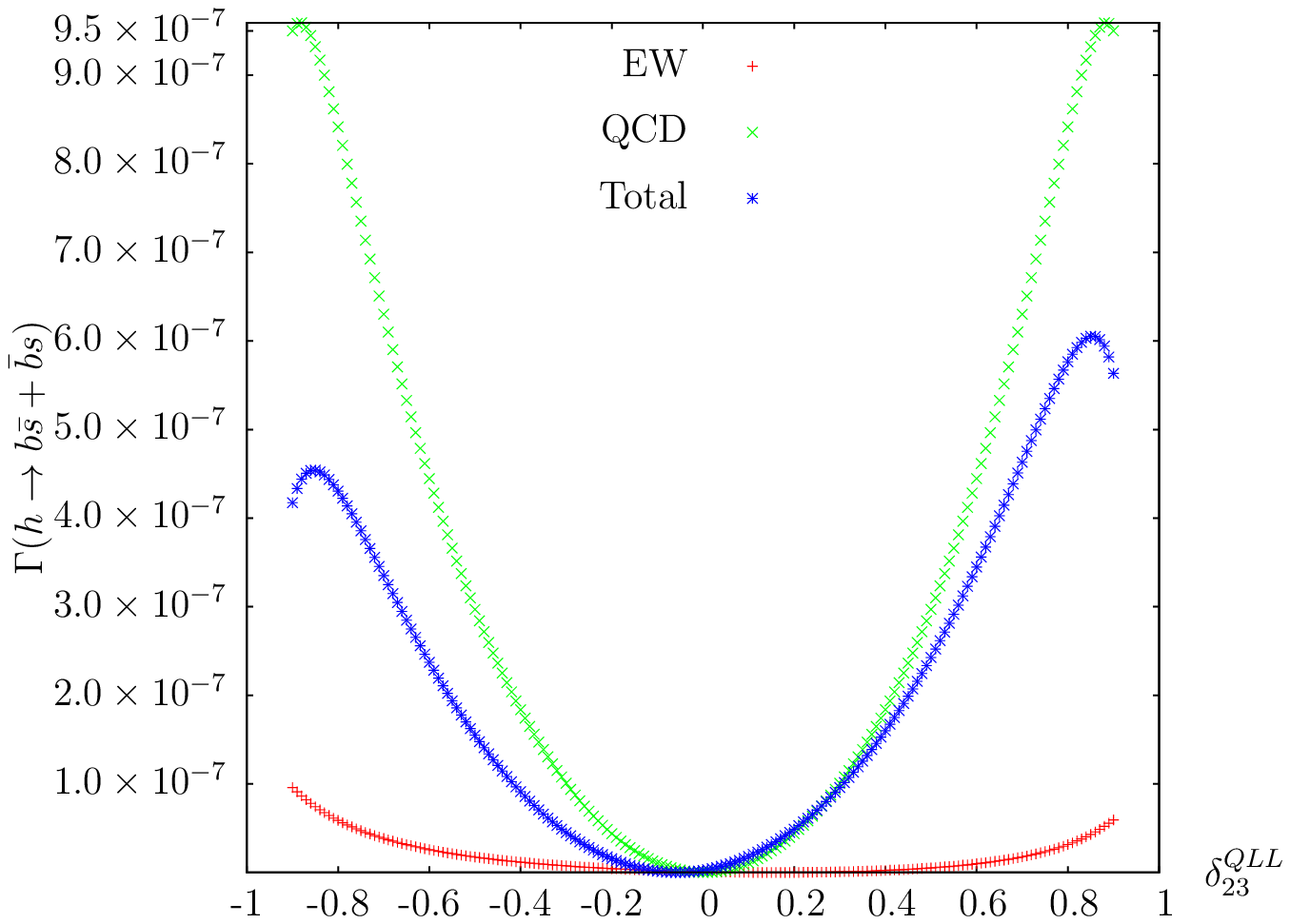,scale=0.56,angle=0,clip=}
\psfig{file=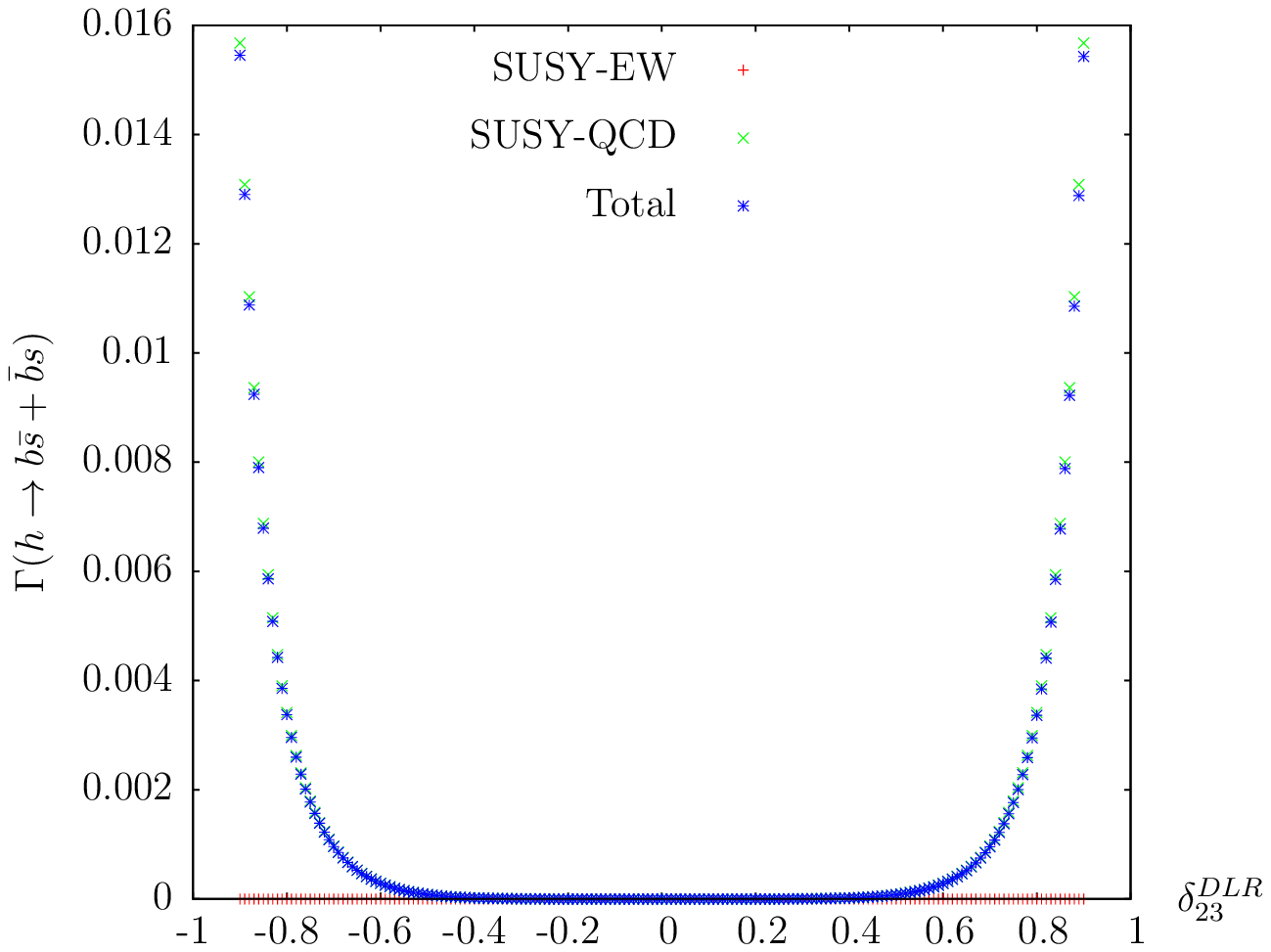,scale=0.56,angle=0,clip=}\\
\vspace{2.0cm}
\psfig{file=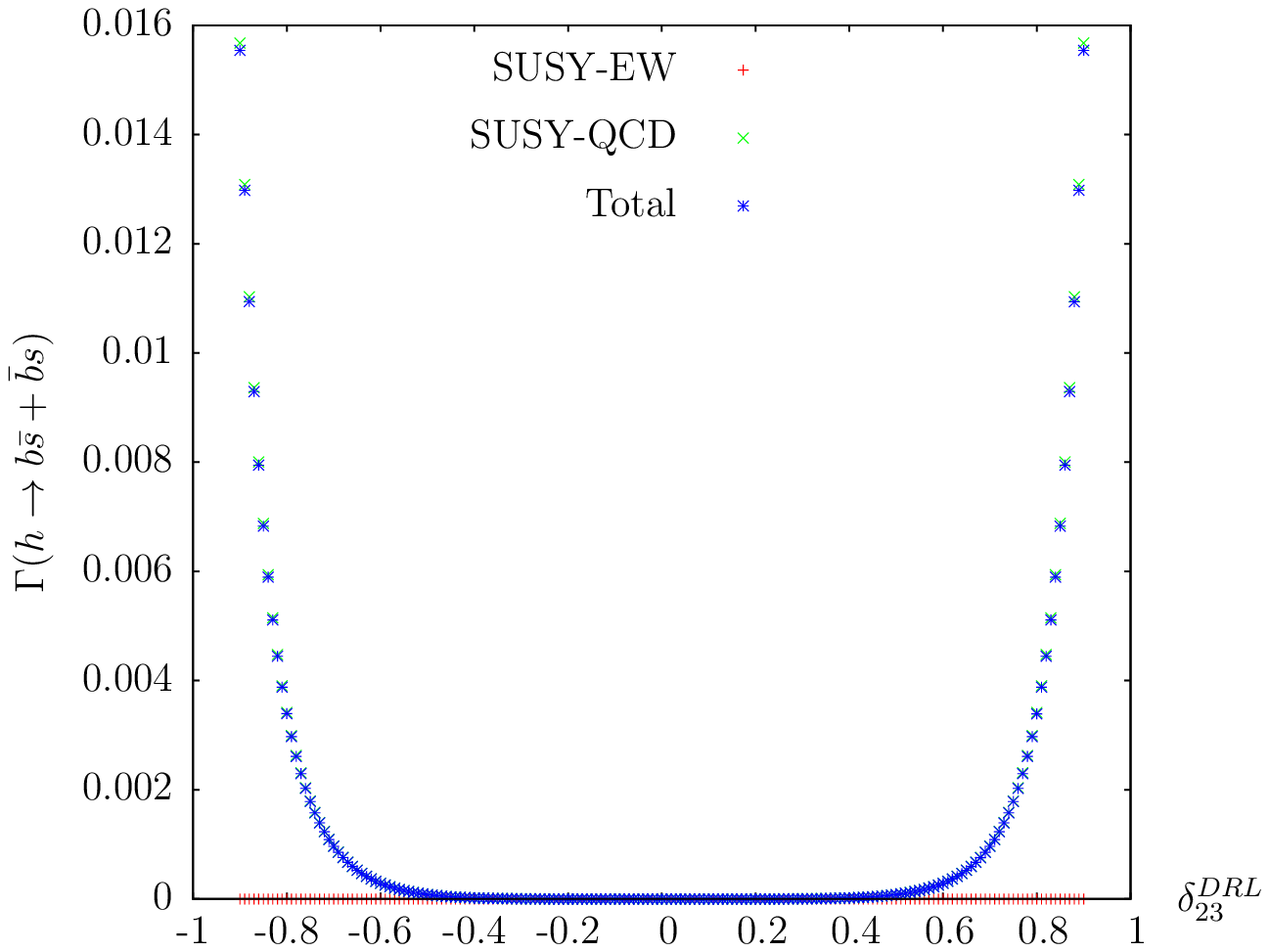,scale=0.56,angle=0,clip=}
\psfig{file=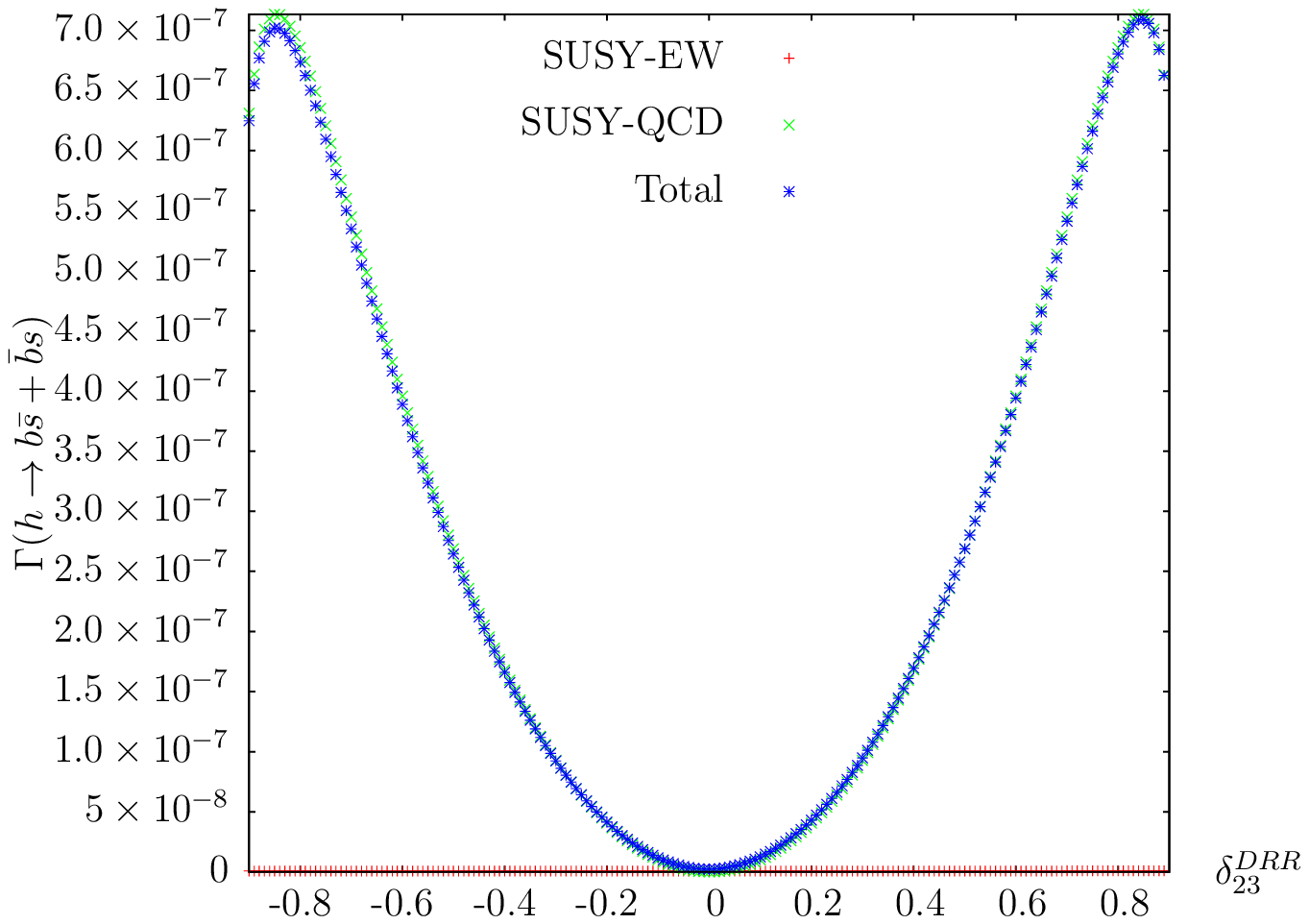,scale=0.56,angle=0,clip=}
\vspace{0.2cm}
\end{center}
\caption{$\Ga(\hbs)$ as a function of $\del{QLL}{23}$ (upper left),
$\del{DLR}{23}$ (upper right), $\del{DRL}{23}$ (lower left) and
$\del{DRR}{23}$ (lower right). 
}     
\label{SUSY-CONT}
\end{figure}

In order to illustrate the contributions from different diagrams we show
in \reffi{SUSY-CONT} the SUSY-EW, SUSY-QCD and total SUSY contribution
to $\Ga(\hbs)$ as a function of
$\del{QLL}{23}$ (upper left), $\del{DLR}{23}$ (upper right),
$\del{DRL}{23}$ (lower left) and $\del{DRR}{23}$ (lower
right). These four $\deFABij$ are the only relevant ones, since we
are mainly concerned with the down-type sector, and mixing with the
first generation does not play a role.
 
In order to compare our results with the literature, we have used
the same set of input parameters as in \cite{SUSY-EW}:
\begin{align}
\mu &= 800 \gev,\; \msusy = 800 \gev,\;  A_f = 500 \gev, \nonumber\\
\MA &= 400 \gev,\; M_2 = 300 \gev,\; \tb = 35 \, ,
\end{align}
where we have chosen, for simplicity, $\msusy$ as a common value for the
soft SUSY-breaking squark mass parameters, 
$\msusy = M_{\tilde Q,q} = M_{\tilde U,(c,t)} = M_{\tilde D,(s,b)}$,
and all the various trilinear parameters to be universal,
$A_f=A_t=A_b=A_c=A_s$. The value of the $\deFABij$'s are varied from 
-0.9~to~0.9, and GUT relations are used to calculate $M_1$ and
$M_3$. 
In \citere{SUSY-EW}, only LL mixing was considered. In this limit we find
results in qualitative agreement with \citere{SUSY-EW}. This analysis has 
been done just to illustrate the different contributions and we do not
take into account any experimental constraints. A detailed analysis for
realisitic SUSY scenerios (defined in \refta{tab:spectra}) constrained
by BPO and EWPO can be found below.

As can be seen in \reffi{SUSY-CONT}, for the decay width $\Ga(\hbs)$ the
SUSY-QCD contribution is dominant in all the cases. For LL mixing shown
in the upper left plot, the SUSY-QCD contribution reaches up
to \order{10^{-6}}, while the SUSY-EW contribution 
reach up to \order{10^{-7}}, resulting in a total contribution ``in
between'', due to the negative interference between SUSY-EW and SUSY-QCD
contribution. 
For LR and RL mixing, shown in the upper right and lower left plot,
respectively, the SUSY-QCD contribution reach up to the
maximum value of \order{10^{-2}}, while the SUSY-EW contribution reach
only up to \order{10^{-7}}. In this case total contriution is almost equal to
SUSY-QCD contribution as SUSY-EW contibution (and thus the interference)
is relatively neglible.
For RR mixing, shown in the lower right plot, the SUSY-EW
contribution of \order{10^{-10}} is again neglible compared to SUSY-QCD
contribution of \order{10^{-7}}.  

\begin{figure}[ht!]
\begin{center}
\psfig{file=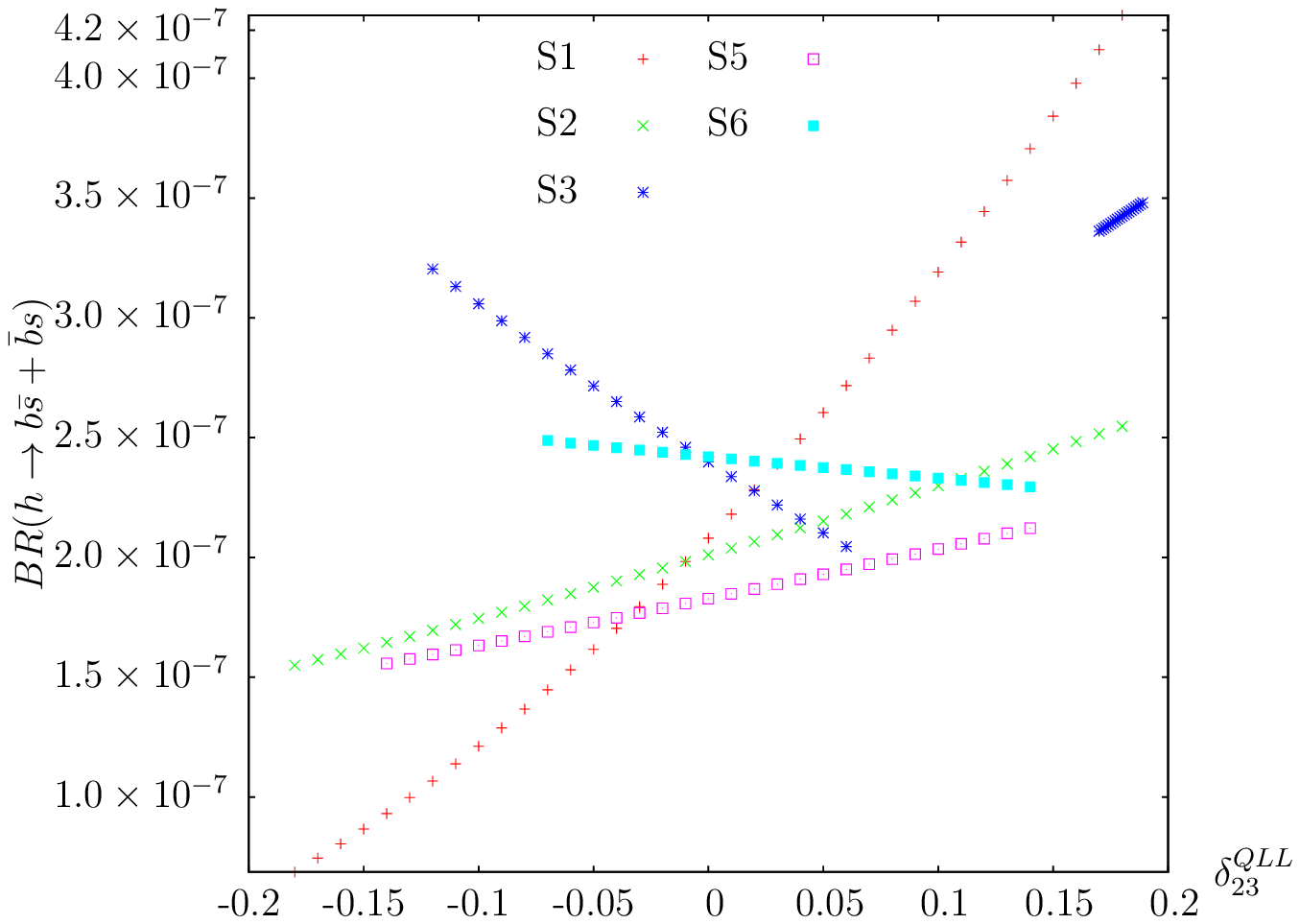,scale=0.54,angle=0,clip=}
\psfig{file=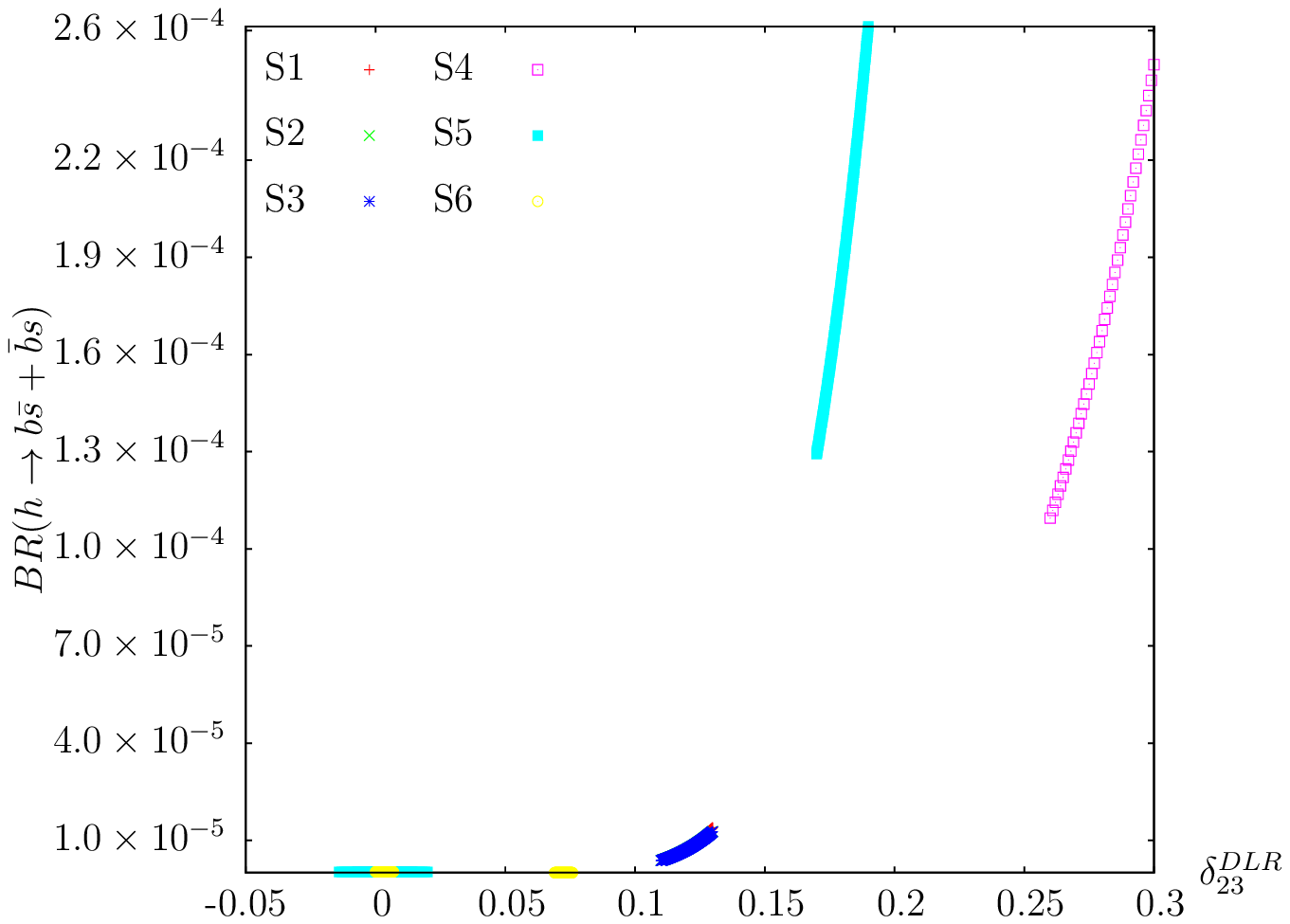,scale=0.54,angle=0,clip=}\\
\vspace{0.2cm}
\psfig{file=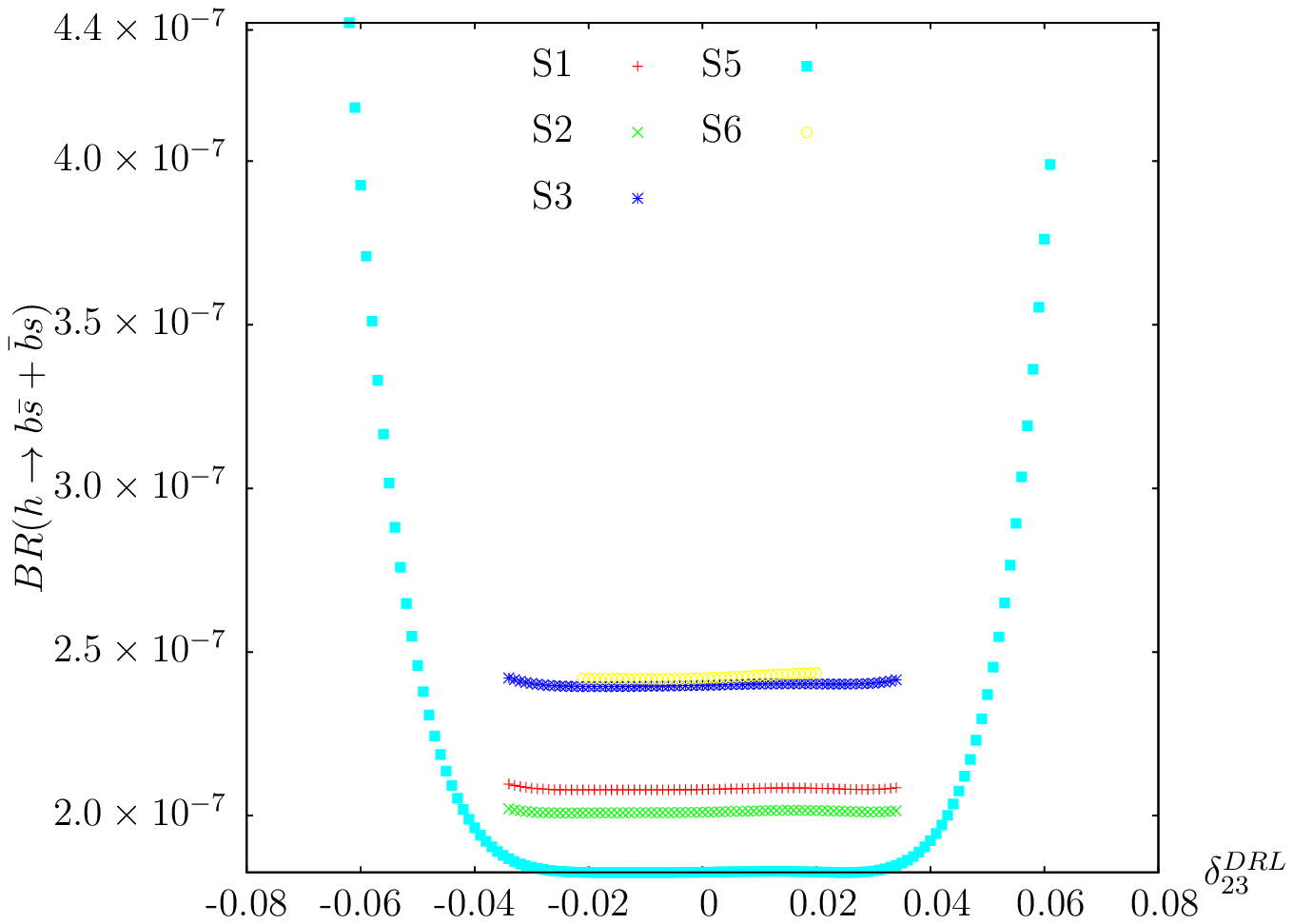,scale=0.54,angle=0,clip=}
\psfig{file=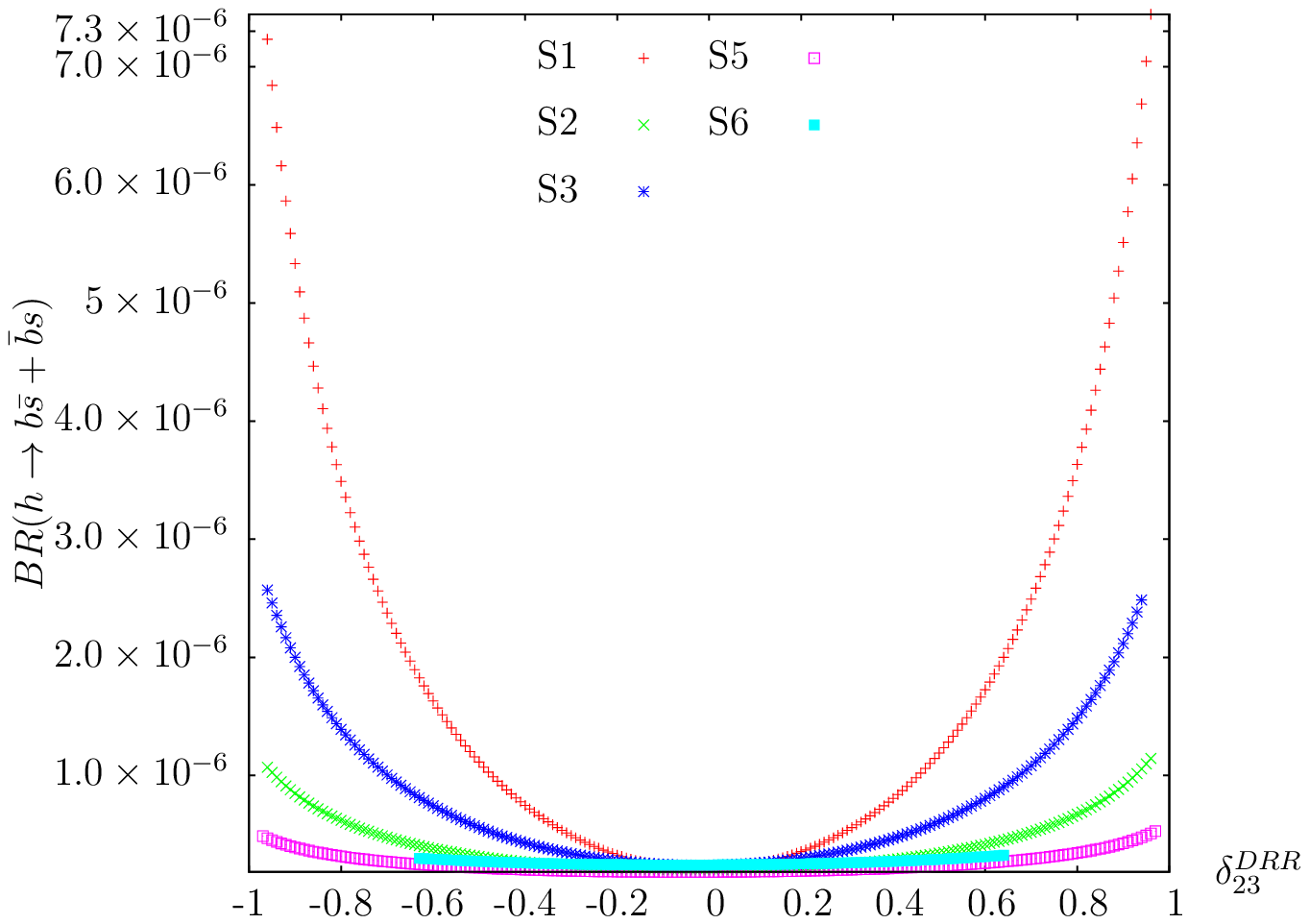,scale=0.54,angle=0,clip=}\\
\end{center}
\caption{\brhbs\ as a function of  $\del{QLL}{23}$ (upper left), 
$\del{DLR}{23}$ (upper right), $\del{DRL}{23}$ (lower left) and 
$\del{DRR}{23}$ (lower right).}
\label{Fig:QFVHD}
\end{figure}

\medskip
Now we turn to realistic scenarios that are in agreement with
experimental data from BPO and EWPO. Starting point are the scenarios
S1\ldots S6 defined in \refta{tab:spectra}, where we vary the flavor
violating $\deFABij$ within the experimentally allowed ranges following
the results given in \reftas{tab:boundsS1S6}, \ref{tab:EWPOboundsS1S6}.
We start with the scenarios in which we allow one of the $\deFABij$ to
be varied, while the others are set to zero. 
In \reffi{Fig:QFVHD} we show \brhbs as a function of $\del{QLL}{23}$ 
(upper left), $\del{DLR}{23}$ (upper right), $\del{DRL}{23}$ (lower
left) and $\del{DRR}{23}$ (lower right), i.e.\ for the same set of
$\deFABij$ that has been analyzed in \reffi{SUSY-CONT}. 
It can be seen that allowing only one $\deFABij \neq 0$ results in
rather small values of \brhbs. LL (upper left) and RL (lower left plot)
mixing results in \order{10^{-7}} values for \brhbs. One order of
magnitude can be gained in the RR mixing case (lower right). The largest
values of \brhbs\ are obtained in the case of $\del{DLR}{23} \neq 0$
(upper right plot). Here in S4 and S5 values of 
$\brhbs \sim 2 \times 10^{-4}$ can be found, possibly in the reach of
future $e^+e^-$ colliders, see \refse{sec:hbs}.


\medskip

So far we have shown the effects of independent variations of
one $\deFABij$. Obviously, a realistic model would include several 
$\deFABij \neq 0$ that may interfere, increasing or decreasing the results
obtained with just the addition of independent contributions.
GUT based MFV models that induce the flavor violation
via RGE running automatically generate several $\deFABij \neq 0$ at the
EW scale.
In the following we will present results with two or three 
$\deFABij \neq 0$, where we combined the ones that showed the largest
effects. 

\begin{figure}[ht!]
\begin{center}
\psfig{file=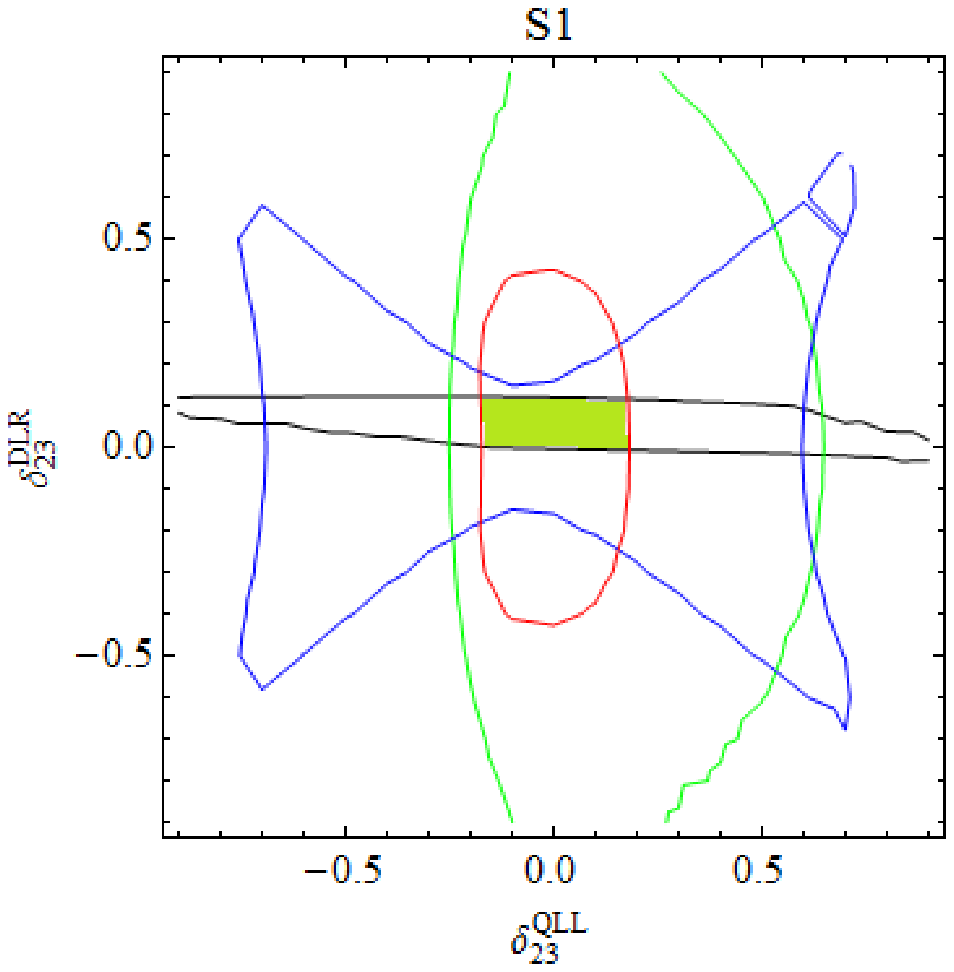,scale=0.72,angle=0,clip=}
\psfig{file=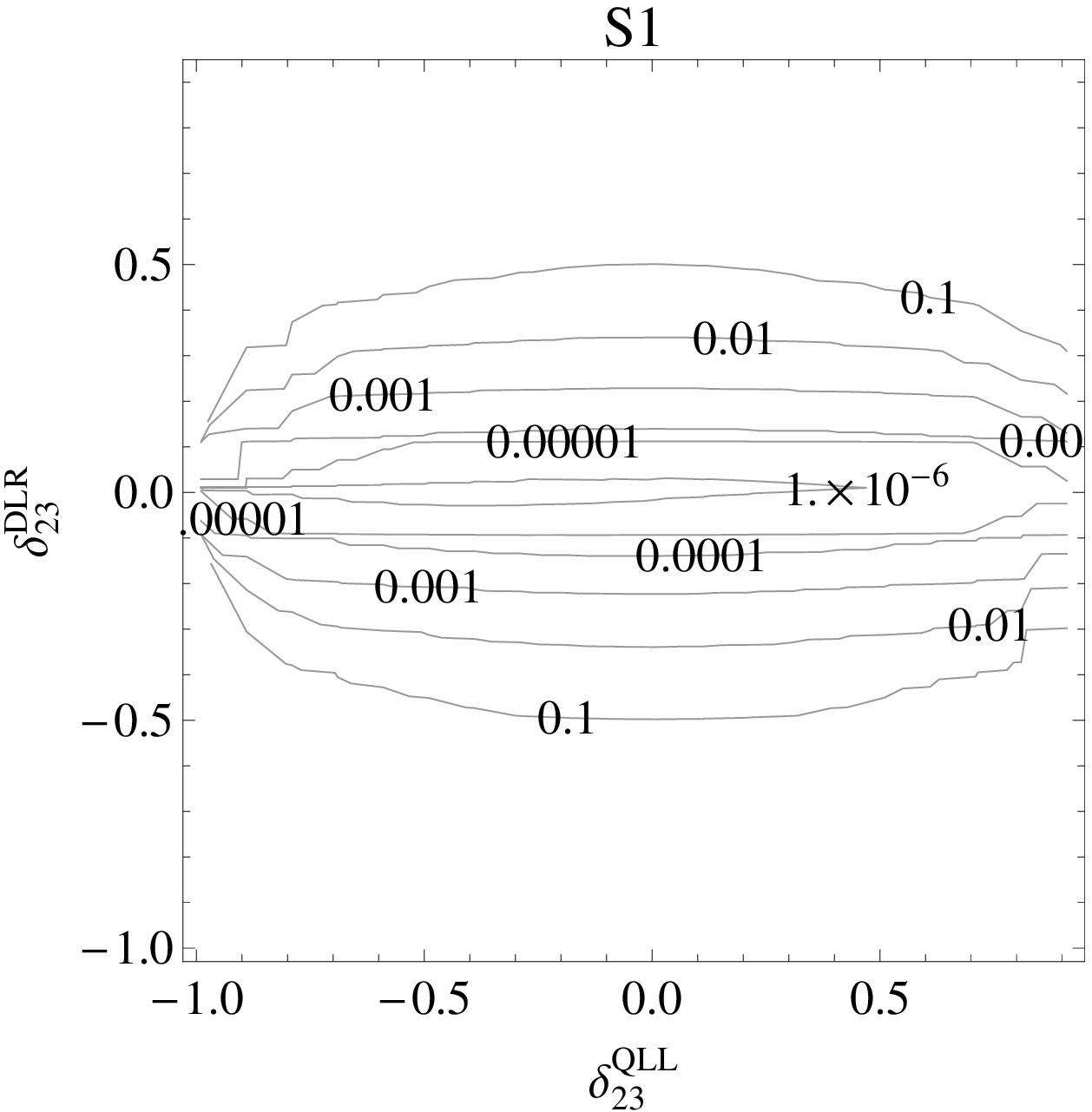,scale=0.55,angle=0,clip=}\\
\vspace{0.2cm}
\psfig{file=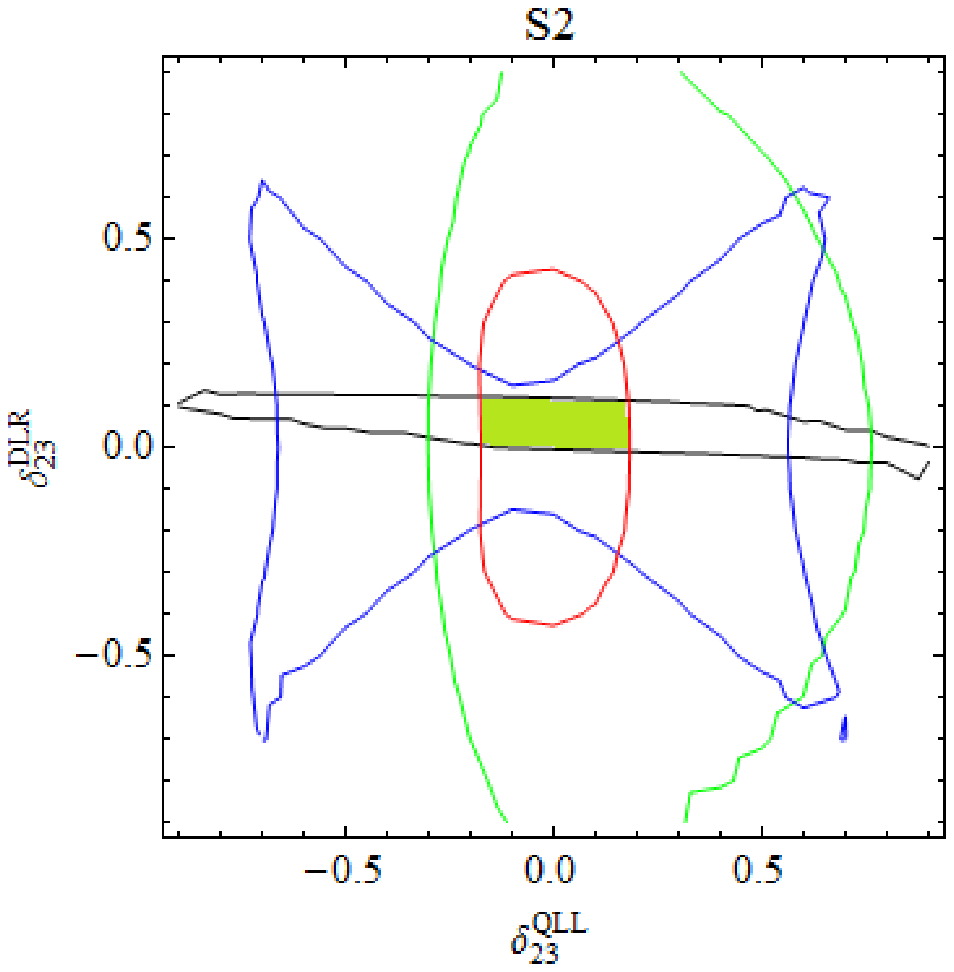,scale=0.72,angle=0,clip=}
\psfig{file=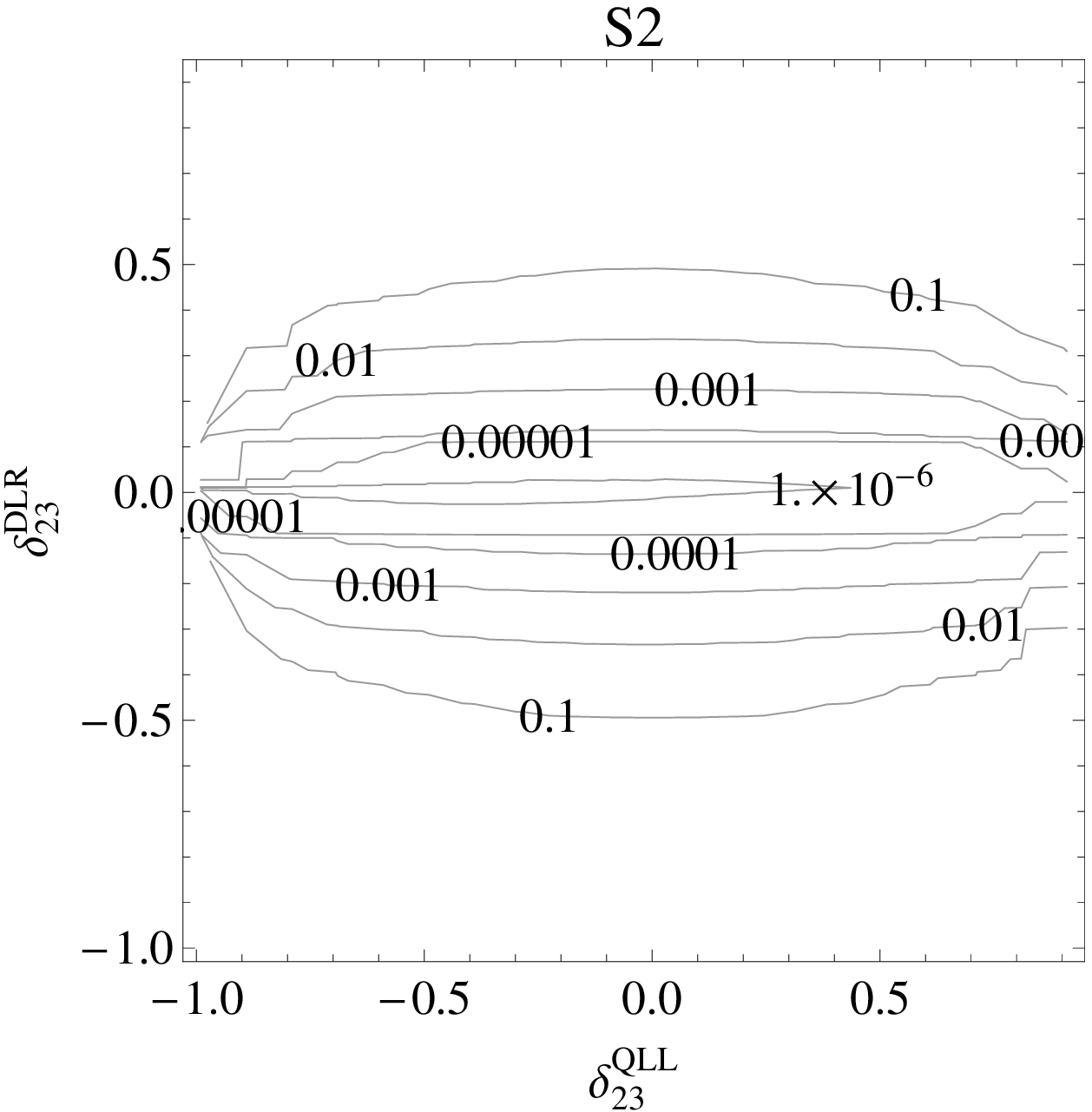,scale=0.54,angle=0,clip=}\\
\vspace{0.2cm}
\psfig{file=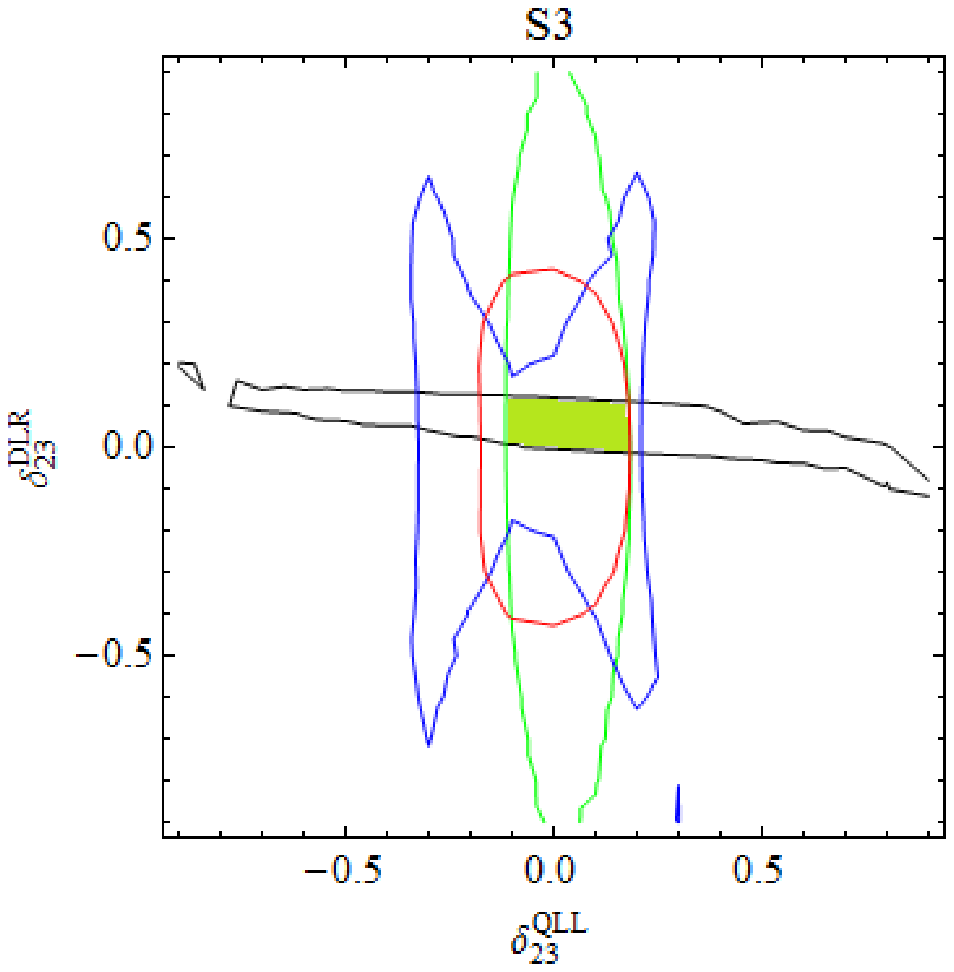,scale=0.72,angle=0,clip=}
\psfig{file=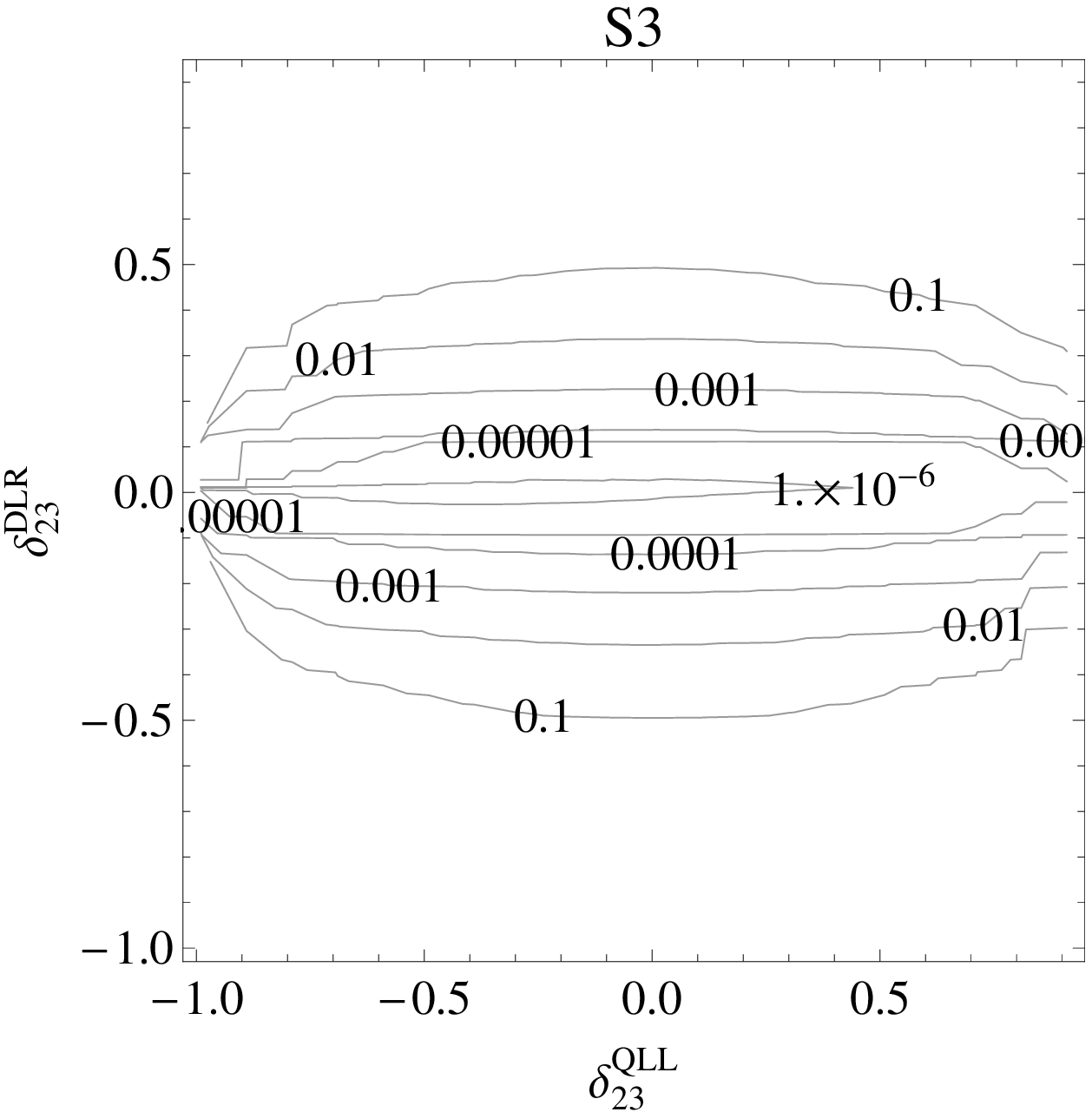,scale=0.54,angle=0,clip=}
\end{center}
\vspace{-1em}
\caption{Left: Contours of \bsg\ (Black), \bmm\ (Green), \dmbs (Blue) and
$\MW$ (Red) in ($\del{QLL}{23}$ , $\del{DLR}{23}$) plane for
points S1-S3. The shaded area shows the range of values allowed by all
cronstraints. Right: corresponding countours for \brhbs.}     
\label{Fig:S1S3QLLDLR23}
\end{figure}

\begin{figure}[ht!]
\begin{center}
\psfig{file=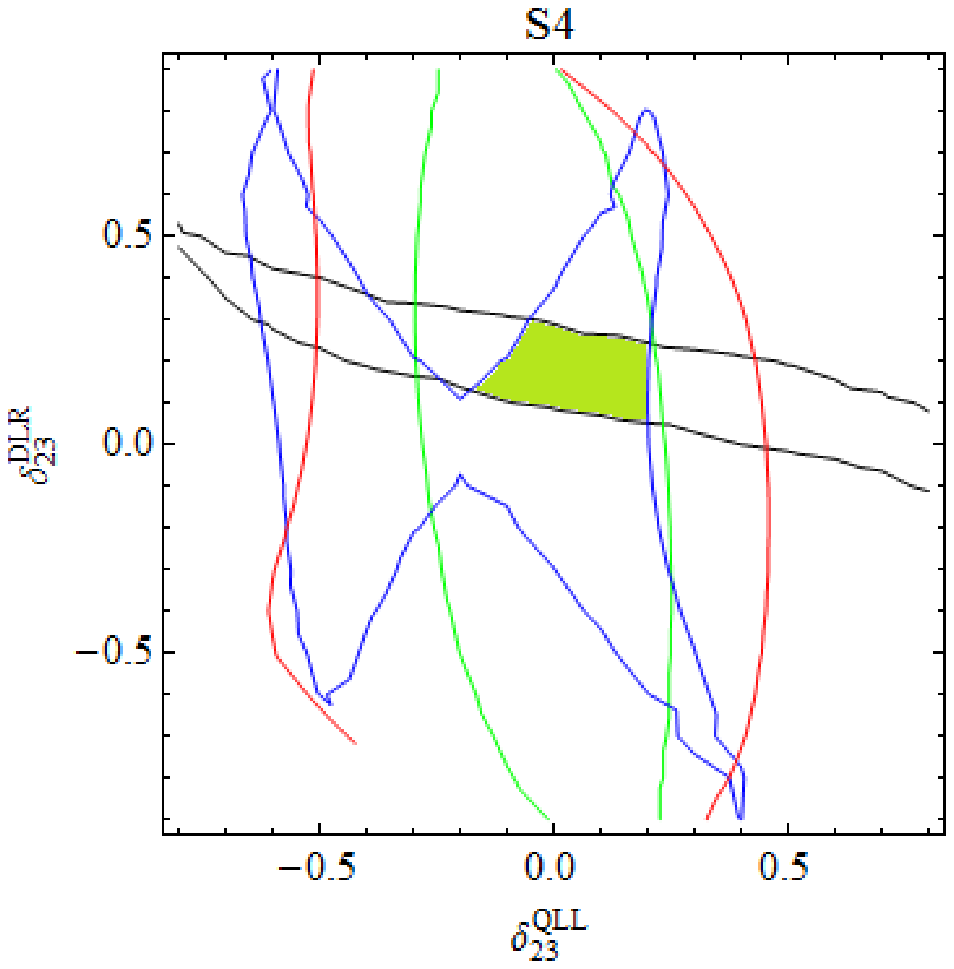,scale=0.72,angle=0,clip=}
\psfig{file=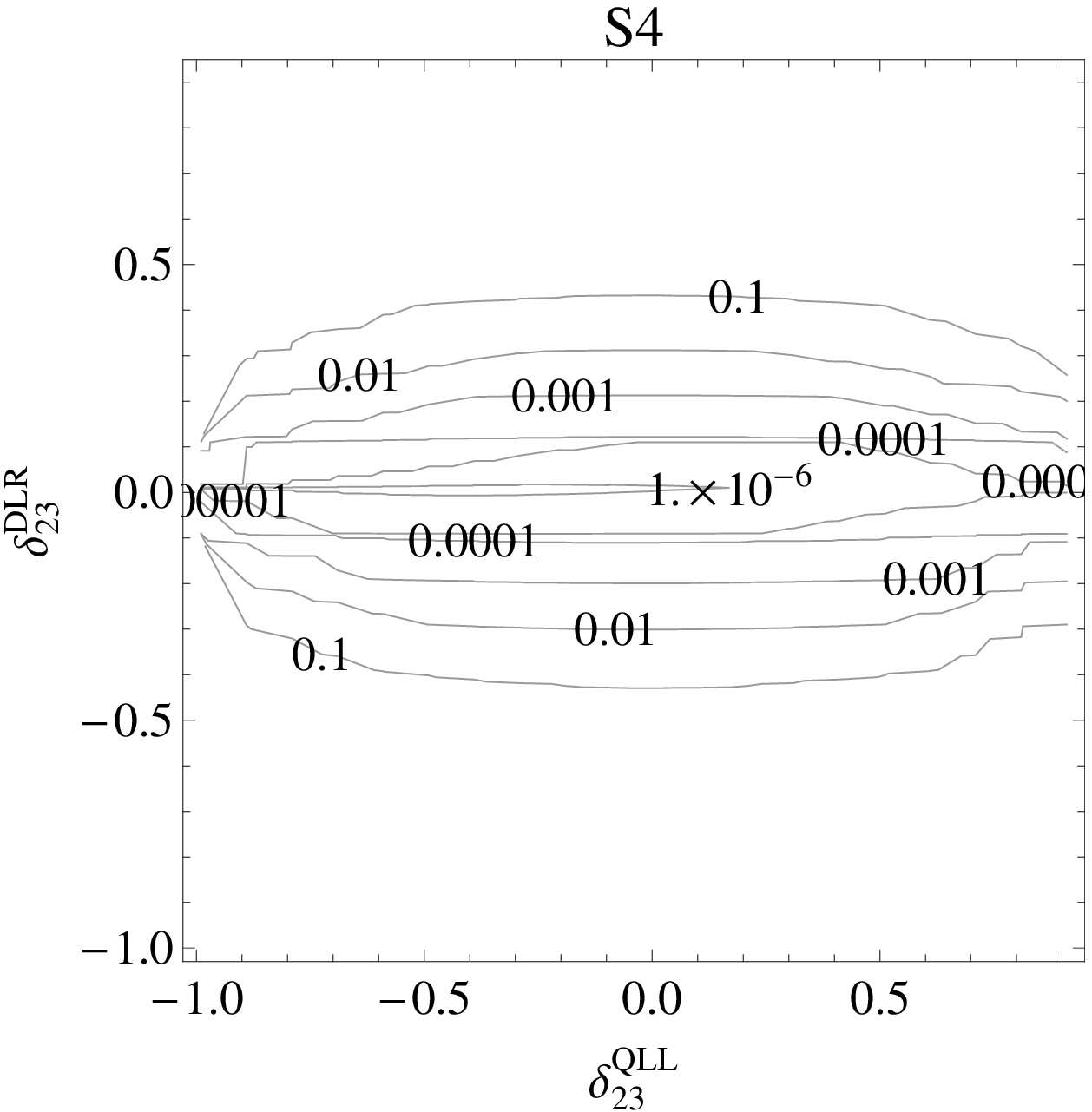,scale=0.55,angle=0,clip=}\\
\vspace{0.2cm}
\psfig{file=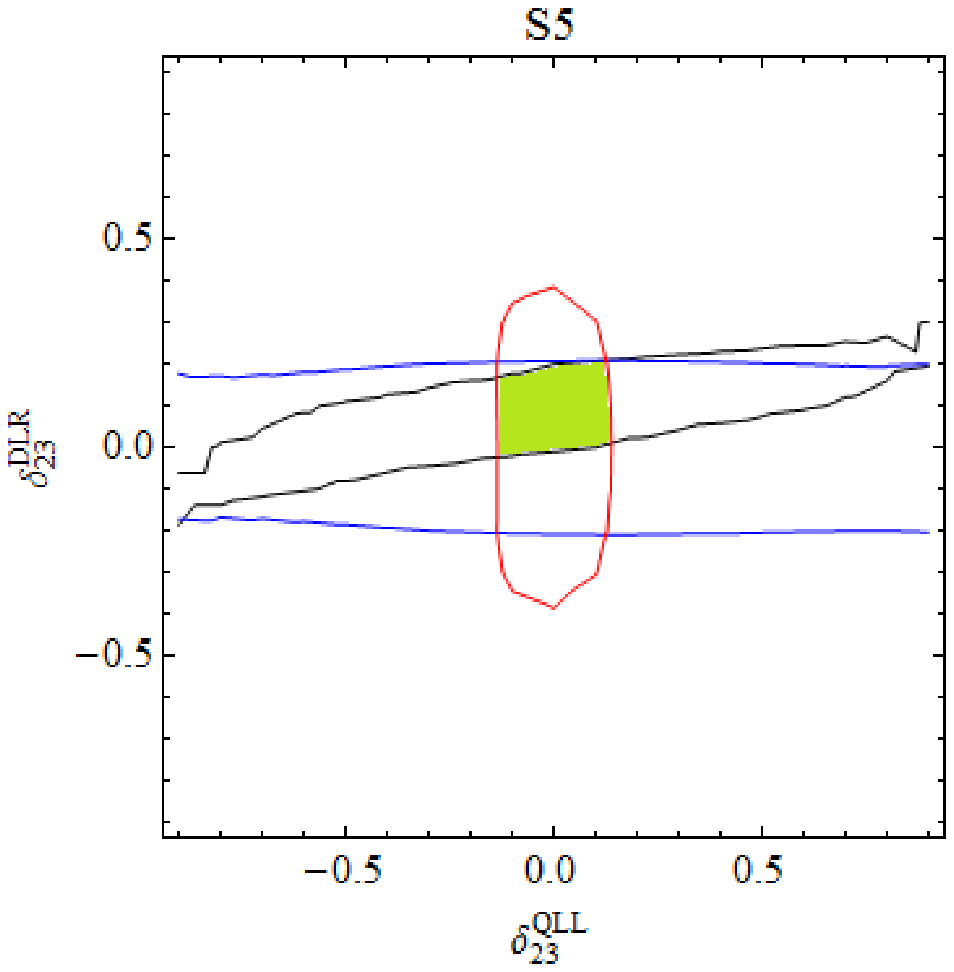,scale=0.71,angle=0,clip=}
\psfig{file=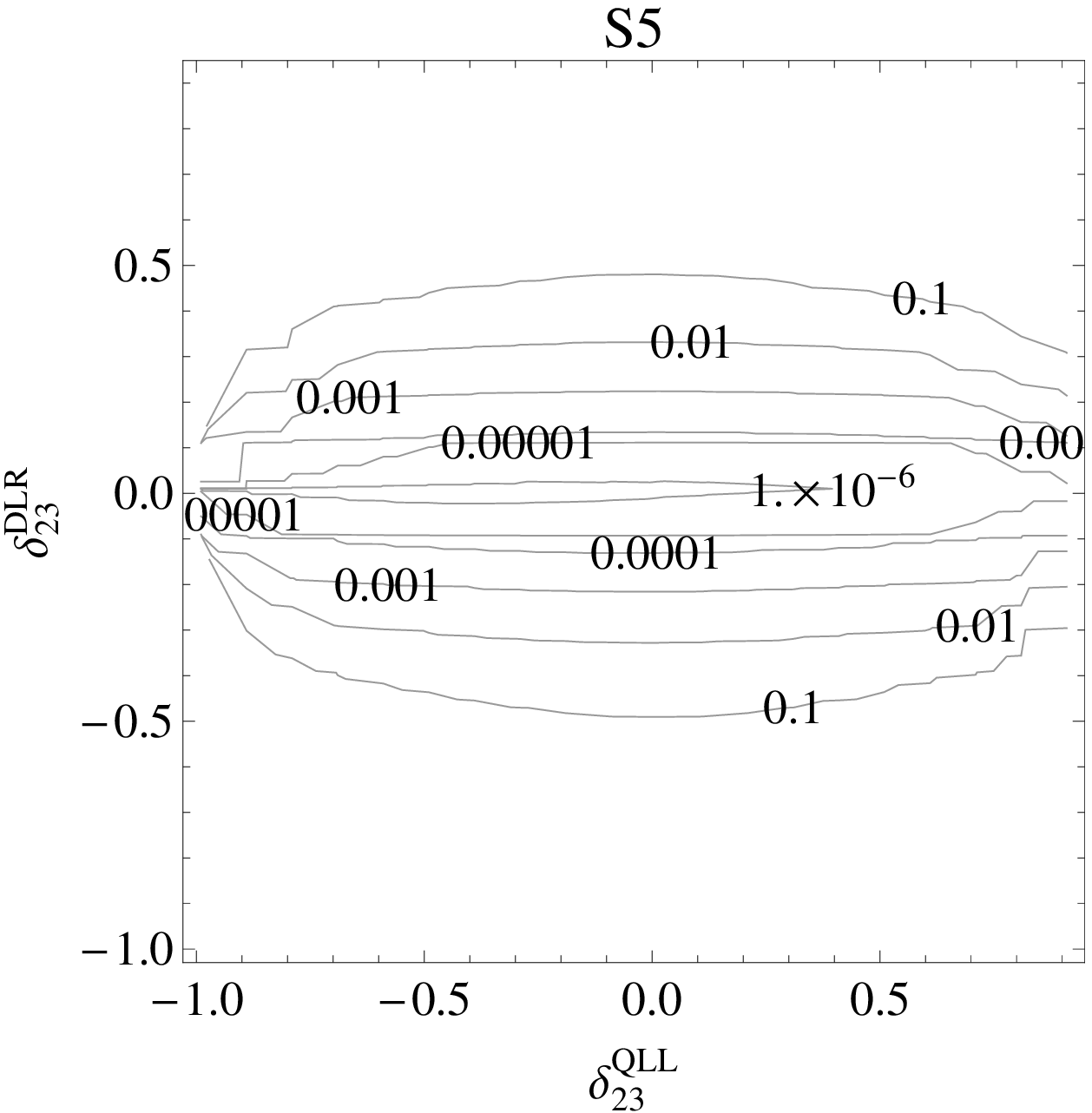,scale=0.54,angle=0,clip=}\\
\vspace{0.2cm}
\psfig{file=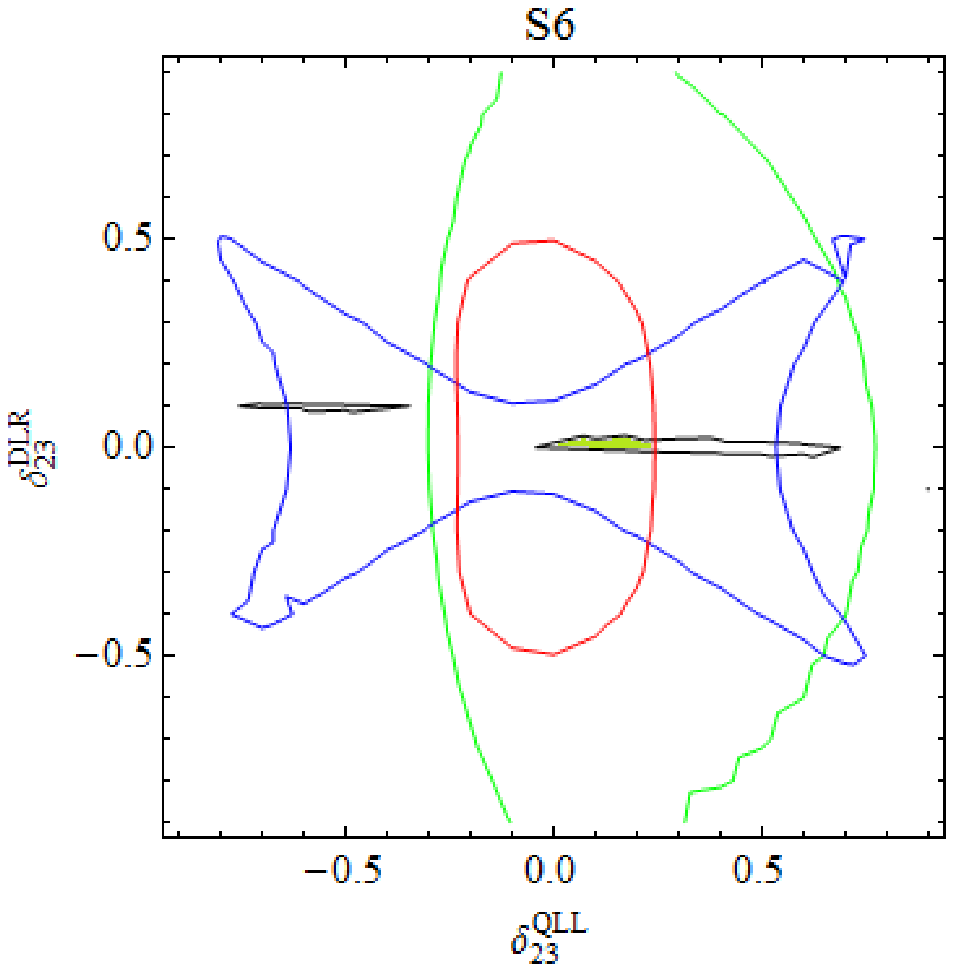,scale=0.71,angle=0,clip=}
\psfig{file=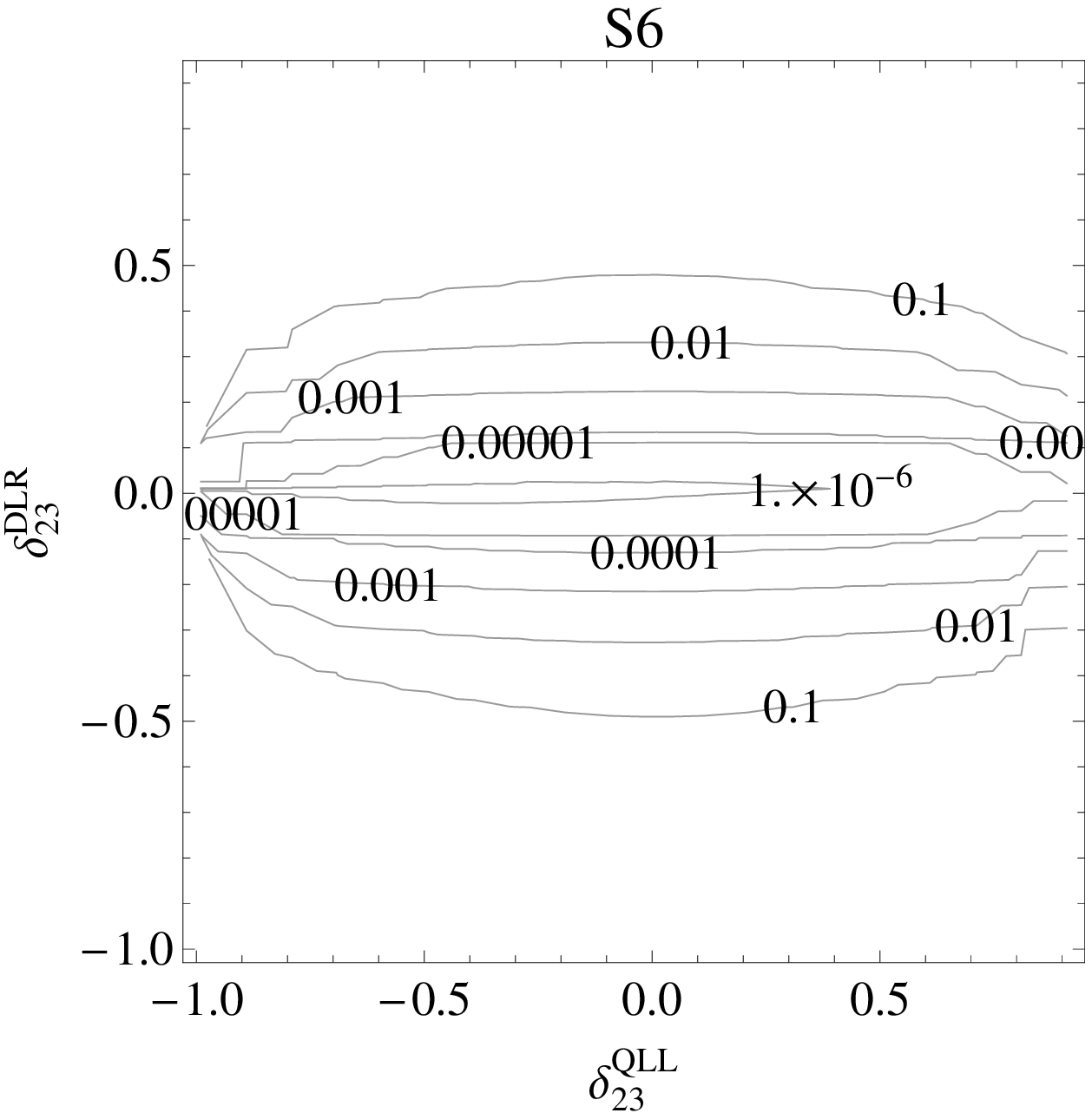,scale=0.54,angle=0,clip=}
\vspace{-1em}
\end{center}
\caption{Left: Contours of \bsg\ (Black), \bmm\ (Green), \dmbs\ (Blue) and
$\MW$ (Red) in ($\del{QLL}{23}$ , $\del{DLR}{23}$) plane for
points S4-S6. The shaded area shows the range of values allowed by all
cronstraints. Right: corresponding countours for \brhbs.}    
\label{Fig:S4S6QLLDLR23}
\end{figure}

\begin{figure}[ht!]
\begin{center}
\psfig{file=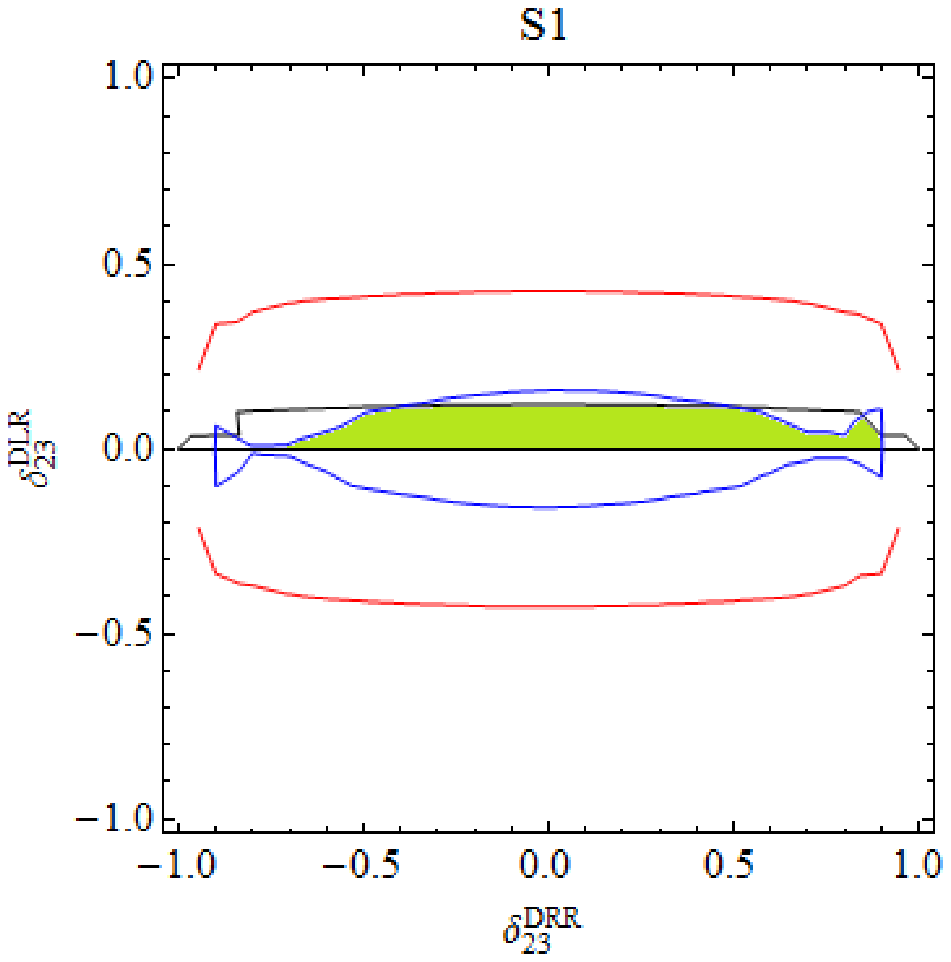,scale=0.72,angle=0,clip=}
\psfig{file=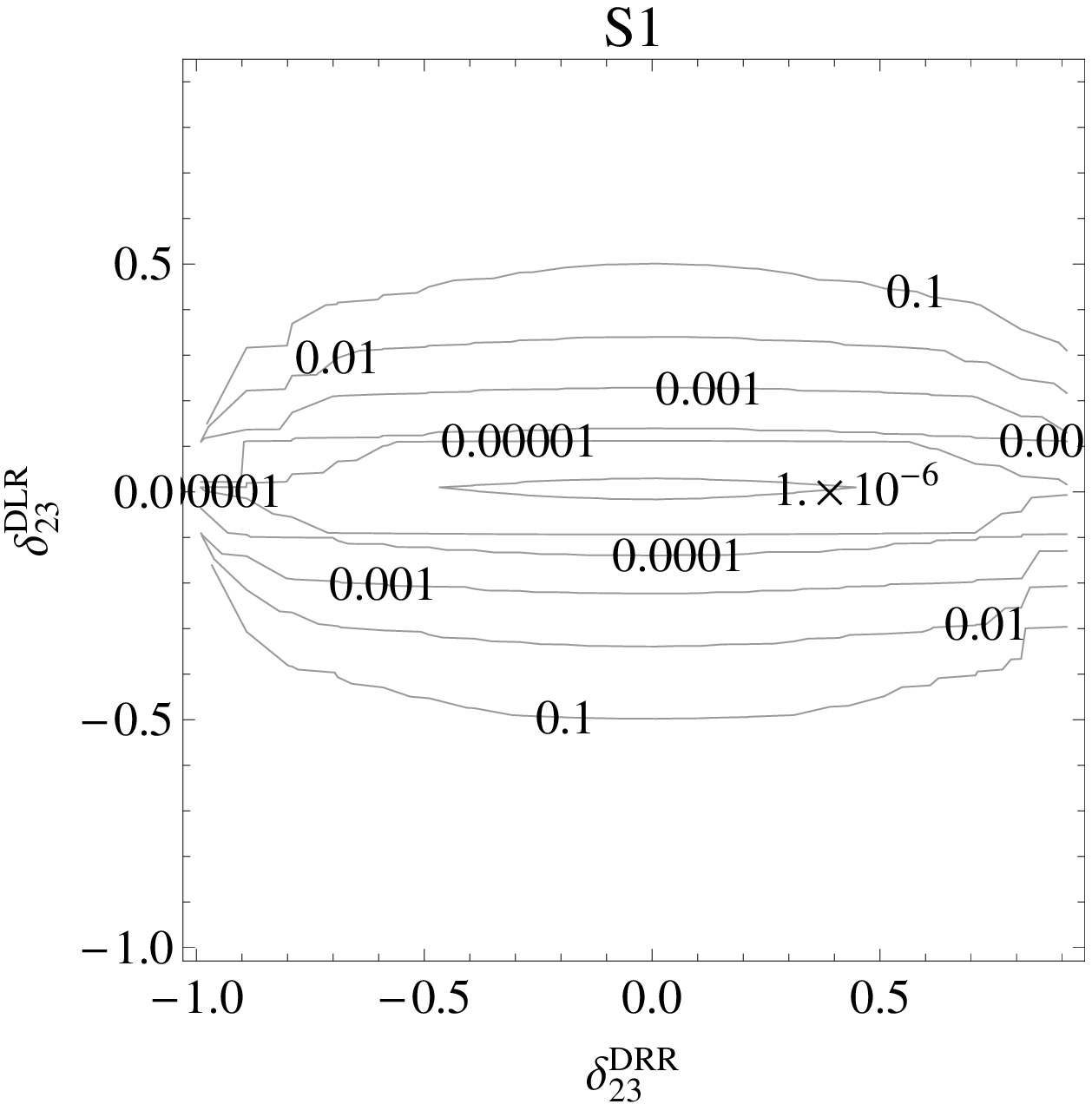,scale=0.55,angle=0,clip=}\\
\vspace{0.2cm}
\psfig{file=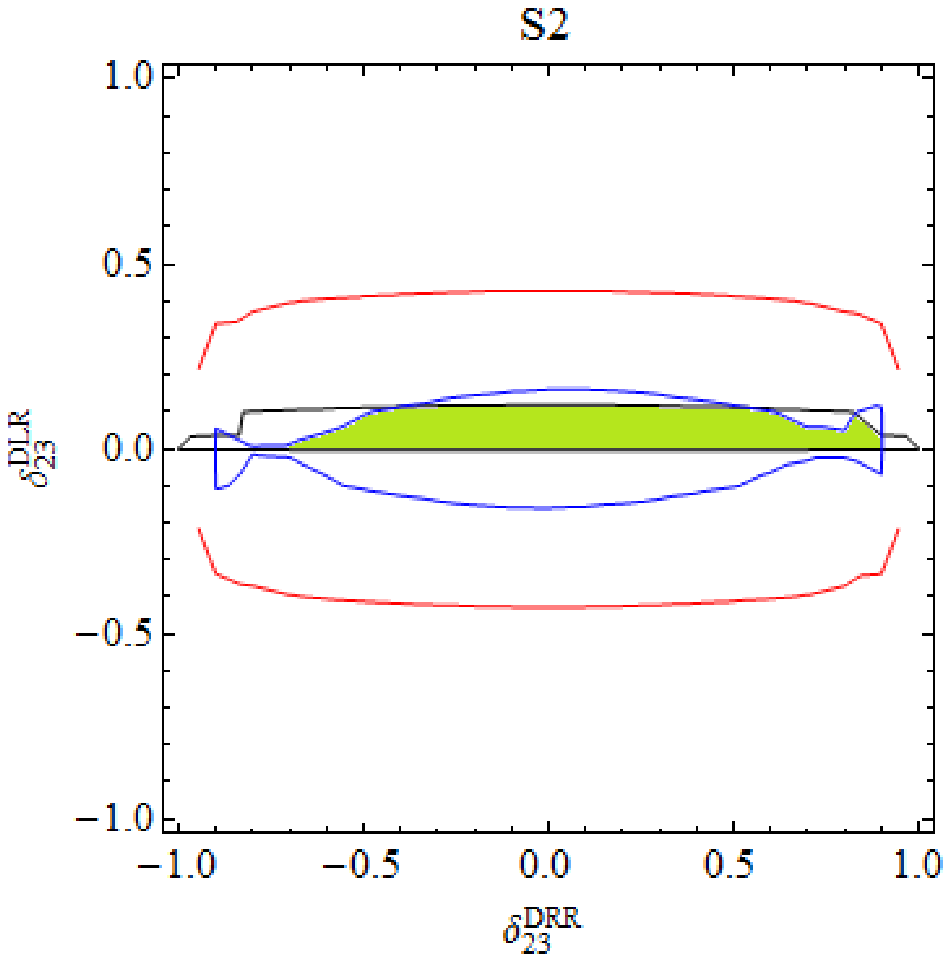,scale=0.71,angle=0,clip=}
\psfig{file=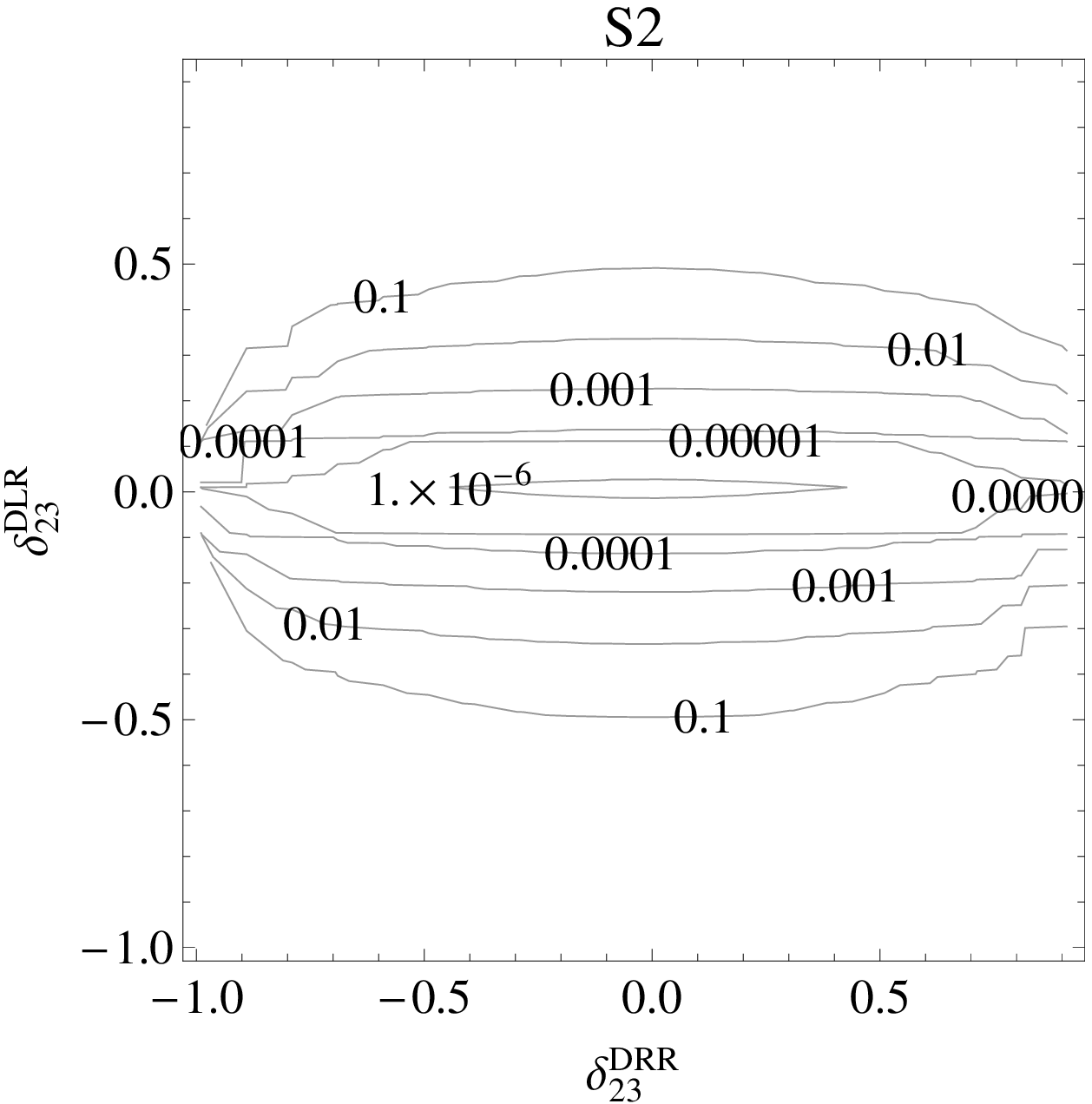,scale=0.54,angle=0,clip=}\\
\vspace{0.2cm}
\psfig{file=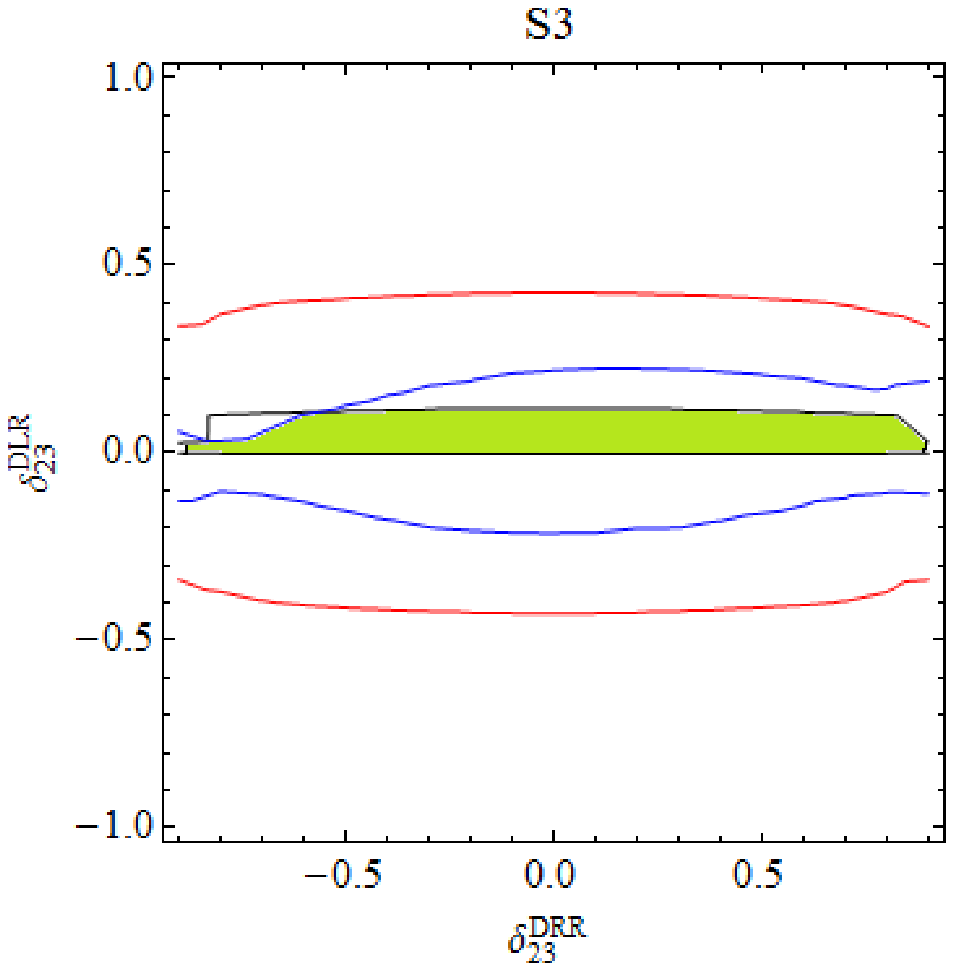,scale=0.71,angle=0,clip=}
\psfig{file=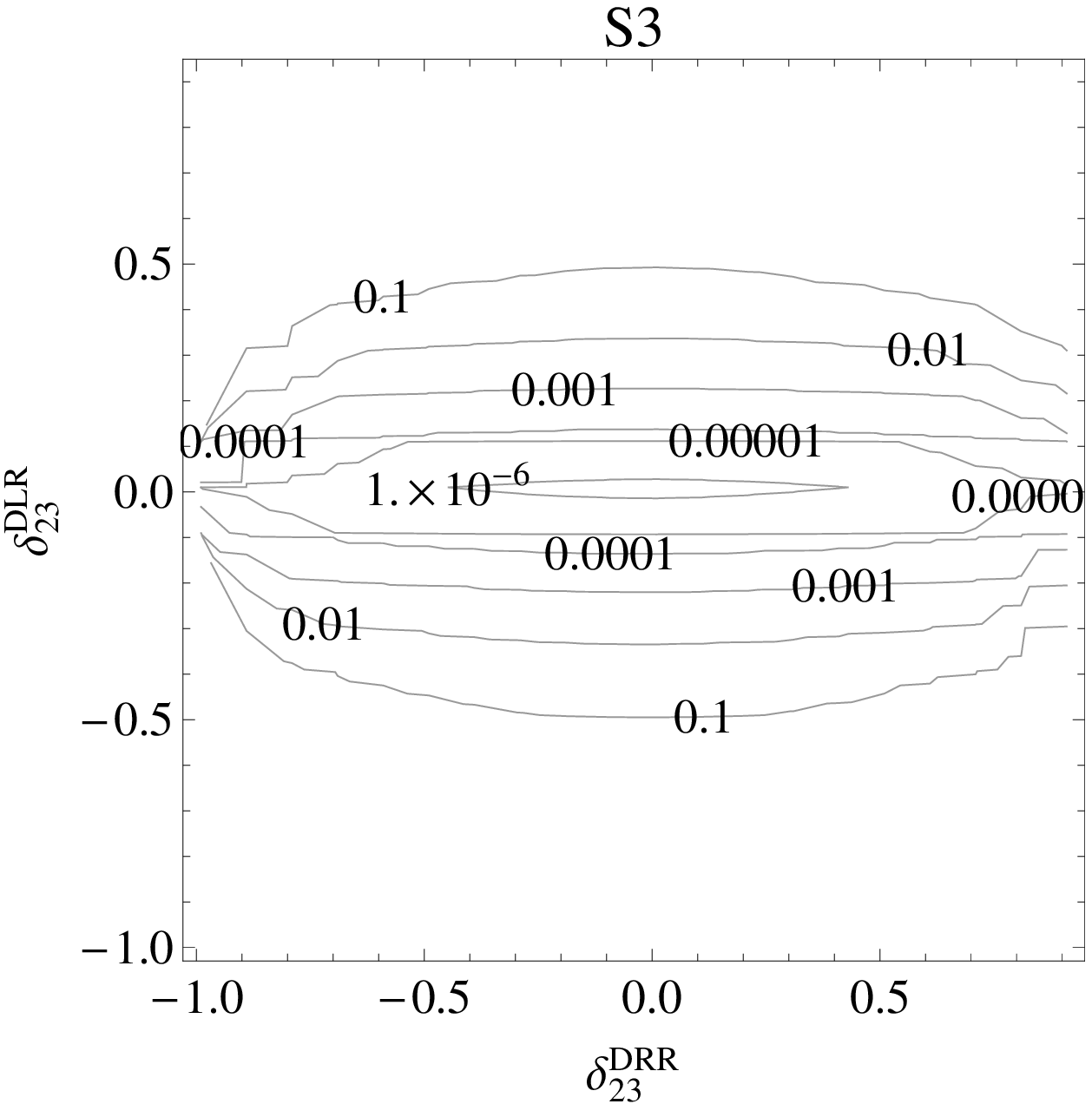,scale=0.54,angle=0,clip=}
\vspace{-1em}
\end{center}
\caption{Left: Contours of \bsg\ (Black), \bmm\ (Green), \dmbs\ (Blue) and
$\MW$ (Red) in ($\del{DRR}{23}$ , $\del{DLR}{23}$) plane for points S1-S3. 
The shaded area shows the range of values allowed by all
cronstraints. Right: corresponding countours for \brhbs.}   
\label{Fig:S1S3DLRDRR23}
\end{figure}

\begin{figure}[ht!]
\begin{center}
\psfig{file=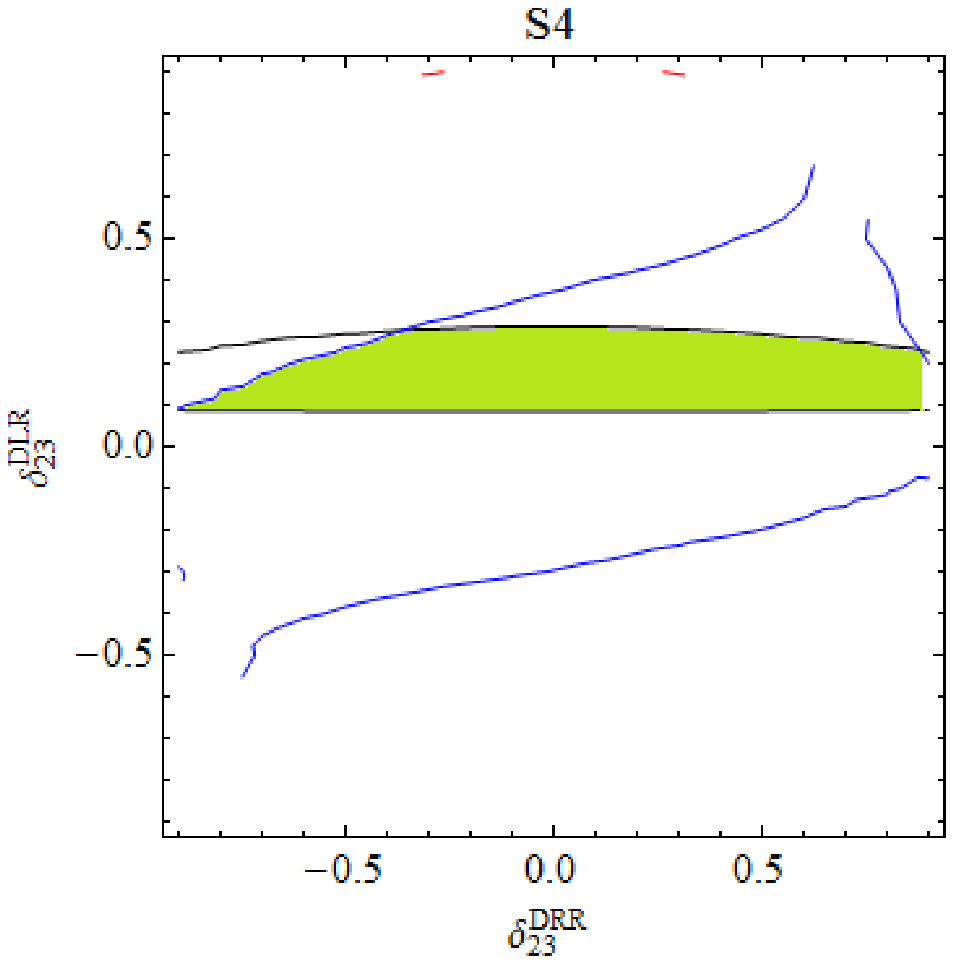,scale=0.72,angle=0,clip=}
\psfig{file=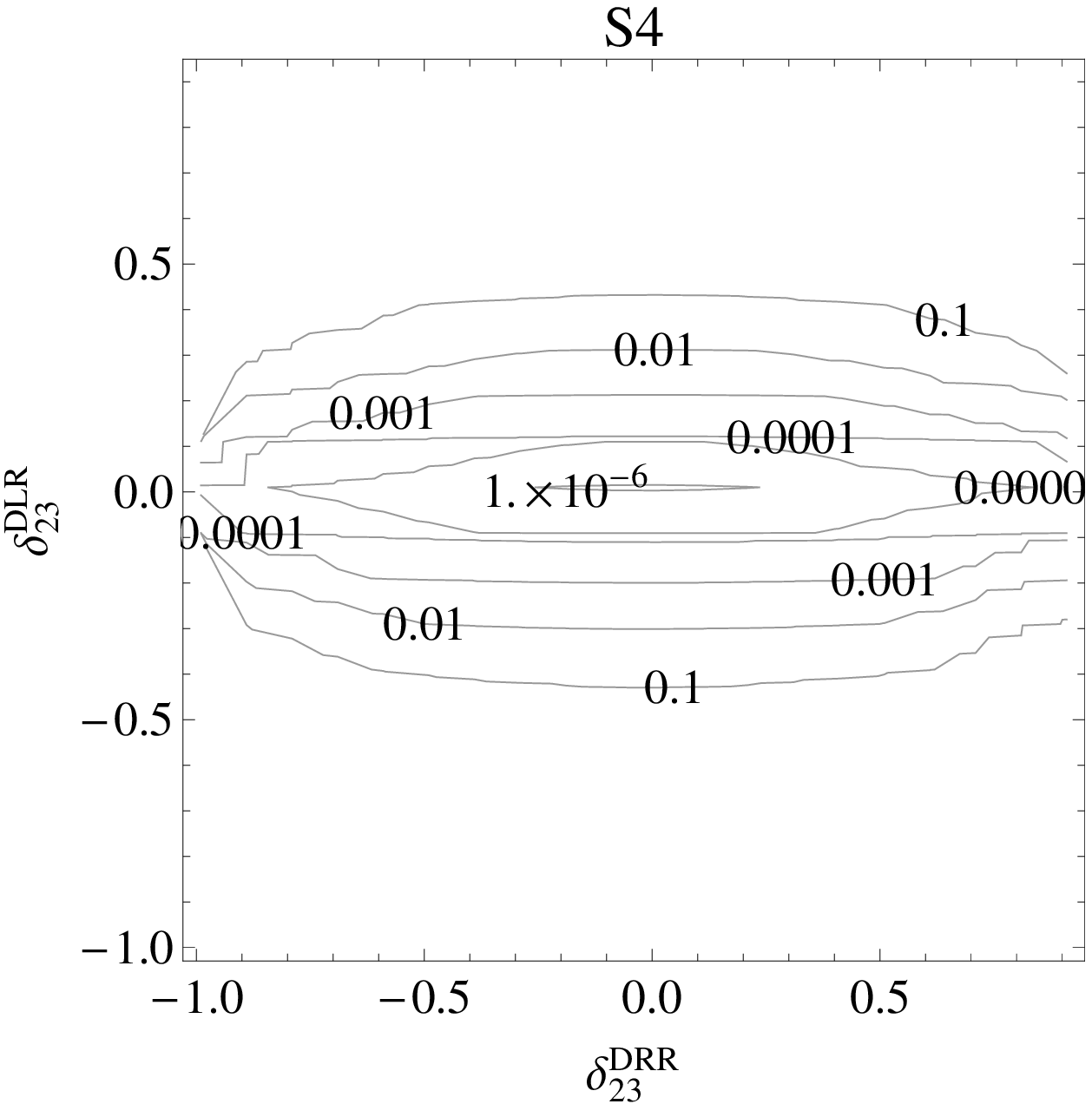,scale=0.55,angle=0,clip=}\\
\vspace{0.2cm}
\psfig{file=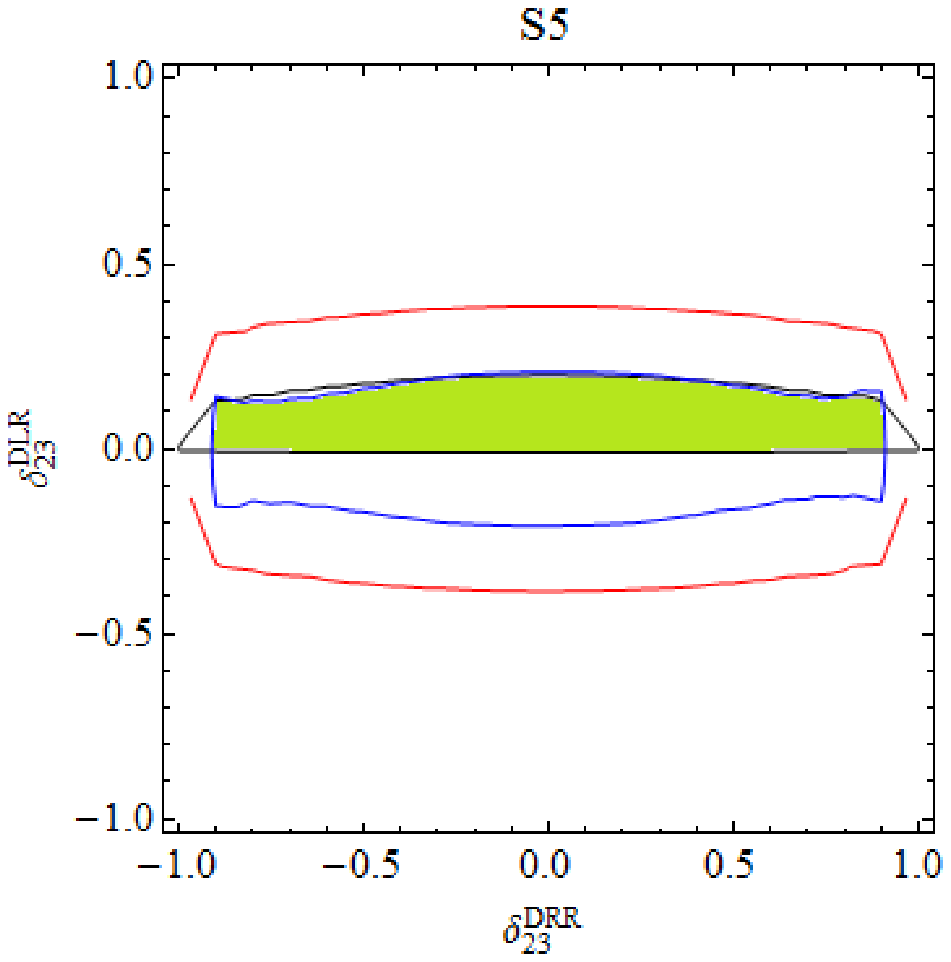,scale=0.71,angle=0,clip=}
\psfig{file=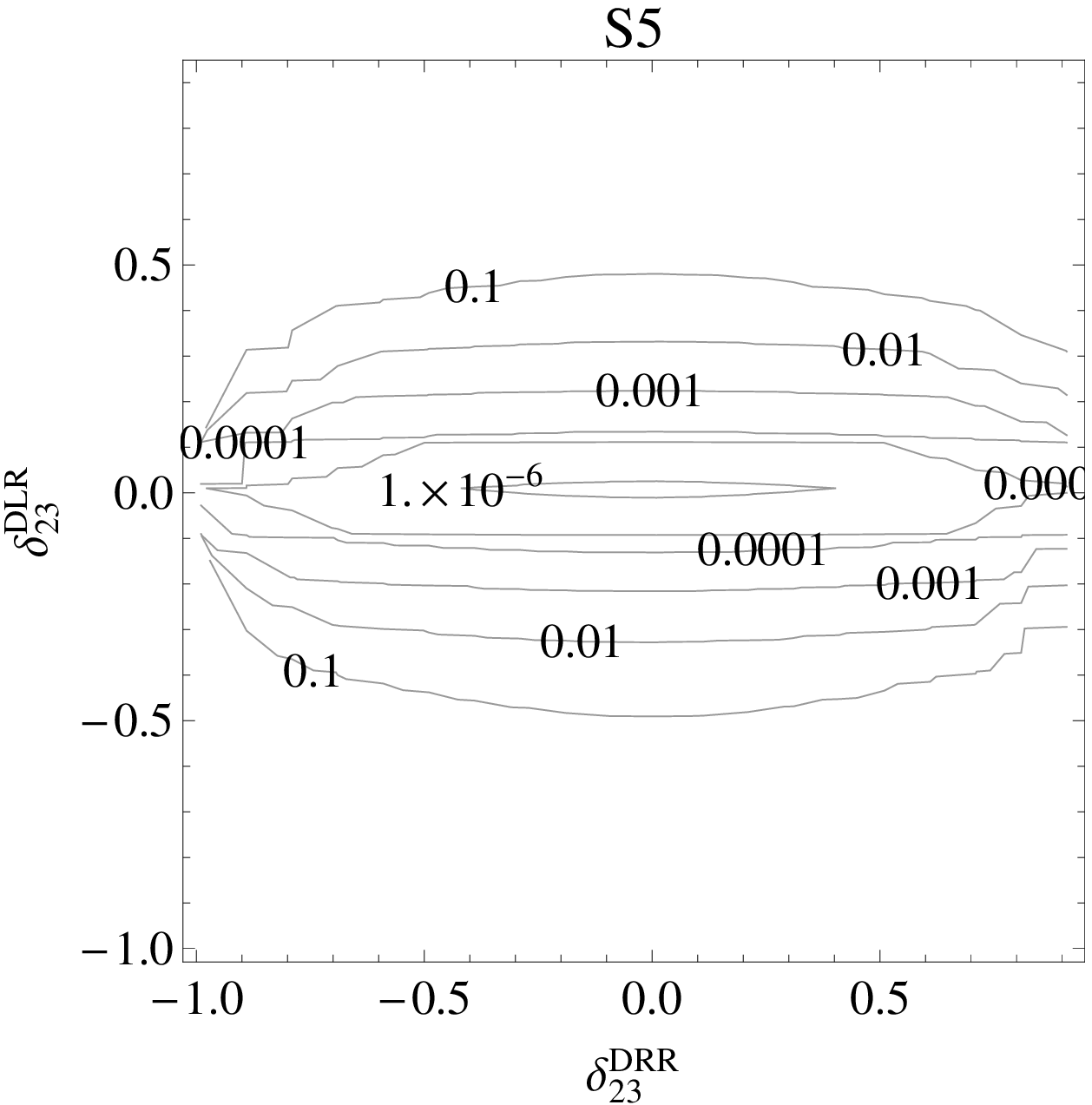,scale=0.54,angle=0,clip=}\\
\vspace{0.2cm}
\psfig{file=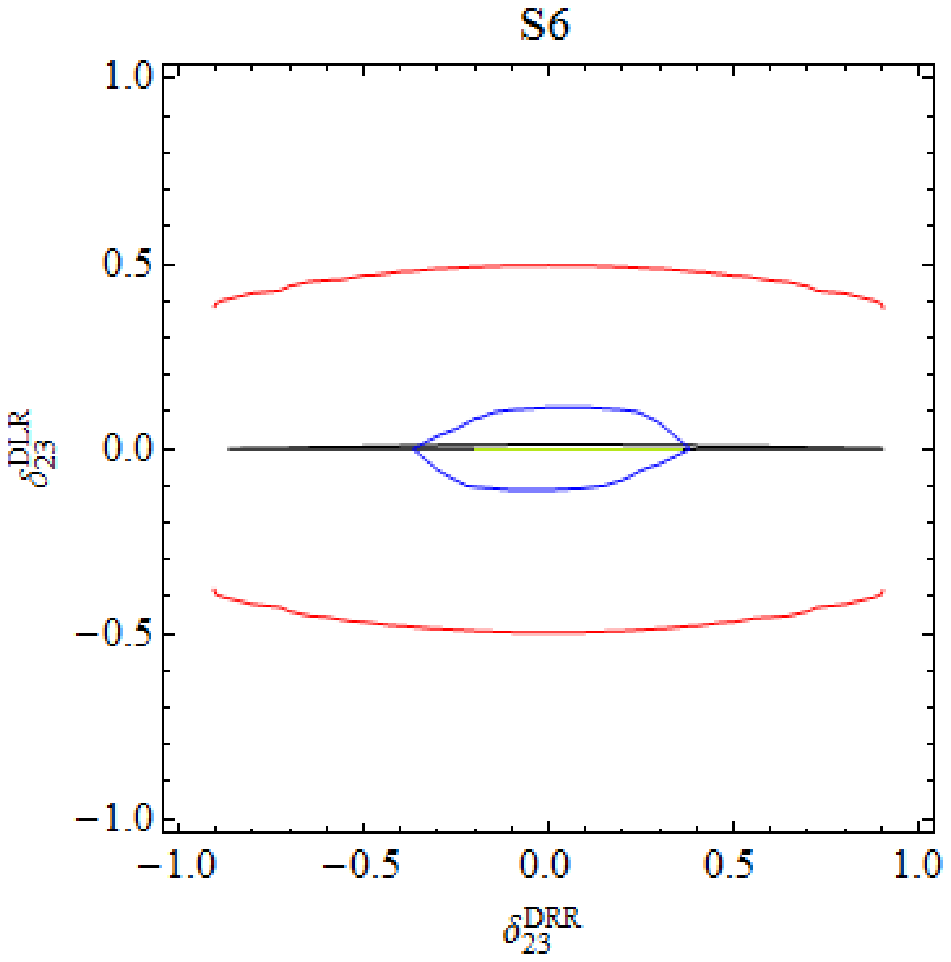,scale=0.71,angle=0,clip=}
\psfig{file=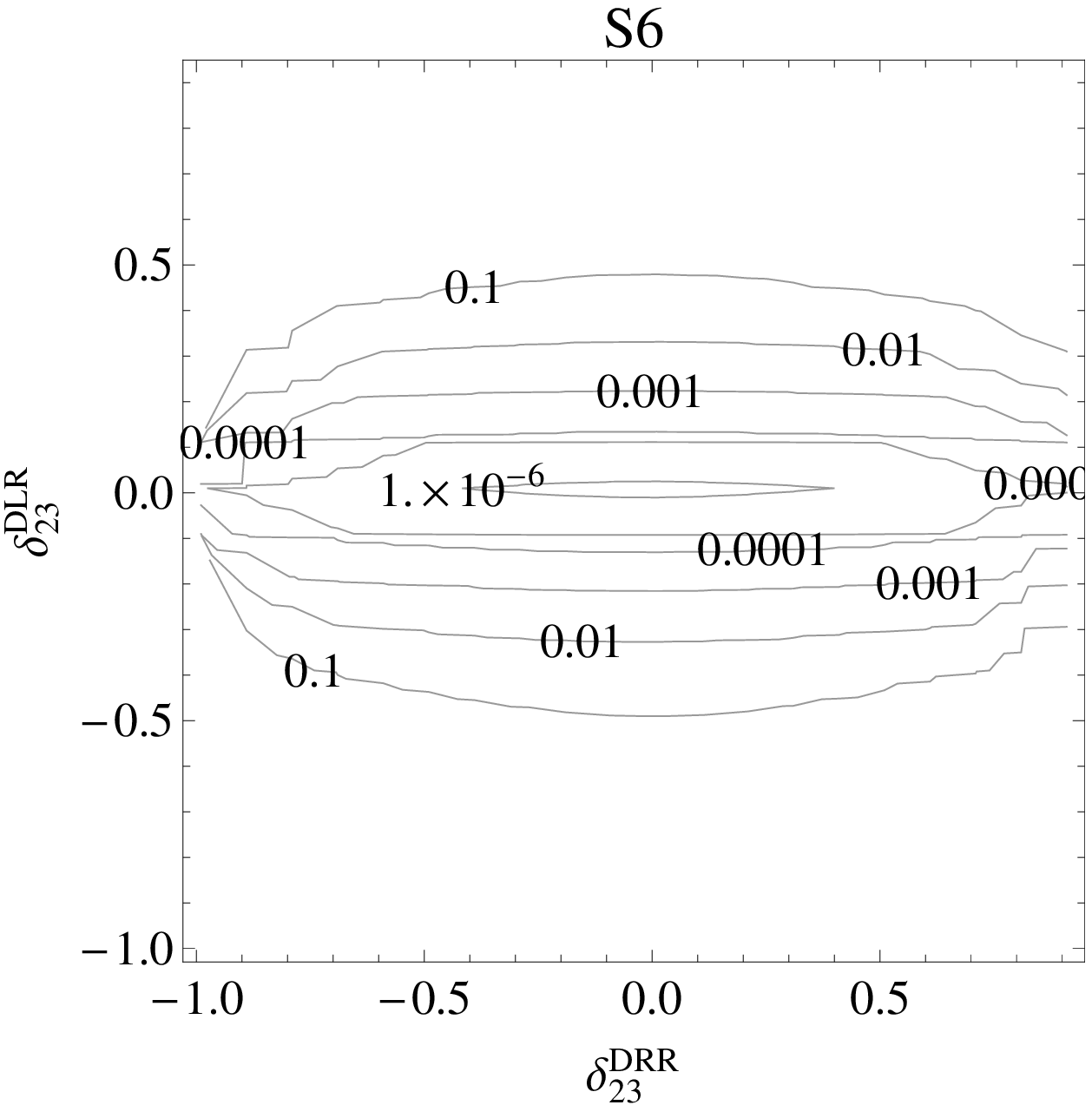,scale=0.54,angle=0,clip=}
\vspace{-1em}
\end{center}
\caption{Left: Contours of \bsg\ (Black), \bmm\ (Green), \dmbs\ (Blue) and
$\MW$ (Red) in ($\del{DRR}{23}$ , $\del{DLR}{23}$) plane for points S4-S6. 
The shaded area shows the range of values allowed by all
cronstraints. Right: corresponding countours for \brhbs.}   
\label{Fig:S4S6DLRDRR23}
\end{figure}

In \reffis{Fig:S1S3QLLDLR23}-\ref{Fig:S4S6DLRDRR23}, in the left columns
we show the $3\,\si$ contours (with experimental and theory
uncertainties added linearly) of 
\bsg\ (Black), \bmm\ (Green), \dmbs\ (Blue) and $\MW$ (Red). 
For non-visible contours the whole plane is allowed by that constraint. 
The right columns show, for the same parameters, the results for \brhbs. 
In \reffis{Fig:S1S3QLLDLR23} and \ref{Fig:S4S6QLLDLR23} we present the
results for the plane (\del{QLL}{23}, \del{DLR}{23}) for S1\ldots
S3 and for S4\ldots S6, respectively. Similarly,
in \reffis{Fig:S1S3DLRDRR23} and \ref{Fig:S4S6DLRDRR23} we show the
(\del{DRR}{23}, \del{DLR}{23}) plane.
The shaded area in the left columns indicates the area that is allowed
by all experimental constraints. 
In the (\del{QLL}{23}, \del{DLR}{23}) planes one can see that the
large values for $\del{QLL}{23}$ are not allowed by $\MW$, on the
other hand, \bsg\ mostly restricts the value of
$\del{DLR}{23}$. 
The largest values for \brhbs\ in each plane in the arrea allowed by the
BPO and the EWPO are summarized in the upper part of \refta{tab:brhbs-2d}. 
One can see that 
in most cases we find $\br(\hbs) \sim \order{10^{-5}}$, 
which would render the observation difficult at current and future
colliders. However, in the ($\del{QLL}{23},\del{DLR}{23}$) plane in the
scenarios S4 and S5 maximum values of \order{3 \times 10^{-4}} can be
observed, which could be detectable at future ILC measurements.
In the (\del{DRR}{23}, \del{DLR}{23}) plane for these two scenarios
even values of \order{10^{-3}} are reached, which would make a
measurement of the flavor violating Higgs decay relatively easy at the
ILC.

\renewcommand{\arraystretch}{1.2}
\begin{table}[htb!]
\begin{center}
\resizebox{11.0cm}{!} {
\begin{tabular}{|c|c|c|c|} \hline
 Plane & MSSM point & Maximum possible value & Figure \\ \hline
($\del{QLL}{23},\del{DLR}{23}$) & \begin{tabular}{c}  S1 \\ S2 \\ S3 \\ S4 \\ S5 \\ S6 \end{tabular} &  
\begin{tabular}{c} 
$1.38 \times 10^{-5}$ \\ $1.39 \times 10^{-5}$ \\ $1.43 \times 10^{-5}$
  \\ $3.34 \times 10^{-4}$ \\ $2.74 \times 10^{-4}$ \\ $1.36 \times 10^{-8}$ 
\end{tabular} &
\begin{tabular}{c} 
\reffi{Fig:S1S3QLLDLR23} \\ \reffi{Fig:S1S3QLLDLR23} \\ \reffi{Fig:S1S3QLLDLR23} \\
\reffi{Fig:S4S6QLLDLR23} \\ \reffi{Fig:S4S6QLLDLR23} \\ \reffi{Fig:S4S6QLLDLR23}
\end{tabular}
\\ \hline
($\del{DRR}{23},\del{DLR}{23}$)  & \begin{tabular}{c}  S1 \\ S2 \\ S3 \\ S4 \\ S5 \\ S6 \end{tabular}    
& \begin{tabular}{c} 
$4.41 \times 10^{-6}$ \\ $3.32 \times 10^{-6}$ \\ $3.07 \times 10^{-5}$
    \\ $1.66 \times 10^{-3}$ \\ $1.97 \times 10^{-3}$ \\ $6.03 \times 10^{-8}$ 
\end{tabular} & 
\begin{tabular}{c} 
\reffi{Fig:S1S3DLRDRR23} \\ \reffi{Fig:S1S3DLRDRR23} \\ \reffi{Fig:S1S3DLRDRR23} \\
\reffi{Fig:S4S6DLRDRR23} \\ \reffi{Fig:S4S6DLRDRR23} \\ \reffi{Fig:S4S6DLRDRR23}
\end{tabular}
\\ \hline\hline
\begin{tabular}{c}($\del{QLL}{23},\del{DLR}{23}$) \\ with $\del{DRR}{23}= 0.5$ \end{tabular} & \begin{tabular}{c}  S1 \\ S2 \\ S3 \\ S4 \\ S5 \\ S6 \end{tabular}    & 
\begin{tabular}{c} 
$7.49 \times 10^{-6}$ \\ $7.33 \times 10^{-6}$ \\$3.50 \times 10^{-6}$
  \\ Excluded \\ Excluded \\ Excluded \end{tabular} &
\begin{tabular}{c} 
\reffi{Fig:S1S3QLLDLRDRR23} \\ \reffi{Fig:S1S3QLLDLRDRR23} \\ \reffi{Fig:S1S3QLLDLRDRR23} \\
\reffi{Fig:S4S6QLLDLRDRR23} \\ \reffi{Fig:S4S6QLLDLRDRR23} \\ \reffi{Fig:S4S6QLLDLRDRR23}
\end{tabular}  \\ \hline
\end{tabular}}  
\end{center}
\caption{Maximum possible value for $\brhbs$ for two and three $\deFABij \neq 0$ case for the selected S1-S6 MSSM points defined in
\refta{tab:spectra}. 
}
\label{tab:brhbs-2d}
\end{table}
\renewcommand{\arraystretch}{1.55}

\begin{figure}[ht!]
\begin{center}
\psfig{file=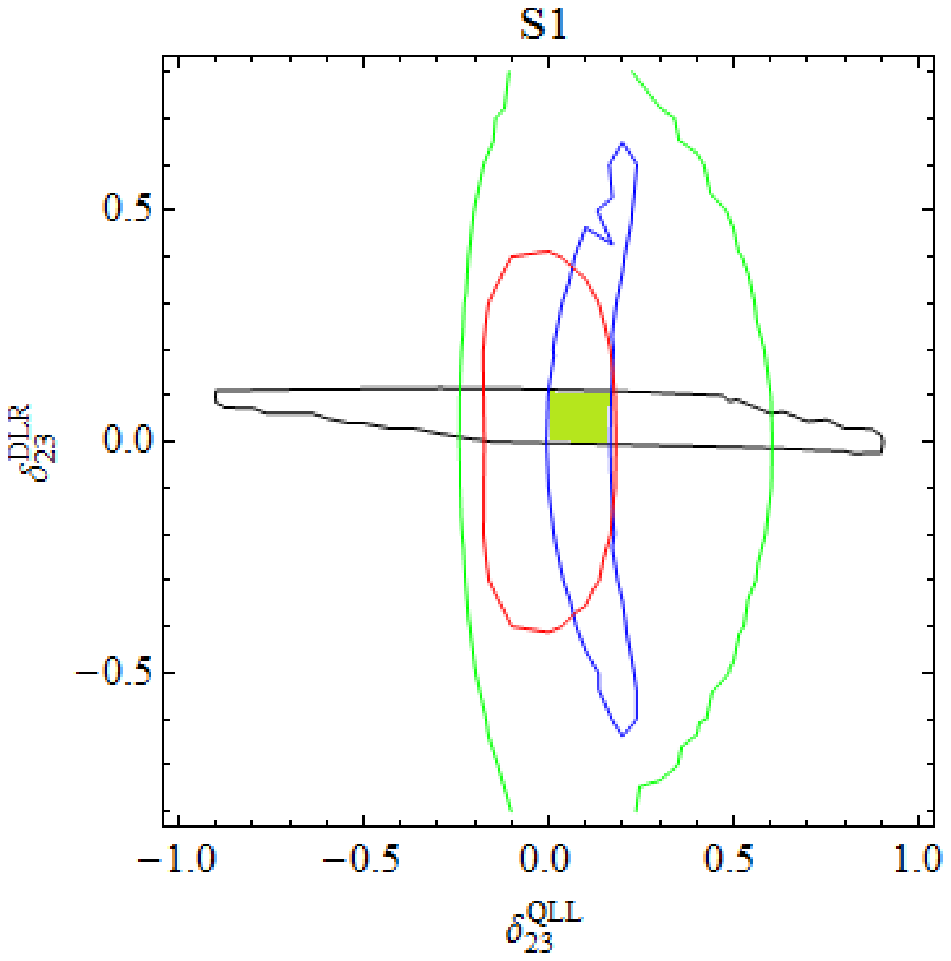,scale=0.72,angle=0,clip=}
\psfig{file=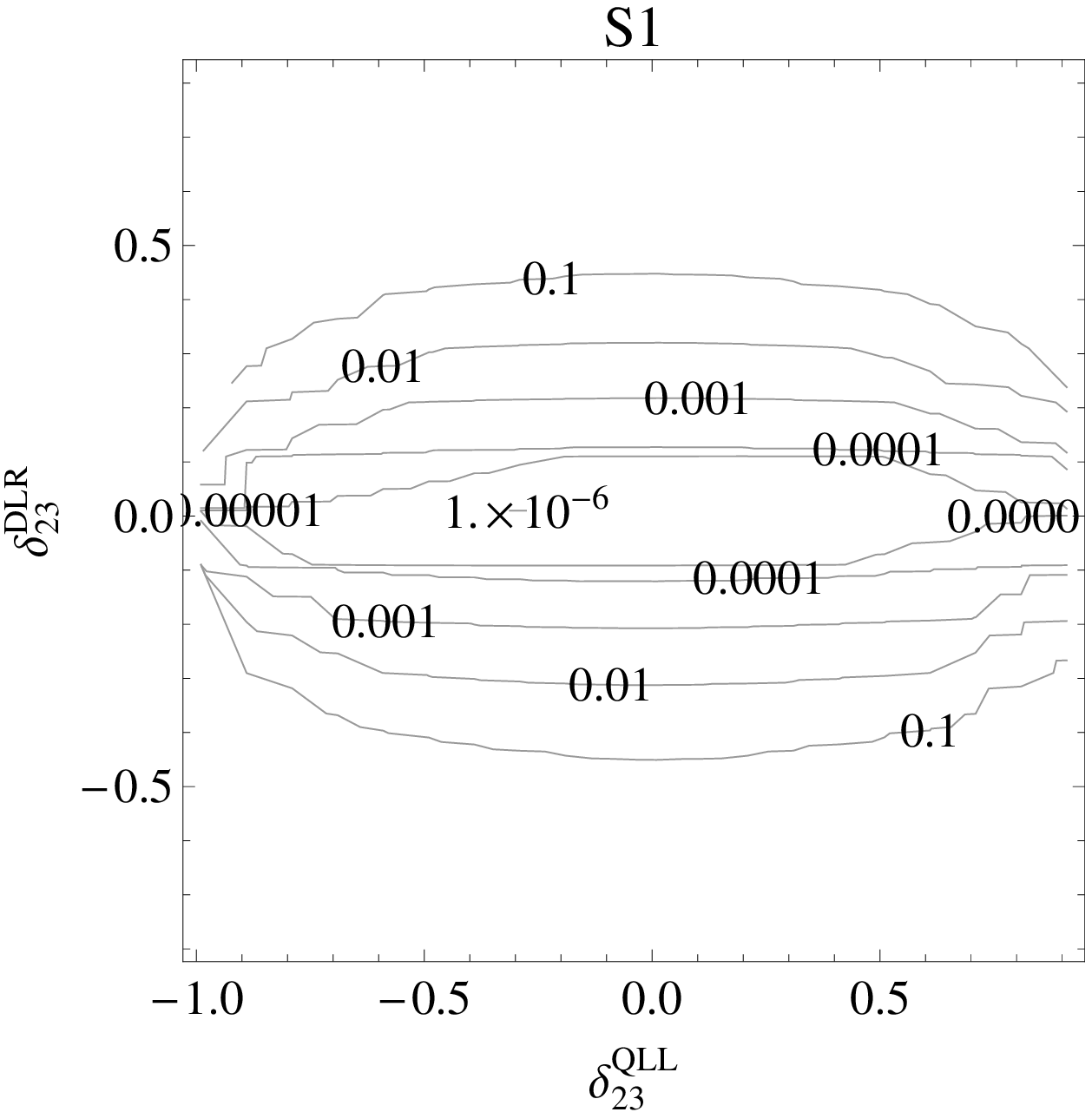,scale=0.55,angle=0,clip=}\\
\vspace{0.2cm}
\psfig{file=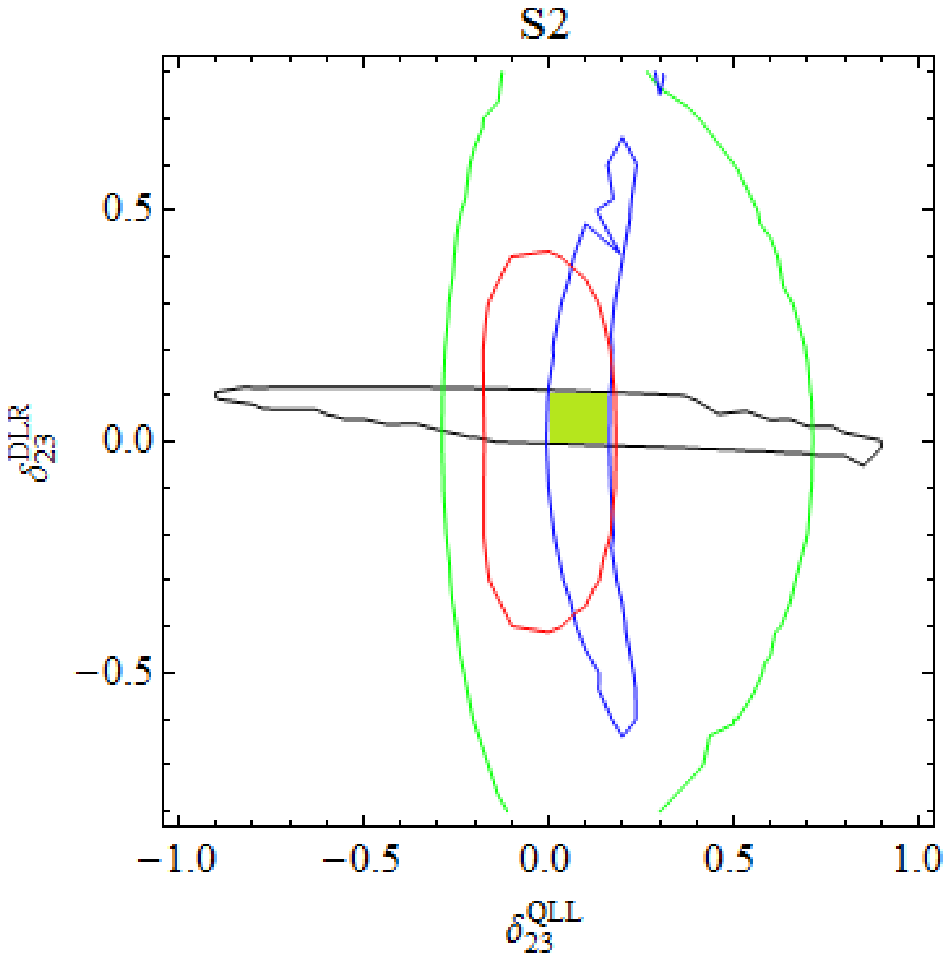,scale=0.71,angle=0,clip=}
\psfig{file=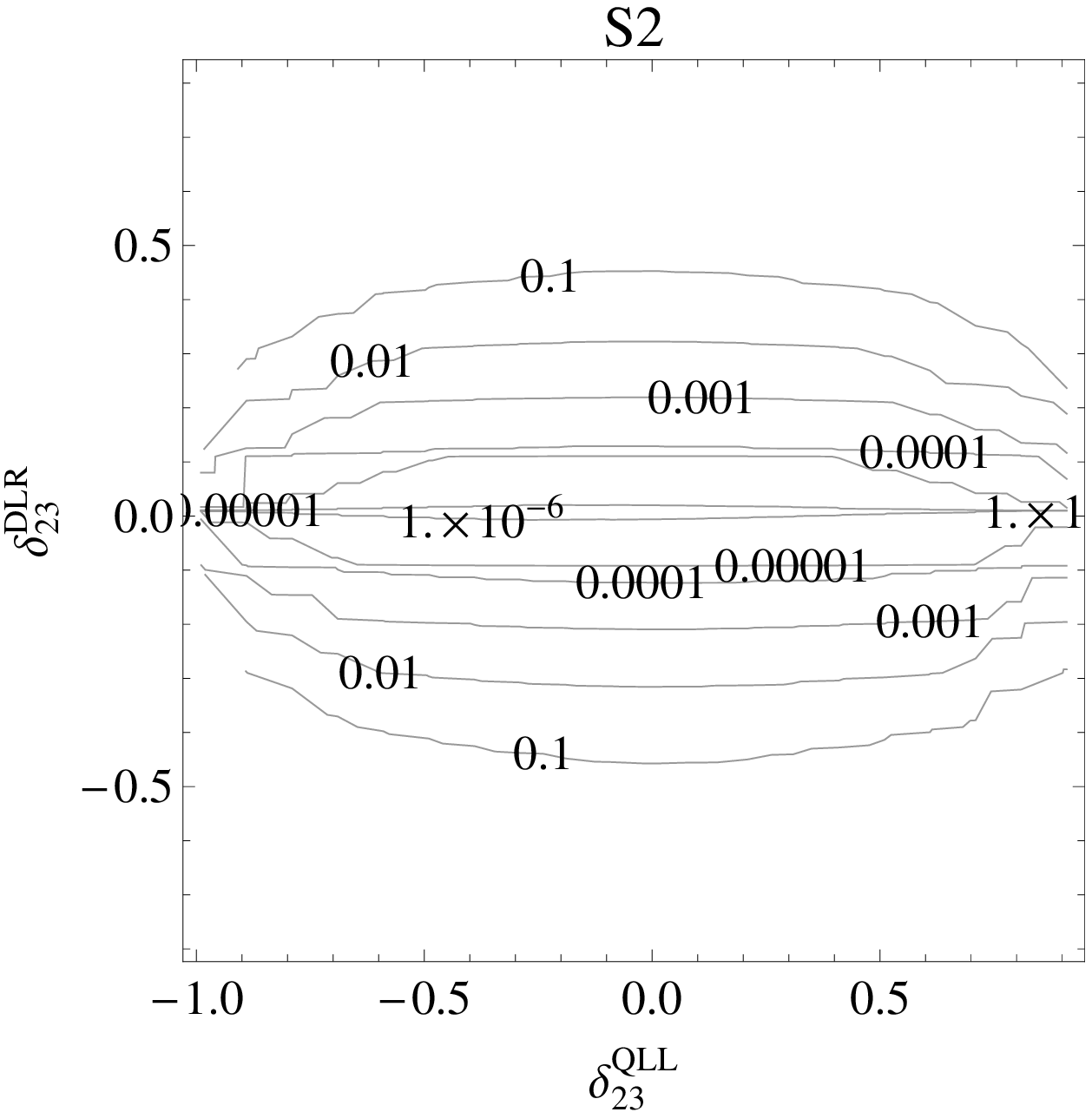,scale=0.52,angle=0,clip=}\\
\vspace{0.2cm}
\psfig{file=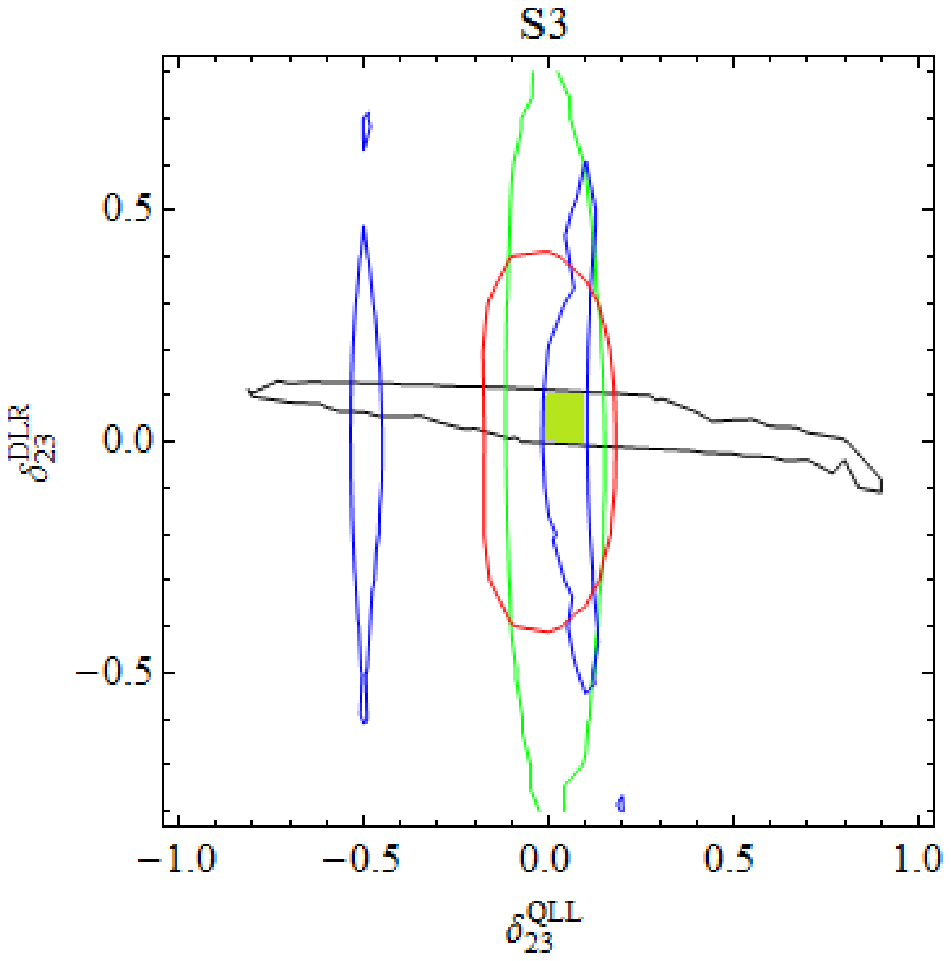,scale=0.71,angle=0,clip=}
\psfig{file=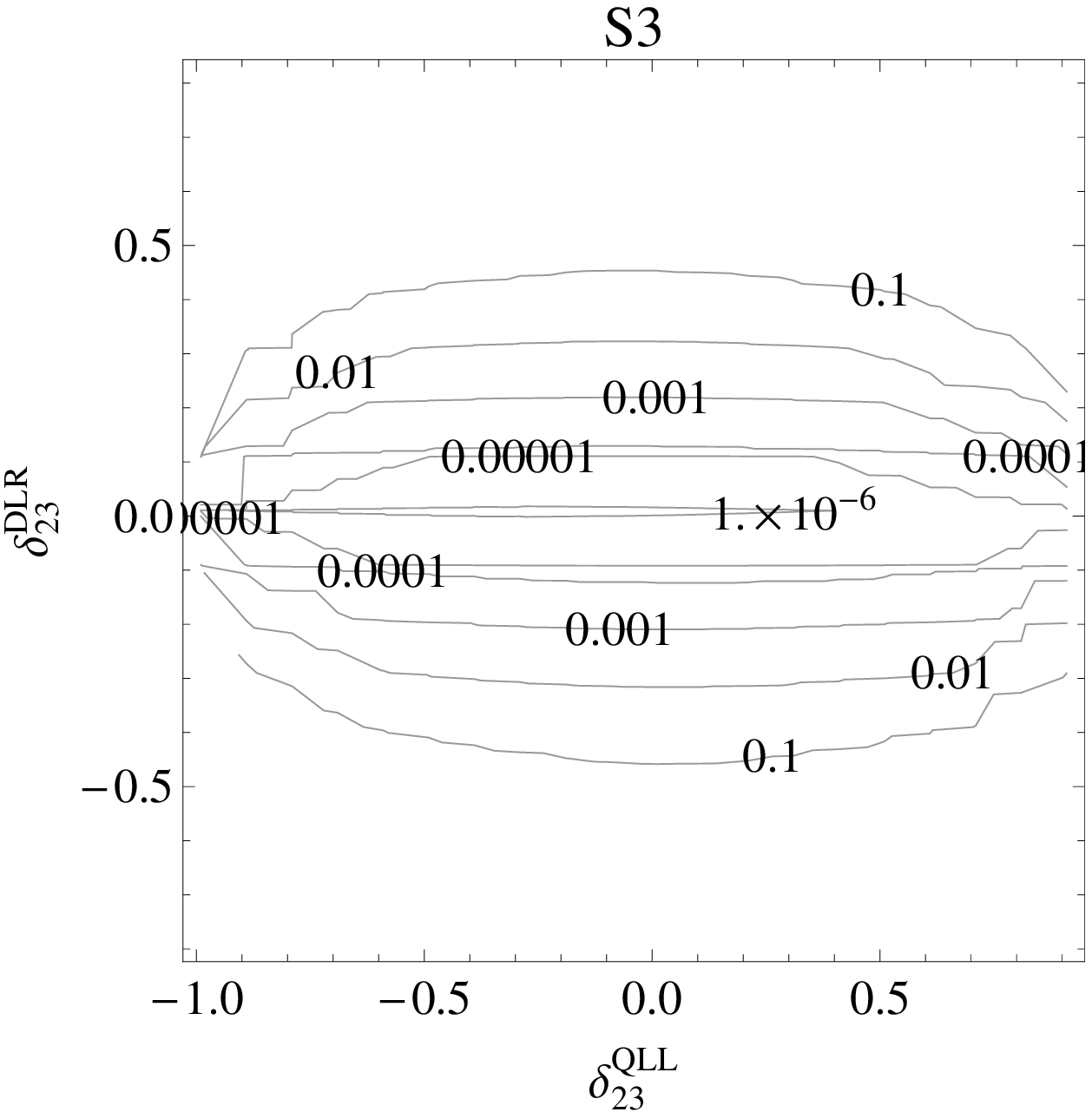,scale=0.54,angle=0,clip=}
\vspace{-1em}
\end{center}
\caption{Left: Contours of \bsg\ (Black), \bmm\ (Green), \dmbs\ (Blue) and
$\MW$ (Red) in the ($\del{QLL}{23}$ , $\del{DLR}{23}$) plane with 
$\del{DRR}{23} = 0.5$ for points S1-S3. The shaded area shows the range
    of values allowed by all 
cronstraints. Right: corresponding countours
for \brhbs.}    
\label{Fig:S1S3QLLDLRDRR23}
\end{figure}

\begin{figure}[ht!]
\begin{center}
\psfig{file=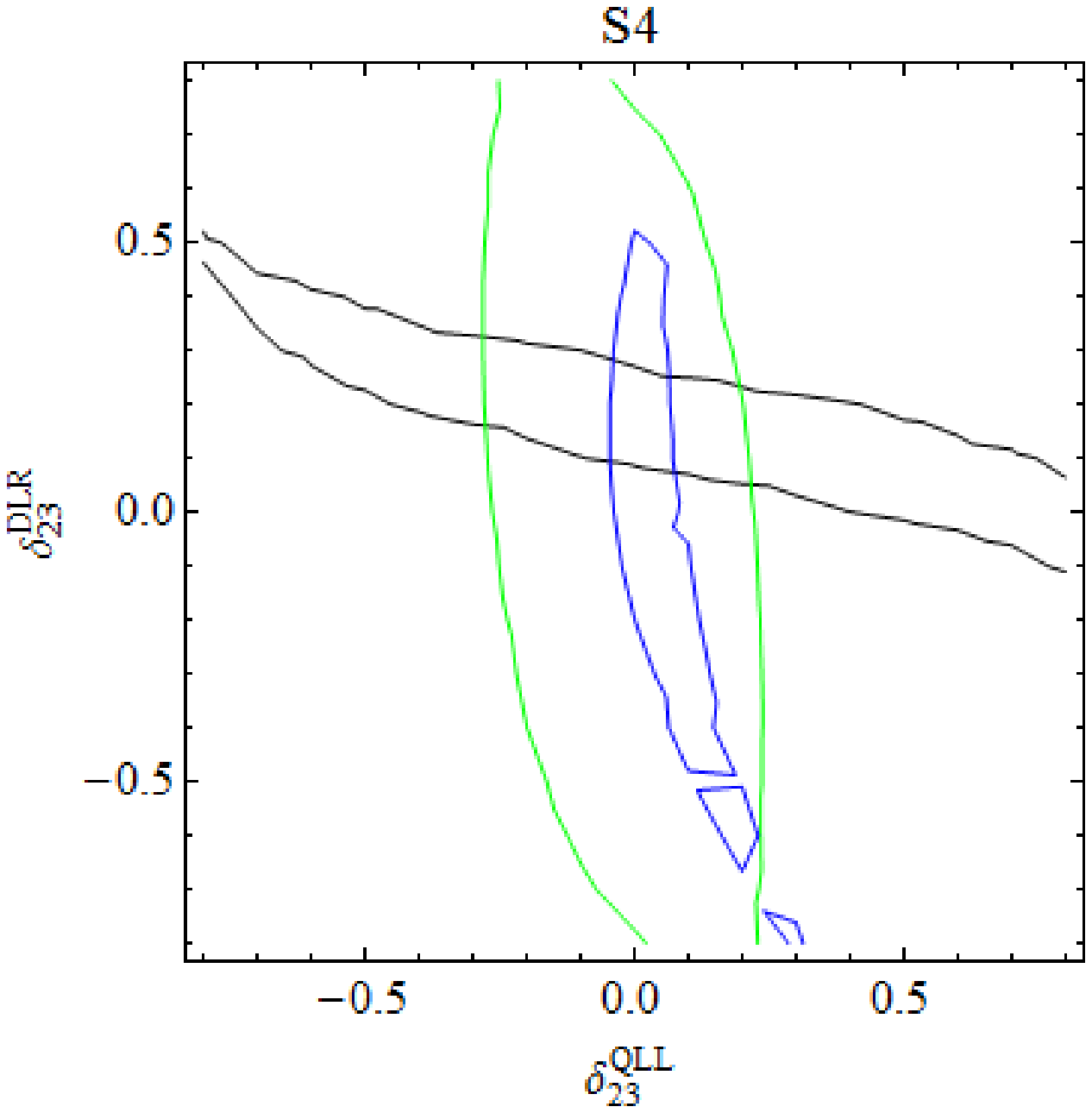,scale=0.55,angle=0,clip=}
\psfig{file=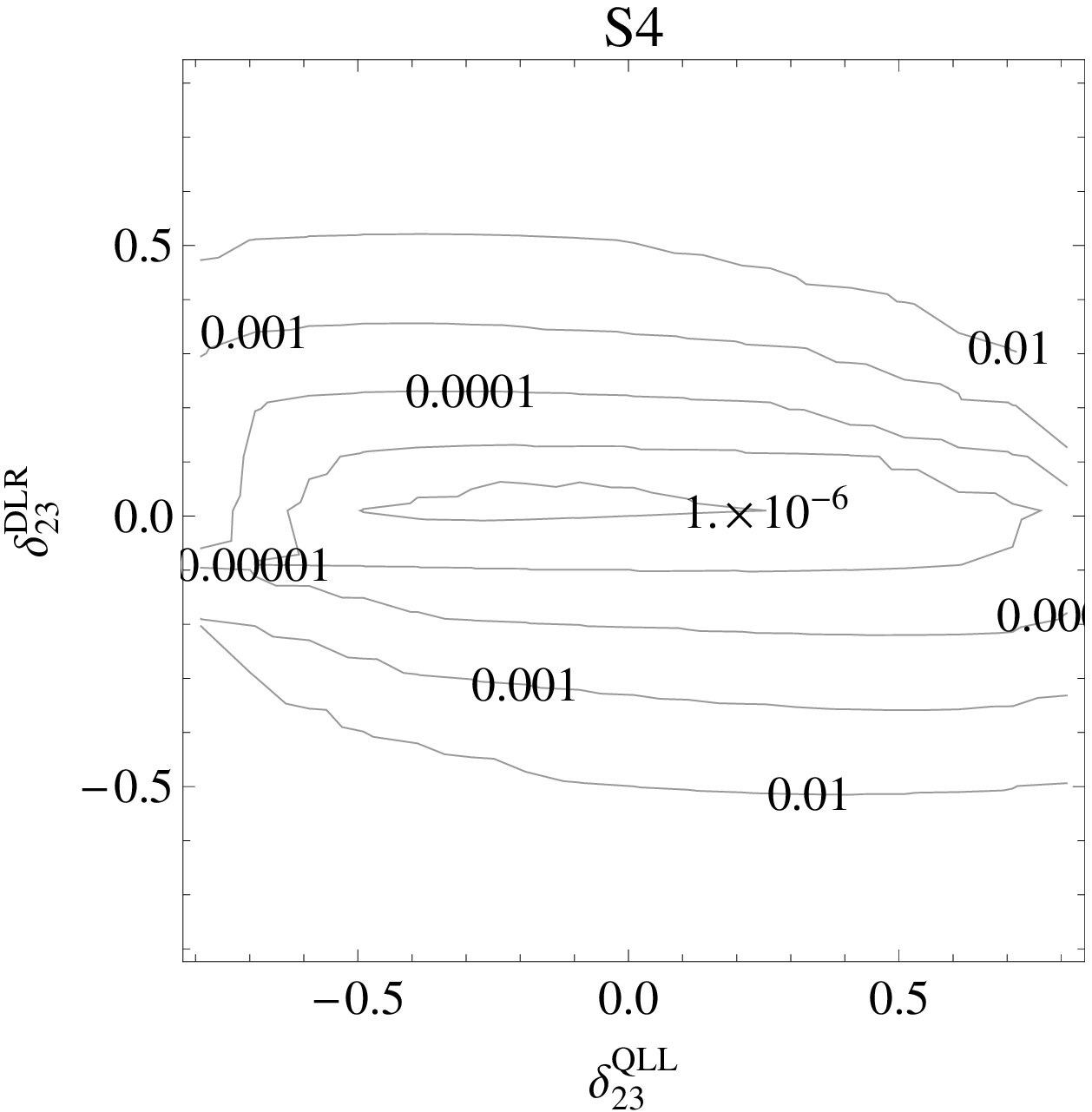,scale=0.55,angle=0,clip=}\\
\vspace{0.2cm}
\psfig{file=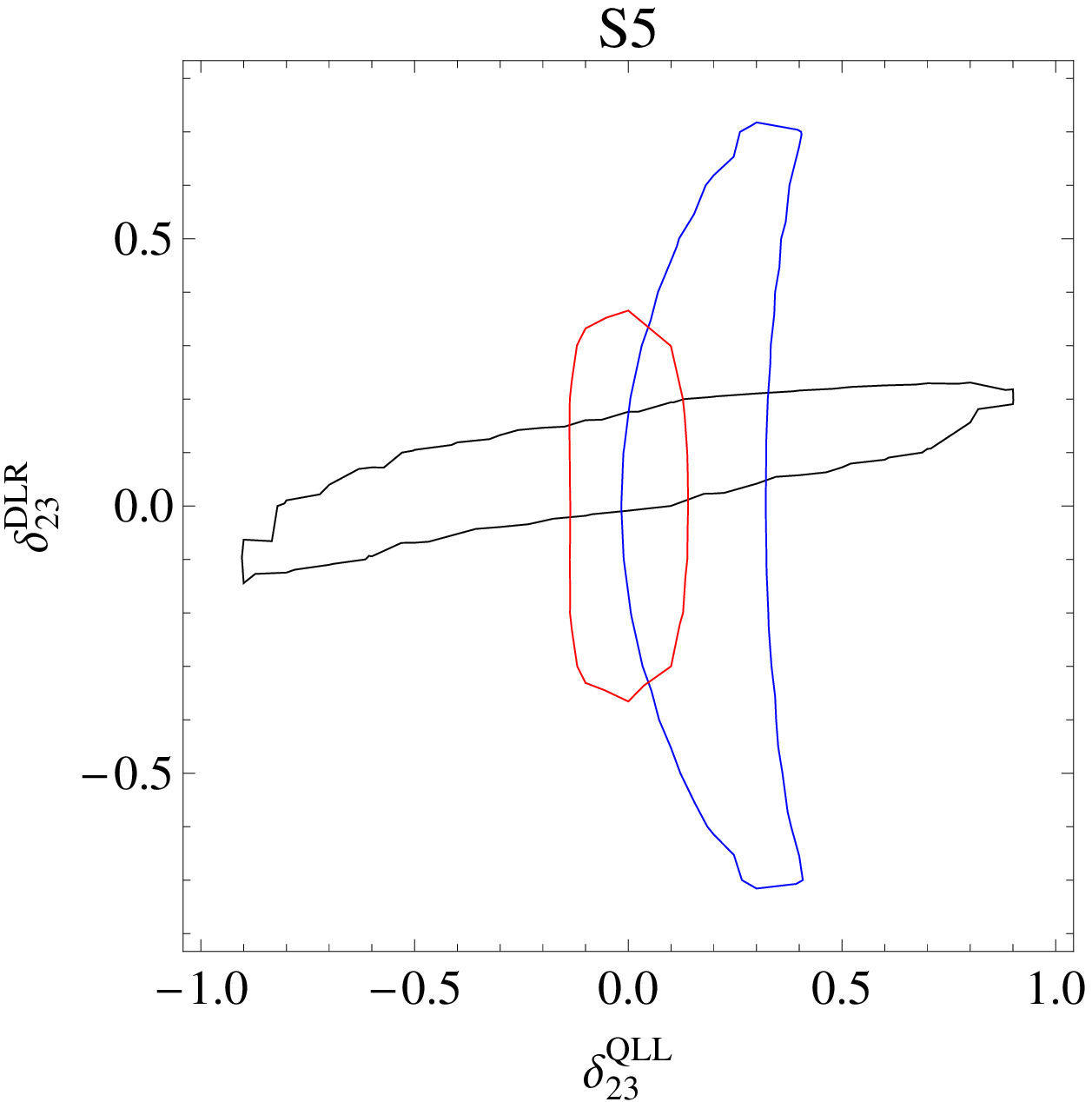,scale=0.54,angle=0,clip=}
\psfig{file=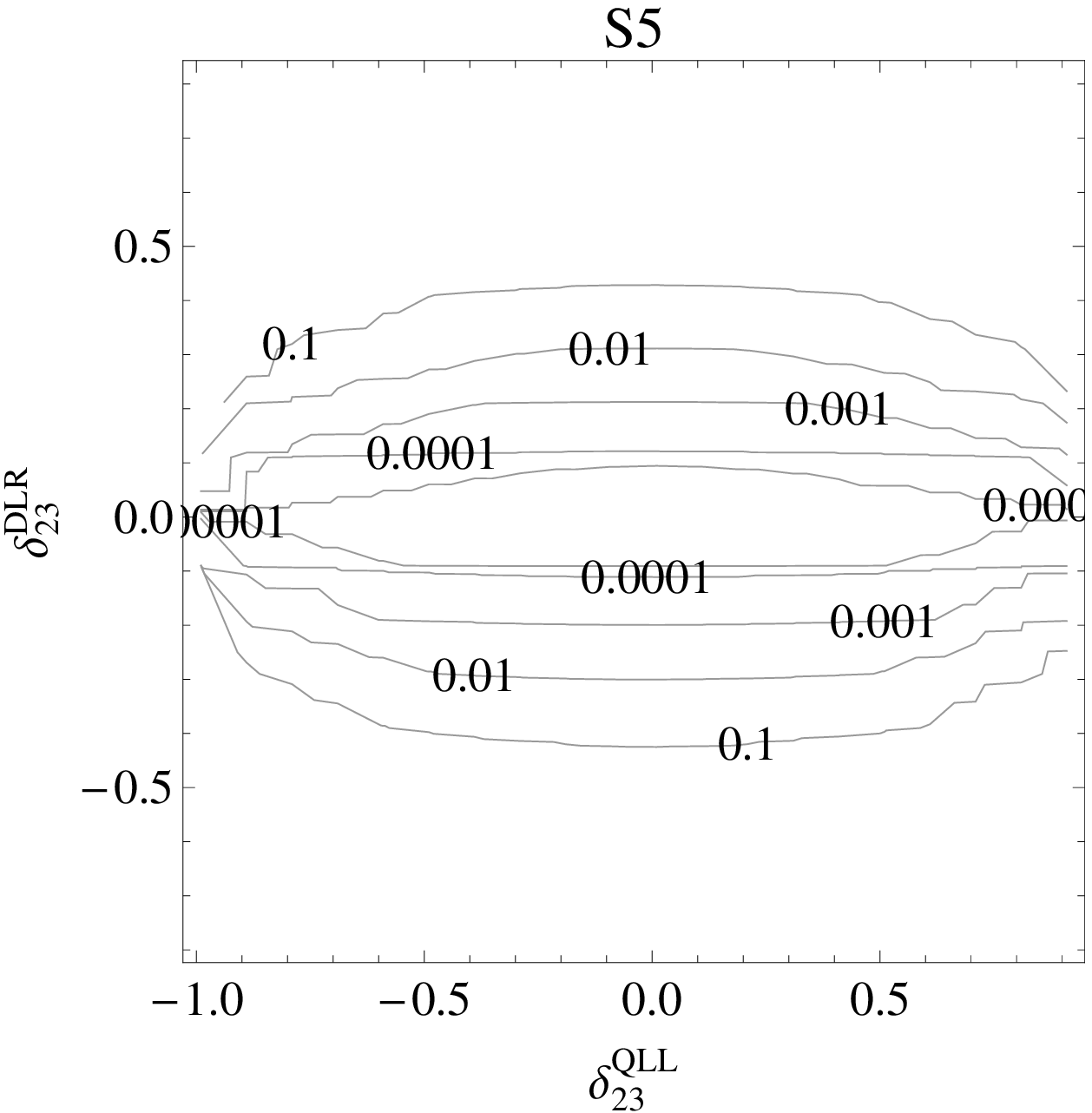,scale=0.54,angle=0,clip=}\\
\vspace{0.2cm}
\psfig{file=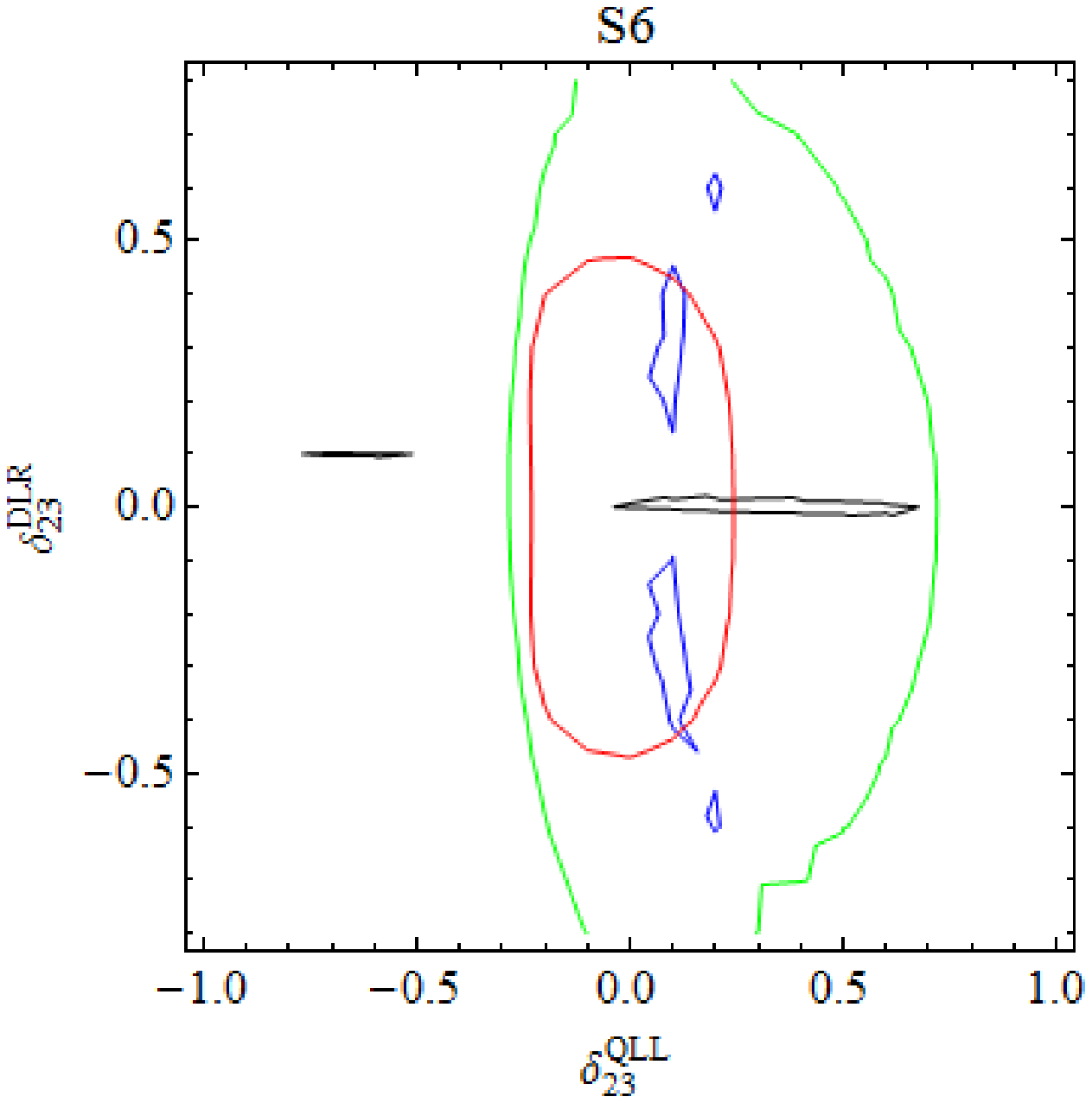,scale=0.55,angle=0,clip=}
\psfig{file=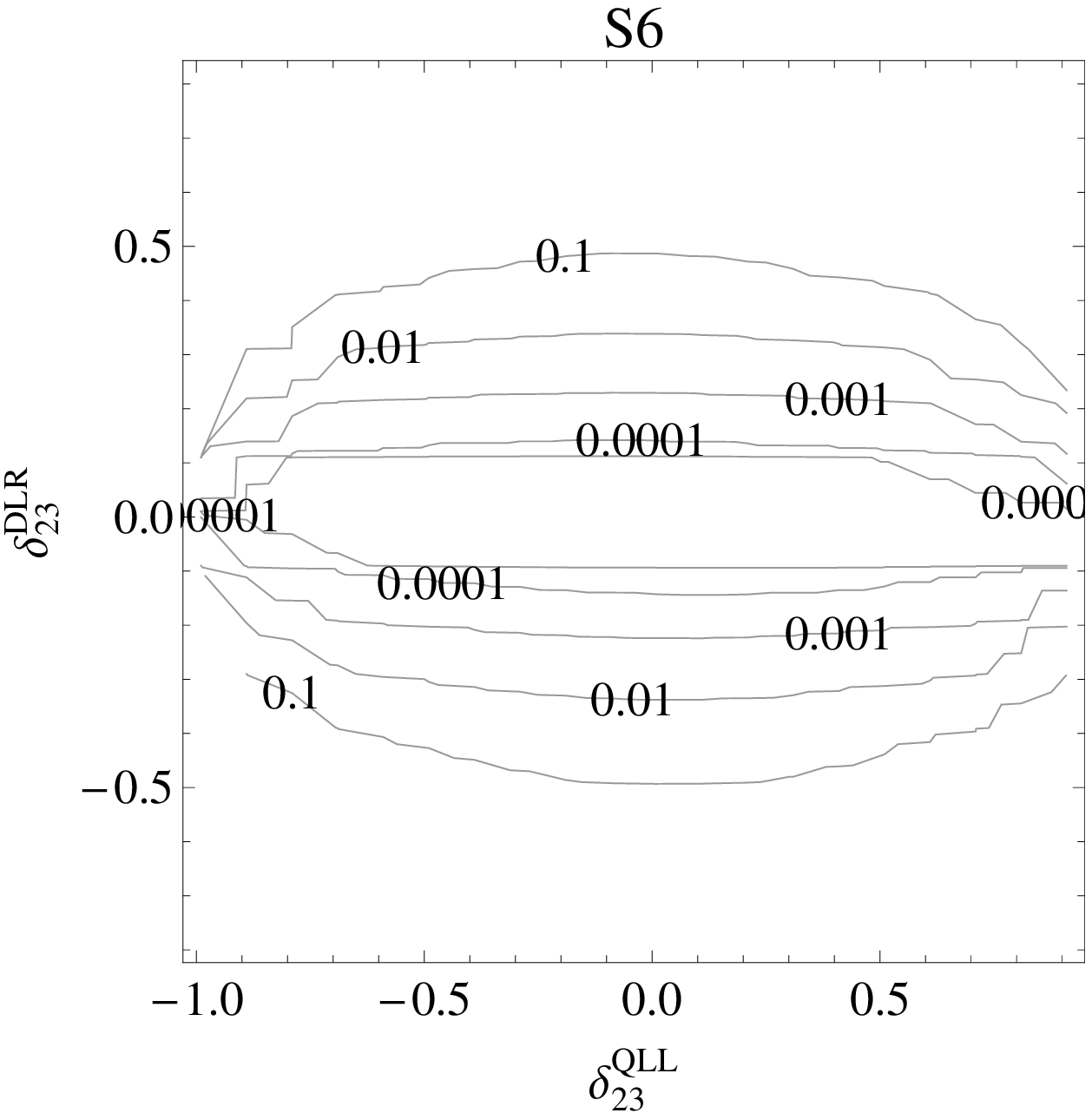,scale=0.54,angle=0,clip=}
\vspace{-1em}
\end{center}
\caption{Left: Contours of \bsg\ (Black), \bmm\ (Green), \dmbs\ (Blue) and
$\MW$ (Red) in the ($\del{QLL}{23}$ , $\del{DLR}{23}$) plane with 
$\del{DRR}{23} = 0.5$ for points S4-S6. The shaded area shows the range
    of values allowed by all 
cronstraints. Right: corresponding countours
for \brhbs.}    
\label{Fig:S4S6QLLDLRDRR23}
\end{figure}

\medskip
As a last step in model independent analysis, we consider the case of
three $\deFABij \neq 0 $ at a time. For this purpose we 
scan the parameters in the ($\del{QLL}{23}$, $\del{DLR}{23}$)
plane and set $\del{DRR}{23} = 0.5$. For reasons of practicability
we choose {\em one} intermediate value for $\del{DRR}{23}$; a very small
value will have no additional effect, and a very large value of
$\del{DRR}{23}$  leads to large excluded areas in the 
($\del{QLL}{23}$, $\del{DLR}{23}$) plane.
We show our results in \reffis{Fig:S1S3QLLDLRDRR23} and 
\ref{Fig:S4S6QLLDLRDRR23} in the scenarios S1-S3 and S4-S6,
respectively. Colors and shadings are chosen as in the previous
analysis.
Here it should be noted that in S4 the whole plane is excluded by
$\MW$, and in S5 by \bmm\ (both contours are not visible). In S6 no
overlap between the four constraints is found, and again this scenario
is excluded. We have checked that also a smaller value of
$\del{DRR}{23} = 0.2$ does not qualitatively change the picture for
S4, S5 and S6.
The highest values that can be reached for \brhbs\ in the three
remaining scenarios in the experimentally
allowed regions are shown in the lower part of \refta{tab:brhbs-2d}. 
One can see only very small valus or \order{5 \times 10^{-6}} are
found, i.e.\ choosing $\del{DRR}{23} \neq 0$ did not lead to
observable values of \brhbs.

\medskip
To summarize, in our model independent analysis, allowing for more than
one $\deFABij \neq 0$ we find 
that the additional freedom resulted in somewhat larger values of
\brhbs\ as compared to the case of only one non-zero $\deFABij$.
In particular in the two scenarios S4 and S5 values of 
$\brhbs \sim 10^{-3} - 10^{-4}$ can be reached, allowing the
detection of the flavor violating Higgs decay at the ILC. The other
scenarios always yield values that are presumably too low for
current and future colliders.

\clearpage
\newpage

\subsection{Numerical results in MFV CMSSM}
\label{sec:cmssm-ana}

In this final step of our numerical analysis we investigate the CMSSM as
described in \refse{sec:cmssm}. Here the MFV hypothesis is realized by
demanding no flavor violation at the GUT scale, and the various flavor
violating $\deFABij$ are induced by the RGE running to the EW scale. 
For this analysis the SUSY spectra have been generated with the code 
{\tt SPheno 3.2.4}~\cite{Porod:2003um}. 
We started with the definition of the (MFV) SLHA file~\cite{SLHA} at the
GUT scale. In a first step within {\tt SPheno}, gauge and
Yukawa couplings at $\MZ$ scale are calculated using tree-level
formulas. Fermion masses, the $Z$~boson pole mass, the fine structure constant
$\alpha$, the Fermi constant $G_F$ and the strong coupling constant
$\alpha_s(\MZ)$ are used as input parameters. The gauge and Yukawa
couplings, calculated at $\MZ$, are then used as 
input for the one-loop
RGE's to obtain the corresponding values at the GUT scale
which is calculated from the requirement that $g_1 = g_2$
(where $g_{1,2}$ denote the gauge couplings of the $U(1)$ and
$SU(2)$, respectively). The CMSSM boundary
conditions (with the numerical values from the SLHA file) are then
applied to the complete set of two-loop RGE's and are evolved to the EW scale.  
At this point the SM and SUSY radiative
corrections are applied to the gauge and Yukawa couplings, and the 
two-loop RGE's are again evolved to GUT scale. 
After applying the CMSSM  boundary conditions again
the two-loop RGE's are run down to EW scale to get SUSY spectrum. This
procedure is iterated until the required precision is achieved. 
The output is given in the form of an SLHA, file which is used as 
input for {\tt FeynHiggs} to calculate low energy observables discussed above. 

In order to get an overview about the size of the effects in the CMSSM
parameter space, the relevant parameters $m_0$, $m_{1/2}$ have been
scanned as, or in case of $A_0$ and
$\tb$ have been set to all combinations of 
\begin{align}
m_0 &\eq 500 \gev \ldots 5000 \gev~, \non \\
m_{1/2} &\eq 1000 \gev \ldots 3000 \gev~, \non \\
A_0 &\eq -3000, -2000, -1000, 0 \gev~, \non \\
\tb &\eq 10, 20, 35, 45~,
\end{align}
with $\mu > 0$.  

\begin{figure}[ht!]
\begin{center}
\vspace{3.0cm}
\psfig{file=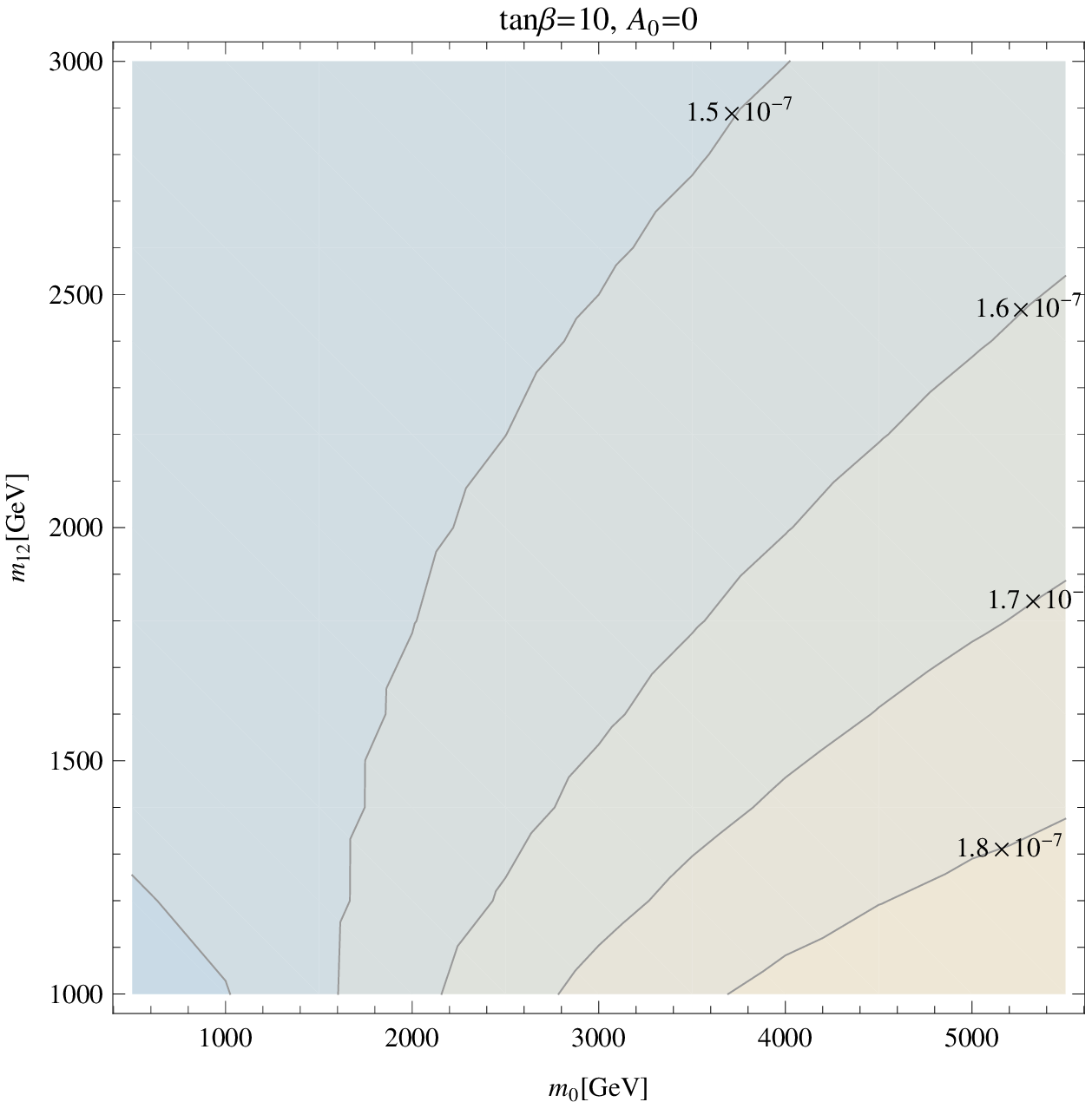,scale=0.57,angle=0,clip=}
\psfig{file=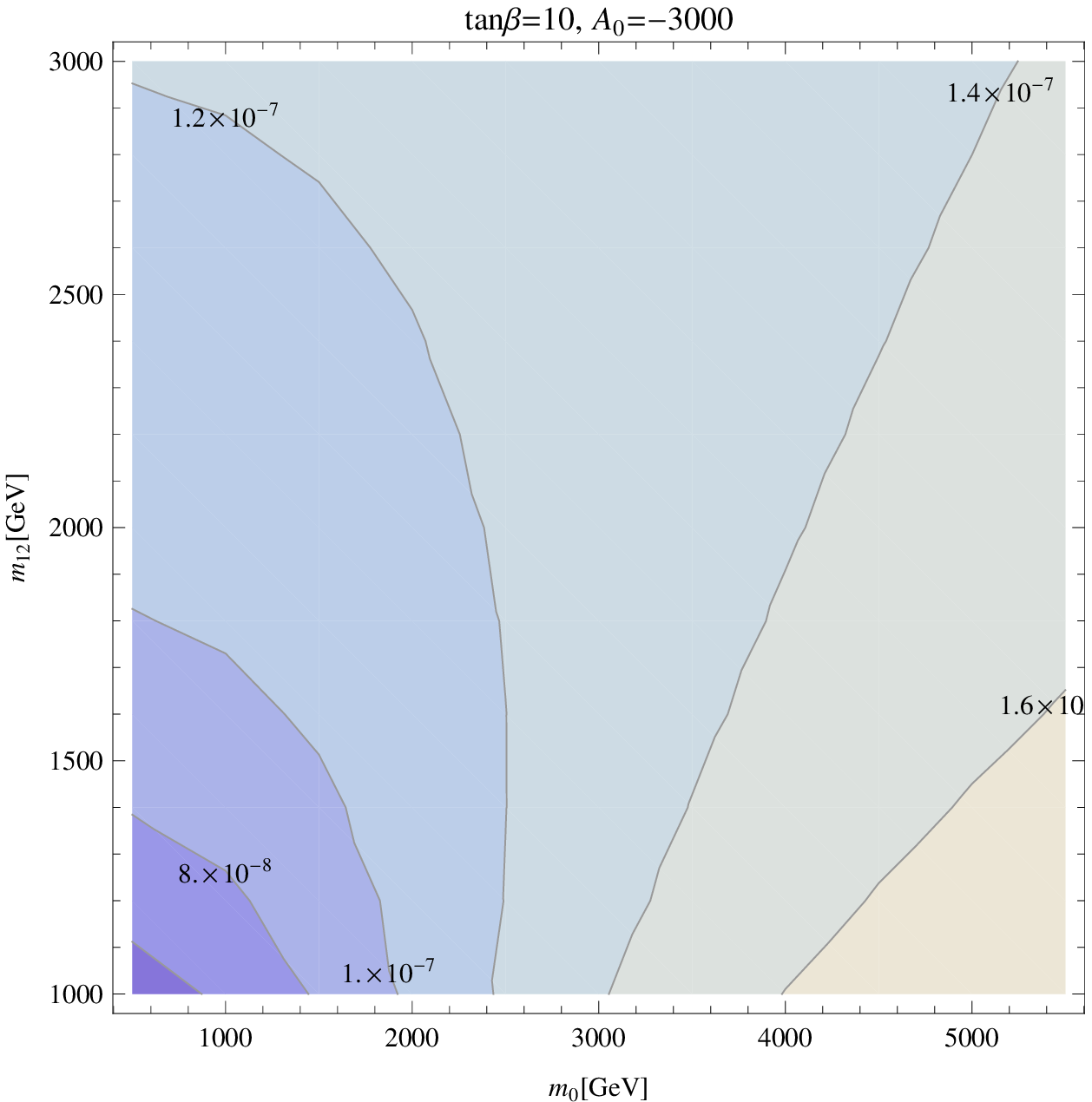,scale=0.57,angle=0,clip=}\\
\vspace{2.0cm}
\psfig{file=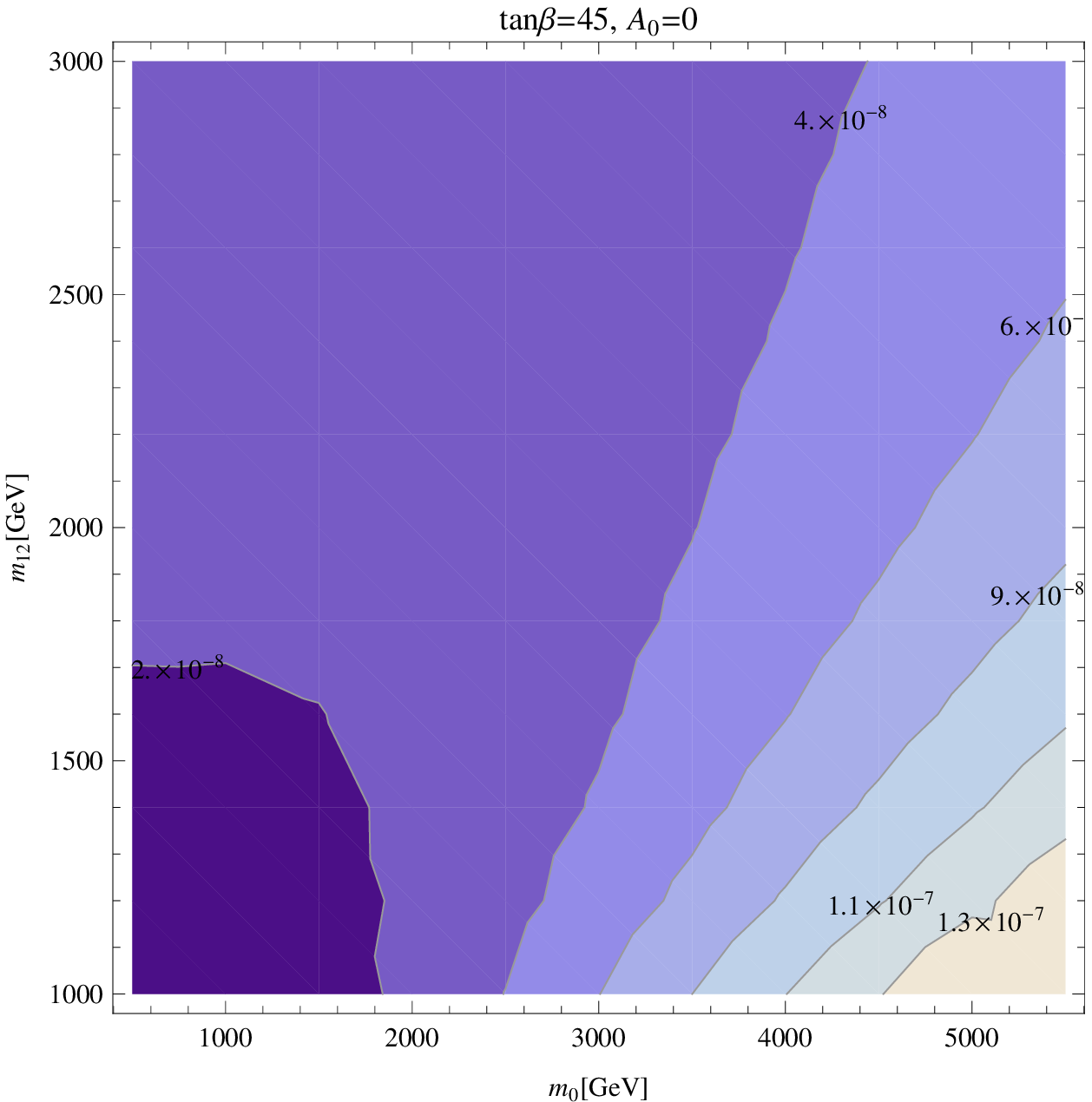,scale=0.56,angle=0,clip=}
\psfig{file=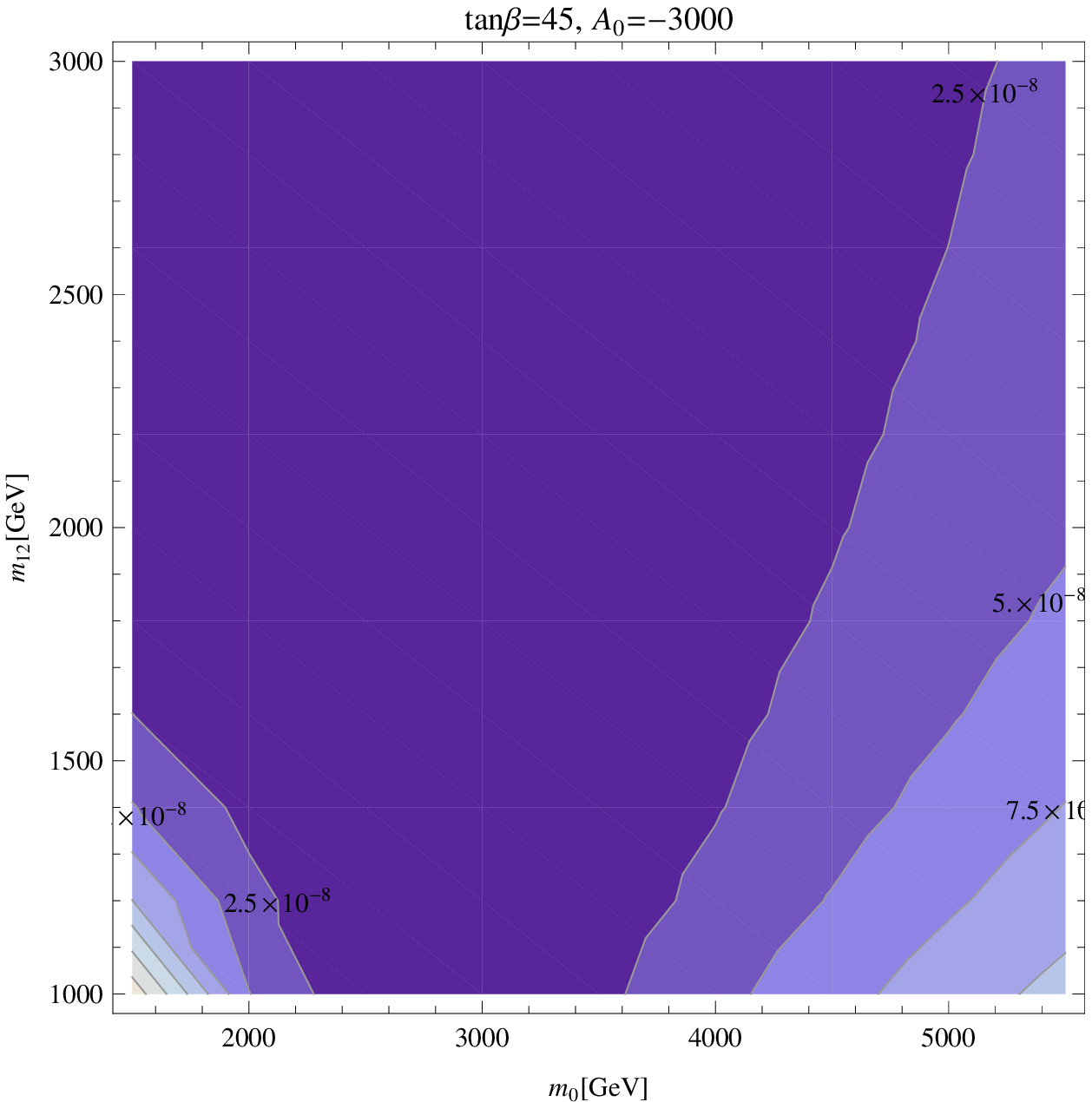,scale=0.56,angle=0,clip=}\\
\vspace{0.2cm}
\end{center}
\caption{Contours of $\br(h \rightarrow b \bar{s}+\bar{b}s)$  in the
  $(m_0$, $m_{1/2})$ plane for different values of $\tb$ and $A_0$ in
  the CMSSM.}   
\label{fig:BRhbs}
\end{figure}

The results are shown in \reffi{fig:BRhbs}, where we display the
contours of \brhbs\ in the ($m_0$, $m_{1/2}$) plane for $\tb=10$,
$A_0=0$ (upper left), $\tb=10$, $A_0=-3000 \gev$ (upper right), 
$\tb=45$, $A_0=0$ (lower left) and $\tb=45$, $A_0=-3000 \gev$ (lower
right). By comparison with planes for other $\tb$-$A_0$ combinations we
have varyfied that these four planes constitute a representative
example. The allowed parameter space can be deduced by comparing to the
results presented in \citeres{MFV-CMSSM,mc9}. While not all the planes are in
agreement with current constraints, large parts, in particular for
larger values of $m_0$ and $m_{1/2}$ are compatible with a combination
of direct searches, flavor and electroweak precision observables as well
as astrophysical data. Upper bounds on $m_0$ at the few~TeV level
could possibly be set by including the findings of \citere{MFV-CMSSM}
into a global CMSSM analysis.

In \reffi{fig:BRhbs} one can see that for most of parameter space values
of \order{10^{-7}} are found for \brhbs, i.e.\ outside the reach of
current or future collider experiments. Even for the ``most
  extreme'' set of parameters we have analyzed, $\tb = 45$ and 
$A_0 = -3000 \gev$, no detectable rate has been found. Turning the
argument around, any observation of the decay \hbs\ at the (discussed)
future experiments would exclude the CMSSM as a possible model.

\section{Conclusions}
\label{sec:conclusions}

We have investigated the flavor violating Higgs boson decay \hbs\ in the
MSSM. This evaluation improves on existing analyses in various ways. We
take into account the full set of SUSY QCD and SUSY EW corrections,
allowing for LL, RL, LR and RR mixing simultaneously. The parameter
space is restricted not only by $B$-physics observables, but also by
electroweak precision observables, in particular the mass of the
$W$~boson. Here we have shown that $\MW$ can
yield non-trivial, additional restrictions on the parameter space of the
flavor violating $\deFABij$. 

From the technical side we have (re-)caculated the decay \hbs\ in the
\fa\ and \fc\ setup. The BPO and EWPO constraints have been evalated
with the help of (a private version of) \fh, taking into account the
full flavor violating one-loop corrections to $\MW$ and to the relevant
$B$-physics observables (supplemented with further MSSM higher-order
corrections). In the GUT based models the low-energy spectra have been
evaluated with the help of {\tt Spheno}. 

The first part of the numerical analysis used a model independent
approach. In six representative scenarios, which are allowed by current
searches for SUSY particles and heavy Higgs bosons, we have evaluated
the allowed parameter space for the various $\deFABij$ by applying BPO
and EWPO constraints. Within these allowed ranges we have then evaluated
\brhbs. In the case of only one $\deFABij \neq 0$ we have found that
only relatively large values of \del{DLR}{23} could lead to rates of 
$\brhbs \sim 10^{-4}$, which could be in the detectable range of future
$e^+e^-$ colliders. Allowing two $\deFABij \neq 0$
simultaneously lead to larger values up to $\brhbs \sim 10^{-3}$,
which would make the observation at the ILC relatively easy. Allowing
for a third $\deFABij \neq 0$, on the other hand, did not lead to
larger values of the flavor violating branching ratio.

In the final step of the numerical analysis we have evaluated \brhbs\ in
the MFV Constrained MSSM. In this model the flavor violation is induced
by CKM effects in the RGE running from the GUT to the EW scale. Here we
have found that also for the ``most extreme'' set of parameters we
have analyzed, $A_0 = -3000 \gev$ and $\tb = 45$, only negligible
effects can be expected. Turning the argument around, detecting a
non-zero value for \brhbs\ at (the discussed) future experiments would
exclude the CMSSM as a viable model.

\vspace{-0.5em}
\subsection*{Acknowledgments}

The work of S.H.\ and M.R.\ was partially supported by CICYT (grant FPA
2013-40715-P). 
M.G., S.H.\ and M.R.\ were supported by 
the Spanish MICINN's Consolider-Ingenio 2010 Programme under grant
MultiDark CSD2009-00064. 
M.E.G.\  acknowledges further support from the
MICINN project FPA2011-23781 and FPA2014-53631-C2-2-P.






\begin{thebibliography}{99} 


\bibitem{GIM} S.~Glashow, J.~Iliopoulos and L.~Maiani, 
              {\em Phys. Rev.} {\bf D2} (1970) 1285.

\bibitem{mssm} H.~Nilles, 
               {\em Phys.\ Rept.} {\bf 110} (1984) 1; \\ 
               H.~Haber and G.~Kane, 
               {\em Phys.\ Rept.} {\bf 117} (1985) 75; \\  
               R.~Barbieri, 
               {\em Riv.\ Nuovo Cim.} {\bf 11} (1988) 1. 

\bibitem{HFAgroup}  Y.~Amhis et al.\ [Heavy Flavor Averaging Group], 
                    arXiv:1412.7515v1 [hep-ex] 

\bibitem{MFV1} R.~Chivukula and H.~Georgi, 
               {\em Phys. Lett.} {\bf B 188} (1987) 99;\\
 L.~Hall and L.~Randall, 
 {\em Phys. Rev. Lett.} {\bf 65} (1990) 2939;\\ 
 A.~Buras et al., 
 {\em Phys. Lett.} {\bf B 500} (2001) 161.

\bibitem{MFV2} G.~D'Ambrosio et al., 
               {\em Nucl. Phys.} {\bf B 645} (2002) 155.

\bibitem{HdecNMFV} S.~Bejar, F.~Dilme, J.~Guasch and J.~Sola,
                   {\em JHEP} {\bf 0408} (2004) 018
                   [arXiv:hep-ph/0402188].

\bibitem{SUSY-QCD} A.~Curiel, M.~Herrero and D.~Temes,
                   {\em Phys.\ Rev.} {\bf D 67} (2003) 075008
                   [arXiv:hep-ph/0210335].

\bibitem{Demir} D.~Demir,
                {\em Phys.~Lett.} {\bf B 571} (2003) 193
                [arXiv:hep-ph/0303249].

\bibitem{SUSY-EW}   A.~Curiel, M.~Herrero, W.~Hollik, F.~Merz and 
                    S.~Pe{\~n}aranda, 
                    {\em Phys. Rev.} {\bf D 69} (2004) 075009
                    [arXiv:hep-ph/0312135].

\bibitem{SUSY-EW-RR}
  G.~Barenboim, C.~Bosch, J.~Lee, M.~López-Ibáñez and O.~Vives,
  arXiv:1507.08304 [hep-ph].

\bibitem{drhoLFV} M.~G{\'o}mez, T.~Hahn, S.~Heinemeyer, M.~Rehman, 
                  {\em Phys.\ Rev.} {\bf D 90} (2014) 074016 
                  [arXiv:1408.0663 [hep-ph]]  

\bibitem{arana-LFV} M.~Arana-Catania, S.~Heinemeyer and M.~Herrero,
  {\em Phys.\ Rev.} {\bf D 88} (2013) 1,  015026
  [arXiv:1304.2783 [hep-ph]].

\bibitem{arana} M.~Arana-Catania, S.~Heinemeyer, M.~Herrero and S.~Pe\~naranda,
                {\em JHEP} {\bf 1205} (2012) 015
                [arXiv:1109.6232 [hep-ph]];
                arXiv:1201.6345 [hep-ph].

\bibitem{arana-NMFV2} M.~Arana-Catania, S.~Heinemeyer and M.~Herrero,
  {\em Phys.\ Rev.} {\bf D 90} (2014) 075003
  [arXiv:1405.6960 [hep-ph]].

\bibitem{MFV-CMSSM} M.~G{\'o}mez, S.~Heinemeyer and M.~Rehman, 
                    {\em Eur. Phys. J.} {\bf C} (2015) 9, 434
                    [arXiv:1501.02258 [hep-ph]].

\bibitem{Ibrahim:2007fb}
  T.~Ibrahim and P.~Nath,
  {\rm Rev.\ Mod.\ Phys.}  {\bf 80} (2008) 577
  [arXiv:0705.2008 [hep-ph]].

\bibitem{Pospelov:2005pr}
  M.~Pospelov and A.~Ritz,
  {\em Annals Phys.} {\bf 318} (2005) 119
  [arXvi:hep-ph/0504231].

\bibitem{rge} N.~Falck, 
              {\em Z. Phys.} {\bf C 30} (1986) 247.

\bibitem{bertolini}  S.~Bertolini, F.~Borzumati, A.~Masiero, and G.~Ridolfi,
                     {\em Nucl. Phys.} {\bf B 353} (1991) 591.

\bibitem{feynarts} J.~K\"ublbeck, M.~B\"ohm and A.~Denner, 
                   {\em Comput. Phys. Commun.} {\bf 60} (1990) 165;\\
                   T.~Hahn, 
                   {\em Comput. Phys. Commun.} {\bf 140} (2001) 418
                   [arXiv:hep-ph/0012260].

\bibitem{famssm}   T.~Hahn and C.~Schappacher, 
                   {\em Comput. Phys. Commun.} {\bf 143} (2002) 54
                   [arXiv:hep-ph/0105349].\\
                   The program and the user's guide 
                   are available via {\tt www.feynarts.de} .

\bibitem{formcalc} T.~Hahn and M.~P\'erez-Victoria,
                   {\em Comput. Phys. Commun.} {\bf 118} (1999) 153
                   [arXiv:hep-ph/9807565].

\bibitem{feynhiggs} S.~Heinemeyer, W.~Hollik and G.~Weiglein,
                   {\em Comput. Phys. Commun.} {\bf 124} (2000) 76
                   [arXiv:hep-ph/9812320];\\
                   T.~Hahn, S.~Heinemeyer, W.~Hollik, H.~Rzehak and
                   G.~Weiglein, 
                   {\em Comput.\ Phys.\ Commun.} {\bf 180} (2009) 1426;
                   see {\tt www.feynhiggs.de} .

\bibitem{mhiggslong} S.~Heinemeyer, W.~Hollik and G.~Weiglein,
                    {\em Eur. Phys. J.} {\bf C 9} (1999) 343
                    [arXiv:hep-ph/9812472].

\bibitem{mhiggsAEC} G.~Degrassi, S.~Heinemeyer, W.~Hollik,
                   P.~Slavich and G.~Weiglein,
                   {\em Eur. Phys. J.} {\bf C 28} (2003) 133
                   [arXiv:hep-ph/0212020].

\bibitem{mhcMSSMlong}
                   M.~Frank, T.~Hahn, S.~Heinemeyer, W.~Hollik, 
                   R.~Rzehak and G.~Weiglein,
                   {\em JHEP} {\bf 0702} (2007) 047
                   [arXiv:hep-ph/0611326].

\bibitem{Mh-logresum}  T.~Hahn, S.~Heinemeyer, W.~Hollik, H.~Rzehak and
  G.~Weiglein,
  {\em Phys.\ Rev.\ Lett.} {\bf 112} (2014) 14,  141801
  [arXiv:1312.4937 [hep-ph]].

\bibitem{ILCreview}
  G.~Moortgat-Pick et al.,
  {\em Eur. Phys. J.} {\bf C 75} (2015) 8, 371
  [arXiv:1504.01726 [hep-ph]].

\bibitem{Isidori:2002qe}
  G.~Isidori and A.~Retico,
  {\em JHEP} {\bf 0209} (2002) 063
  [arXiv:hep-ph/0208159].

\bibitem{Chankowski:2000ng}
  P.~Chankowski and L.~Slawianowska,
  {\em Phys.\ Rev.} {\bf D 63} (2001) 054012
  [arXiv:hep-ph/0008046].

\bibitem{Foster:2005wb}
  J.~Foster, K.~Okumura and L.~Roszkowski,
  {\em JHEP} {\bf 0508} (2005) 094
  [arXiv:hep-ph/0506146].

\bibitem{sufla}
   G.~Isidori and P.~Paradisi,
  {\em Phys.\ Lett.} {\bf B 639} (2006) 499
  [arXiv:hep-ph/0605012];\\
  G.~Isidori, F.~Mescia, P.~Paradisi and D.~Temes,
  {\em Phys.\ Rev.} {\bf D 75} (2007) 115019
  [arXiv:hep-ph/0703035], and references therein.

\bibitem{hfag:rad}
See: {\tt https://www.slac.stanford.edu/xorg/hfag/rare/2013/radll/\\
OUTPUT/TABLES/radll.pdf}~.

\bibitem{Misiak:2009nr}
  M.~Misiak,
  {\em Acta Phys.\ Polon.} {\bf B 40} (2009) 2987
  [arXiv:0911.1651 [hep-ph]].

\bibitem{Chatrchyan:2013bka}
  S.~Chatrchyan et al.\  [CMS Collaboration],
  {\em Phys.\ Rev.\ Lett.} {\bf 111} (2013) 101804
  [arXiv:1307.5025 [hep-ex]].

\bibitem{Aaij:2013aka}
  R. Aaij et al.\  [LHCb Collaboration],
  {\em Phys.\ Rev.\ Lett.} {\bf 111} (2013) 101805
  [arXiv:1307.5024 [hep-ex]].

\bibitem{Buras:2012ru}
  A.~Buras, J.~Girrbach, D.~Guadagnoli and G.~Isidori,
  {\em Eur.\ Phys.\ J.} {\bf C 72} (2012) 2172
  [arXiv:1208.0934 [hep-ph]].

\bibitem{hfag:pdg}
See: {\tt https://www.slac.stanford.edu/xorg/hfag/osc/PDG\_2013/}~.

\bibitem{Buras:1990fn}
  A.~Buras, M.~Jamin and P.~Weisz,
  {\em Nucl.\ Phys.} {\bf B 347} (1990) 491.

\bibitem{Golowich:2011cx}
  E.~Golowich, J.~Hewett, S.~Pakvasa, A.~Petrov and G.~Yeghiyan,
  {\em Phys.\ Rev.} {\bf D 83} (2011) 114017
  [arXiv:1102.0009 [hep-ph]].

\bibitem{bmm-CMS-LHCb}
  CMS and LHCb Collaborations, 
  CMS-PAS-BPH-13-007, LHCb-CONF-2013-012, CERN-LHCb-CONF-2013-012.

\bibitem{PomssmRep} S.~Heinemeyer, W.~Hollik and G.~Weiglein,
                    {\em Phys.\ Rept.} {\bf 425} (2006) 265
                    [arXiv:hep-ph/0412214].

\bibitem{LEPEWWG} S.~Schael et al. [ALEPH and DELPHI and L3 and OPAL and LEP
  Electroweak Collaborations], 
  {\em Phys.\ Rept.} {\bf 532} (2013) 119
  [arXiv:1302.3415 [hep-ex]];\\
  see {\tt http://www.cern.ch/LEPEWWG}~.

\bibitem{Baak:2013fwa}
M.~Baak et al., arXiv:1310.6708 [hep-ph];\\
A.~Freitas et al., arXiv:1307.3962 [hep-ph].

\bibitem{fcc-ee-paris} S.~Heinemeyer, 
  talk given at the {\em 8$^{th}$ FCC-ee Physics Workshop}, 
  Paris, France, October 2014, see:
  {\tt
https://indico.cern.ch/event/337673/session/3/contribution/\\
41/material/slides}~.

\bibitem{rho} M.~Veltman, 
              {\em Nucl. Phys.} {\bf B 123} (1977) 89. 

\bibitem{delrhoNMFV} S.~Heinemeyer, W.~Hollik, F.~Merz and S.~Pe\~naranda,
                     {\em Eur. Phys. J.} {\bf C 37} (2004) 481
                     [arXiv:hep-ph/0403228]. 

\bibitem{Drho2LQCD}
  A.~Djouadi, P.~Gambino, S.~Heinemeyer, W.~Hollik, C.~J\"unger and G.~Weiglein,
  {\em Phys.\ Rev.\ Lett.} {\bf 78} (1997) 3626
  [arXiv:hep-ph/9612363];
  {\em Phys.\ Rev.} {\bf D 57} (1998) 4179
  [arXiv:hep-ph/9710438].

\bibitem{Heinemeyer:2006px} S.~Heinemeyer, W.~Hollik, D.~St\"ockinger, 
                            A.~Weber, and G.~Weiglein
			    {\em JHEP} {\bf 08} (2006) 052
			    [arXiv:hep-ph/0604147]

\bibitem{Haestier:2005ja} J.~Haestier, S.~Heinemeyer, D.~St\"ockinger 
                          and G.~Weiglein
			 {\em JHEP} {\bf 0512} (2005) 027
			 [arXiv:hep-ph/0508139]

\bibitem{Haber:1989xc} 
  H.~Haber and Y.~Nir,
  {\em Nucl.\ Phys.} {\bf B 335} (1990) 363.

\bibitem{LHCHiggs}
  M.~D\"uhrssen, talk given at `Rencontres de Moriond EW 2015'', see:\\
{\tt
  https://indico.in2p3.fr/event/10819/session/3/contribution/102/material/}\\
{\tt slides/1.pdf}~.

\bibitem{higgsbounds} P.~Bechtle, O.~Brein, S.~Heinemeyer, G.~Weiglein
  and K.~Williams, 
  {\em Comput.\ Phys.\ Commun.} {\bf 181} (2010) 138
  [arXiv:0811.4169 [hep-ph]];
  {\em Comput.\ Phys.\ Commun.} {\bf 182} (2011) 2605
  [arXiv:1102.1898 [hep-ph]];\\
P.~Bechtle, O.~Brein, S.~Heinemeyer, O.~St{\aa}l, T.~Stefaniak, G.~Weiglein
and K.~Williams,
  {\em Eur.\ Phys.\ J.} {\bf C 74} (2014) 2693
  [arXiv:1311.0055 [hep-ph]].

\bibitem{Porod:2003um} W.~Porod, 
  {\em Comput. Phys. Commun.} {\bf 153} (2003) 275
  [arXiv:hep-ph/0301101];\\
  W.~Porod and F.~Staub,
  {\em Comput.\ Phys.\ Commun.} {\bf 183} (2012) 2458
  [arXiv:1104.1573 [hep-ph]].

\bibitem{SLHA} P.~Skands et al.,
  {\em JHEP} {\bf 0407} (2004) 036
  [arXiv:hep-ph/0311123];\\
  B.~Allanach et al.,
  {\em Comput.\ Phys.\ Commun.} {\bf 180} (2009) 8
  [arXiv:0801.0045 [hep-ph]].

\bibitem{mc9}
  O.~Buchmueller et al.,
  {\em Eur.\ Phys.\ J.} {\bf C 74} (2014) 6,  2922
  [arXiv:1312.5250 [hep-ph]];
  arXiv:1508.01173 [hep-ph].

\end{thebibliography}
\end{document}